\documentclass[useAMS,usenatbib]{mn2e}
\usepackage{epsfig}

\def\gsim{\;\rlap{\lower 2.5pt\hbox{$\sim$}}\raise 1.5pt\hbox{$>$}\;}
\def\lsim{\;\rlap{\lower 2.5pt\hbox{$\sim$}}\raise 1.5pt\hbox{$<$}\;}
\def\dML{\mbox{$\nabla_{\ell} \Upsilon$}}

\renewcommand{\arcsec}{\mbox{$^{\prime\prime}$}}

\newcommand{\degree}{\mbox{$^\circ$}}
\newcommand{\kms}{\mbox{\,km~s$^{-1}$}}

\renewcommand{\mag}{\mbox{\,mag}}

\newcommand{\othree}{[O~{\sc iii}]}
\newcommand{\otwo}{[O~{\sc ii}]}

\newcommand{\PNS}{PN.S}

\newcommand{\FWHM}{\mbox{\rm FWHM}}

\newcommand{\VD}{\mbox{\rm VD}}

\renewcommand{\Re}{\mbox{$\,R_{\rm e}$}}
\newcommand{\Ang}{\mbox{\,\AA}}

\newcommand{\LCDM}{\mbox{$\Lambda$CDM}}

\def\vir{\mbox{$_{\rm vir}$}}
\def\kpc{\mbox{\,\rm kpc}}
\def\mag{\mbox{\,\rm mag}}
\def\Msun{\mbox{$M_\odot$}}

\def\Ystar{\mbox{$\Upsilon_*$}}

\def\Ysol{\mbox{$\Upsilon_{\odot, B}$}}

\def\Rm{\mbox{$R$}}

\title[PNe and Dark Matter in NGC~4494]{The PN.S Elliptical Galaxy Survey: \\
the dark matter in NGC~4494\thanks{
Based on observations made with the William Herschel Telescope operated on the island of La Palma by the Isaac Newton Group in the Spanish Observatorio del Roque de los Muchachos of the Instituto de Astrofisica de Canarias and MMT Observatory, a joint facility of the Smithsonian Institution and the University of Arizona.}}

\author[Napolitano et al.]{\noindent
N.R.~Napolitano$^{1}$\thanks{E-mail: napolita@na.astro.it (NRN); mc@na.infn.it (MC)},
A.J.~Romanowsky$^{2,3}$,
L.~Coccato$^{4}$,
M.~Capaccioli$^{5,6}$,\and
N.G.~Douglas$^{7}$,
E.~Noordermeer$^{8}$,
O.~Gerhard$^{4}$,
M.~Arnaboldi$^{9,10}$
F.~De Lorenzi$^{4}$,\and
K.~Kuijken$^{11}$,
M.R.~Merrifield$^{8}$,
E.~O'Sullivan$^{12}$,
A. Cortesi$^{8}$,
P.~Das$^{4}$,
K.C.~Freeman$^{13}$
\\~\\
$^1$ INAF-Observatory of Capodimonte, Salita Moiariello, 16, 80131, Naples, Italy\\
$^2$ UCO/Lick Observatory, University of California, Santa Cruz, CA 95064, USA\\
$^3$ Departamento de F\'{i}sica, Universidad de Concepci\'{o}n, Casilla 160-C, Concepci\'on, Chile\\
$^4$ Max-Planck-Institut f\"ur Extraterrestriche Physik, Giessenbachstrasse, D-85748 Garching b. M\"unchen, Germany\\
$^5$ Dipartimento di Scienze Fisiche, Universit`a Federico II, Via Cinthia, 80126, Naples, Italy \\
$^6$ INAF - VSTceN, Salita Moiariello, 16, 80131, Naples, Italy\\
$^7$ Kapteyn Astronomical Institute, Postbus 800, 9700 AV Groningen, The Netherlands\\
$^8$ School of Physics and Astronomy, University of Nottingham, University Park, Nottingham NG7 2RD, UK\\
$^9$ European Southern Observatory, Karl-Schwarzschild-Strasse 2,
D-85748 Garching, Germany\\
$^{10}$ INAF, Osservatorio Astronomico di Pino Torinese, I-10025 Pino
Torinese, Italy\\
$^{11}$ Leiden Observatory, Leiden University, PO Box 9513, 2300RA Leiden, The Netherlands \\
$^{12}$ Harvard-Smithsonian Center for Astrophysics, 60 Garden Street, Cambridge, MA 02138, USA\\
$^{13}$ Research School of Astronomy \& Astrophysics, ANU, Canberra, Australia}

\begin{document}

\date{Accepted  Received }

\pagerange{\pageref{firstpage}--\pageref{lastpage}} \pubyear{2008}

\maketitle

\label{firstpage}

\begin{abstract}
We present new Planetary Nebula Spectrograph observations of the ordinary elliptical galaxy NGC~4494,
resulting in positions and velocities of 255 PNe out to 7 effective radii (25 kpc).
We also present new wide-field surface photometry from MMT/Megacam, and
long-slit stellar kinematics from VLT/FORS2.
The spatial and kinematical distributions of the PNe agree with the field stars in the region of overlap.
The mean rotation is relatively low, with a possible kinematic axis twist outside
$1 \Re$.  The velocity dispersion profile declines with radius, though not very steeply,
down to $\sim 70\kms$ at the last data point.

We have constructed spherical dynamical models of the system, including
Jeans analyses with
multi-component $\Lambda$CDM-motivated galaxies as well as logarithmic potentials.
These models include special attention to orbital anisotropy, which  we constrain
using fourth-order velocity moments.
Given several different sets of modelling methods and assumptions, we find
consistent results for the mass profile within the radial range constrained by the data.
Some dark matter (DM) is required by the data;
our best-fit solution has a radially anisotropic
stellar halo, a plausible stellar mass-to-light ratio, and a DM halo with an
unexpectedly low central density.
We find that this result does not substantially change  with a flattened  axisymmetric model.

Taken together with other results for galaxy halo masses,
we find suggestions for a puzzling pattern wherein most intermediate-luminosity galaxies
have very low concentration halos,
while some high-mass ellipticals have very high concentrations.
We discuss some possible implications of these results for DM and galaxy formation.
\end{abstract}

\begin{keywords}
galaxies: elliptical and lenticular --- galaxies: kinematics and dynamics --- galaxies: structure --- galaxies: individual: NGC~4494 --- dark matter\ --- planetary nebulae: general
\end{keywords}

\section{Introduction}\label{sec:intro}
The idea that some sort of dark matter (DM) dominates the mass of the universe
arose primarily in the 1970s from observations of the outer kinematics of spiral galaxies \citep{1970ApJ...160..811F,1973ApJ...186..467O,1978ApJ...225L.107R,1981AJ.....86.1825B},
and is now well established on cosmological grounds \citep{2008arXiv0803.0732H}.
As part of the ``concordance'' \LCDM\ model---including cold dark matter (CDM)
and a cosmological constant ($\Lambda$)---DM has been invaluable
in describing the formation, evolution, and stability of cosmic structures.

Yet despite the successes of the concordance model on cosmological scales,
on the scales of galaxies and galaxy clusters, several observational
discrepancies continue to challenge cosmological complacency.
For instance, \LCDM\ simulations of the formation of DM haloes predict
a steep central cusp (\citealt{nfw96,nfw97}, hereafter NFW; \citealt{moore99}), at
variance with many observations of late-type galaxies
\citep{2003MNRAS.340..657D,2004MNRAS.351..903G,2005ApJ...634L.145G,2007ApJ...663..948G,2007MNRAS.378...41S,2008MNRAS.383..297S,2008ApJ...676..920K}.
It remains to be seen whether this discrepancy can be traced to observational
problems, to oversimplified predictions of DM halo properties (concerning in
particular the inclusion of baryonic effects), or to a failure of the \LCDM\ paradigm.

The default expectation for baryons is that their dissipation would exacerbate the
central DM density problem,
but it is possible that processes such as feedback in a complete galaxy formation
picture would {\it decrease} the DM densities.

Given the long-standing DM puzzles in late-type galaxies, it is increasingly
important to examine the mass content of
early-type galaxies (ETGs), i.e. ellipticals and lenticulars.
Since these objects are usually free of cold gas and have their stars moving in random
directions, their kinematics are more difficult to sample than in disc galaxies and
their dynamics harder to model.

The first obstacle comes from observations.
Outside $\sim1\mbox{--}2\Re$ (where the projected effective radius \Re\ is that
encircling half the total light of the galaxy),
the decreasing surface brightness makes kinematics measurements very difficult
(e.g. \citealt{2001MNRAS.326..473H,2005MNRAS.363..769S,2008ApJS..175..462C}).
DM constraints from this technique are
therefore possible for only a fraction of galaxies
(e.g. \citealt{2000A&AS..144...53K,2001AJ....121.1936G,2007MNRAS.382..657T}).

Globular clusters have classically been used as mass tracers in the haloes
of ``bright'' galaxies,
but the kinematical samples in ``ordinary'' ETGs \citep{1992MNRAS.259..323C}
have so far been too small for strong constraints
(e.g. \citealt{2004A&A...415..123P}; \citealt{2006A&A...448..155B}).
Similarly, mass studies using X-ray emission are biassed toward massive,
group or cluster-central galaxies, while ordinary ETGs are much more
difficult to probe
(see \citealt{2003ApJ...586..850P}; \citealt{2004MNRAS.349..535O};
\citealt{2006MNRAS.370.1797P}; \citealt{2006ApJ...646..899H};
\citealt{2007ApJ...667..731P}).

Hence it is not surprising that there is no unbiassed, systematic survey of the
detailed DM properties of ETGs, comparable to what is available
for late-types \citep{1996MNRAS.281...27P}.

A way forward is offered by planetary nebulae (PNe),
which have become established probes of the stellar kinematics in ETG haloes,
where their bright 5007\Ang\ \othree\ line stands out against the
faint galaxy background
\citep[e.g.][]{1993ApJ...414..454C,arn96,1998ApJ...507..759A,nap02,2004ApJ...602..685P}.
In addition to their use as large-radius mass probes, PNe have the
additional advantage of tracing the kinematics of the underlying
field stars, and so their orbital properties can be used to constrain
the formation history of the bulk of the host galaxies' stars.
In the last few years, a novel observational technique of ``counter-dispersed imaging''
\citep{1999MNRAS.307..190D} has overtaken the traditional multi-object spectroscopic
follow-up of PNe candidates discovered in narrow-band images.
This efficient approach to collecting large samples of PN velocities relies on
the comparison of pairs of slitless spectroscopic images taken with the dispersion
element rotated by $180\degree$ between the exposures, and has been
successfully implemented with several different variations
\citep{2000MNRAS.316..795D, 2001ApJ...563..135M,2005ApJ...635..290T}.

The Planetary Nebula Spectrograph (\PNS; \citealt{2002PASP..114.1234D})
is the first custom-designed instrument for counter-dispersed imaging.
This instrument's two spectrographic cameras, fed by a beam-splitter
arrangement at the grating module, operate simultaneously, securing image pairs
from the same collimated beam and via the same filter at the same temperature.
The \PNS\ has begun producing large kinematical samples of PNe in a variety
of galaxy types (\citealt{2006MNRAS.369..120M};
\citealt{2003Sci...301.1696R}, hereafter R+03;
\citealt{2007ApJ...664..257D}, hereafter D+07;
\citealt{2008MNRAS.384..943N},
\citealt{Coccato08}).
The first results from a small sample of ordinary ellipticals
yielded a surprisingly pseudo-Keplerian decline in the observed
velocity dispersion profile, more suggestive of galaxies with
a constant mass-to-light ratio ($M/L$) than of DM-dominated systems
(\citealt{1993ApJ...414..454C}; \citealt{2001ApJ...563..135M};
R+03; \citealt{2006AJ....131.2089S}; D+07).

These PN observations have sparked several main interpretations.
The first is that the data are compatible with $\Lambda$CDM expectations
(\citealt{2005MNRAS.363..705M}, hereafter M{\L}05;
\citealt{2005Natur.437..707D}, hereafter D+05;
\citealt{2007MNRAS.376...39O}).
The second is that the central DM concentrations of many ordinary ETGs
are unexpectedly low \citep[hereafter N+05]{2005MNRAS.357..691N}, akin to similar
conclusions among late-types \citep[]{2006ApJ...643..804K,2007ApJ...659..149M}.
The third is that a modified form of gravity, rather than DM, dominates
the outer regions of galaxies
\citep{2003ApJ...599L..25M,2006ApJ...636..721B,2007A&A...476L...1T}.

The issues raised by D+05 in particular (relating to anisotropy,
viewing angle, and star-PN connections) were explored by R+03 and D+07
in the context of NGC~3379---the paragon of an ordinary elliptical,
with so far the best available kinematical data.
They concluded that the $\Lambda$CDM-based interpretations were not convincing,
although the situation may be less clear under more
general dynamical modelling assumptions \citep[hereafter DL+08b]{2008arXiv0804.3350D}.
DL+08b explored flattened galaxy models in detail, and found a fairly wide
range of mass profiles permitted for NGC~3379.
Although they did not directly test $\Lambda$CDM-based mass models, they
made an {\it a posteriori} comparison to the D+05 simulations. They found
that the data permitted total mass models comparable to D+05, but that these
required stronger radial anisotropy than found in the simulations.

With various efforts underway to elucidate the implications of the data,
there also remains the task of producing the observational results for a
broad sample of galaxies.
Here we continue toward this goal by presenting new photometric, kinematical,
and dynamical results from a second galaxy with preliminary results in R+03:
NGC~4494.

NGC~4494\ is a LINER E1 galaxy located in the Coma~I cloud
on the periphery of the Virgo cluster,
at a distance of 15.8~Mpc and with a total luminosity of $M_B\sim-20.5$,
a value close to the $L^*$ characteristic luminosity.
These and other global parameters are listed and referenced
in Table~\ref{tab:galpar}.
The stellar component of NGC~4494 is somewhat discy,
with a steep central cusp and signification rotation,
exhibiting few peculiarities other than
central discs of dust ($\sim$~100 pc) and kinematically-decoupled stars
($\sim$~400 pc),
which may imply a recent merging episode
\citep{1994MNRAS.269..785B,1997ApJ...481..710C,2001AJ....121.2928T,2007ApJ...664..226L}.
X-ray observations reveal very little hot gas in the halo,
implying either a recent interaction which has depleted the gas,
or a very low-mass DM halo
\citep{2004MNRAS.349..535O,2006ApJ...636..698F}.

\begin{table}
\caption{NGC~4494: basic data.}
\vspace{-0.5cm}
\label{tab:galpar}
\begin{center}
\noindent{\smallskip}\\
\begin{tabular}{lll}
\hline
\hline
Parameter & Value  &  Reference \\
\hline
R.A. (J2000)     & 12$^{\rm h}$ 31$^{\rm m}$ 24$^{\rm s}$     & NED$^{1}$ \\
Decl. (J2000)     & +25$\degree$ 46$'$ 30$''$     & NED \\
$v_{\rm sys}$           & 1344\kms               & NED\\
$(m-M)_0$   & 31.0\mag               & \citet{2001ApJ...546..681T}$^{2}$\\
$A_B$       & 0.09\mag               & \citet{1998ApJ...500..525S} \\
$m_V$       & $9.74\pm0.10 \mag $ & Sec.~\ref{sec:spatdist}\\
$m_B$       & $10.55\pm0.10 \mag $ & Sec.~\ref{sec:spatdist}\\
$M_V$       & $-21.26\pm0.10 \mag $ & Sec.~\ref{sec:spatdist}\\
$M_B$       & $-20.5\pm0.10 \mag $ & Sec.~\ref{sec:spatdist}\\
$(B-V)_0$     & 0.81\mag               &  {\citet{1994A&AS..104..179G}} \\
$\Re$           &$48\arcsec.2\pm3\arcsec.0$      & Sec.~\ref{sec:spatdist}\\
$\sigma_0$    & 150 \kms & HyperLeda$^{3}$\\
$\epsilon$   & $0.162 \pm 0.001$ &  App.~\ref{app:photmega}\\
$a_4$        & $0.16 \pm 0.07$ & App.~\ref{app:photmega}\\
PA  & $-0.9^\circ \pm 0.3^\circ$ & App.~\ref{app:photmega}\\
\hline
\end{tabular}
\noindent{\smallskip}\\

\begin{minipage}{9.5cm}

NOTES -- (1):
http://nedwww.ipac.caltech.edu/\\
         (2): corrected by -0.16 mag (see \citealt{2003ApJ...583..712J})\\
     (3): http://leda.univ-lyon1.fr/ \citep{2003A&A...412...45P}.
\end{minipage}
\noindent{\smallskip}\\
\end{center}
\end{table}

Our new study of NGC~4494 follows the pattern of reduction and basic analysis of
NGC~3379 in D+07, but goes further in establishing a stronger framework for
deriving and interpreting the mass profile.
These include the critical aspects of exploiting higher-order velocity
information in the data to break the mass-anisotropy degeneracy,
and fitting a suite of physically-motivated multi-component mass models.

The paper is organized as follows.
Section~\ref{sec:data} presents a standard reduction
of the PN.S observations of NGC~4494.
In Section~\ref{sec:dist} we present the spatial and kinematical properties of the
PN system, comparing them to newly acquired long-slit stellar data.
We analyse the system's dynamics in Section~\ref{sec:dynamics}, and
put the mass results in context with other galaxies in Section~\ref{sec:impl}.
Section~\ref{sec:concl} summarizes.
New surface photometry for NGC~4494 is tabulated in Appendix~\ref{app:photmega}, and
some technical aspects of the dynamics are described in Appendices \ref{app:eqs} and
\ref{app:flatt}.

\section{\PNS\ observations of NGC~4494}\label{sec:data}

We present a new set of \PNS\ data on NGC~4494, starting with a description of
the observations and basic data reduction in \S\ref{sec:obs}, and
addressing the completeness and possible contamination of the sample
in \S\ref{sec:compcont}.

\subsection{Observations and data reduction}\label{sec:obs}

\PNS\ observations of NGC~4494\ were performed during two runs (March 2002 and
February-March 2003; Table~\ref{tab:obs}) under a variety of seeing conditions,
for a total integration time of $\sim 14$~hours (equivalent to 6.2~hours at a
constant seeing of $\FWHM=1\arcsec$,
for a given $S/N$
in the final co-added image\footnote{In background-limited observations,
$S/N\propto \sqrt{\mbox{\sc exptime}}$ and $S/N \propto {\mbox{\sc seeing}}^{-1}$,
so {\sc exptime}~$\propto{\mbox{\sc seeing}}^{-2}$.}).
Preliminary results from the 2002 data were presented in R+03.

\begin{table}
\caption{Log of NGC~4494\ \PNS\ observations }
\label{tab:obs}
\begin{center}
\setlength\tabcolsep{5pt}
\begin{tabular}{lcc}
\hline \hline
\noalign{\smallskip}
Date & Integration & Seeing \\
 & [hours] & [\FWHM] \\
\hline
2002 March 7    & 1.3 & 1.9\arcsec \\
2002 March 8    & 1.0 & 1.6\arcsec \\
2002 March 9    & 2.8 & 1.6\arcsec \\
2002 March 13   & 0.7 & 2.1\arcsec \\
2002 March 14   & 5.3 & 1.4\arcsec\\
2003 February 28    & 0.5 & 1.6\arcsec \\

2003 March 1    & 0.5 & 1.6\arcsec \\
2003 March 2    & 0.5 & 1.6\arcsec \\
2003 March 3    & 1.2 & 1.8\arcsec \\
2003 March 4    & 0.5 & 1.0\arcsec \\
\hline
\end{tabular}
\end{center}
\end{table}

The systemic velocity of NGC~4494\ (see Table~\ref{tab:galpar}) was matched with
our ``B'' narrow-band (\FWHM$=31\Ang$) filter tilted by $6\degree$ to achieve a
shift of $7.16\Ang$ of its nominal central wavelength $\lambda_c=5033.9\Ang$ (see
also \citealt{2002PASP..114.1234D} for further details).
The instrument position angle was set to $0\degree$, such that the grating
dispersed in the North-South direction.
\begin{table*}
\caption{Catalogue of PNe in NGC~4494}
\begin{center}
\setlength\tabcolsep{5pt}
\begin{tabular}{lcccr}
\hline \hline
\noalign{\smallskip}

ID & R.A. & Decl. & Wavelength & $v_{\rm helio}$  \\

 PNS-EPN- & (J2000) & (J2000) & [\Ang] & [\kms] \\
\hline\\
 NGC4494 1 & 12:31:45.3 & 25:47:26.9 & 5029.3 & 1349     \\
 NGC4494 2 & 12:31:45.4 & 25:45:20.5 & 5029.4 & 1358     \\
 NGC4494 3 & 12:31:44.1 & 25:48:50.0 & 5029.8 & 1377     \\
 NGC4494 4 & 12:31:43.7 & 25:49:07.5 & 5029.1 & 1341     \\
 NGC4494 5 & 12:31:43.5 & 25:45:47.4 & 5028.7 & 1314     \\
 NGC4494 6 & 12:31:42.8 & 25:45:46.2 & 5027.6 & 1250     \\
 NGC4494 7 & 12:31:42.5 & 25:47:02.8 & 5028.3 & 1290     \\
 NGC4494 8 & 12:31:41.6 & 25:48:28.9 & 5017.9 & 670             \\
 NGC4494 9 & 12:31:41.2 & 25:46:04.5 & 5028.2 & 1275     \\
 NGC4494 10 & 12:31:40.0 & 25:47:29.4 & 5029.2 & 1346    \\
 NGC4494 11 & 12:31:36.3 & 25:49:59.1 & 5026.8 & 1199    \\
 NGC4494 12 & 12:31:35.2 & 25:47:56.3 & 5030.8 & 1437    \\
\multicolumn{1}{c}{$\vdots$} &
\multicolumn{1}{c}{$\vdots$} &
\multicolumn{1}{c}{$\vdots$} &
\multicolumn{1}{c}{$\vdots$} &
\multicolumn{1}{c}{$\vdots$} \\
\hline\\
\end{tabular}\label{tab:cat}
\end{center}
\begin{minipage}{16 cm}
NOTES - The \PNS\ data follow the usual definitions, and the ID is
chosen to be compatible with proposed IAU standards (EPN standing for
``Extragalactic Planetary Nebula").
Velocities include the heliocentric correction of 5.7 \kms.
\end{minipage}
\end{table*}

\begin{figure}
\epsfig{file=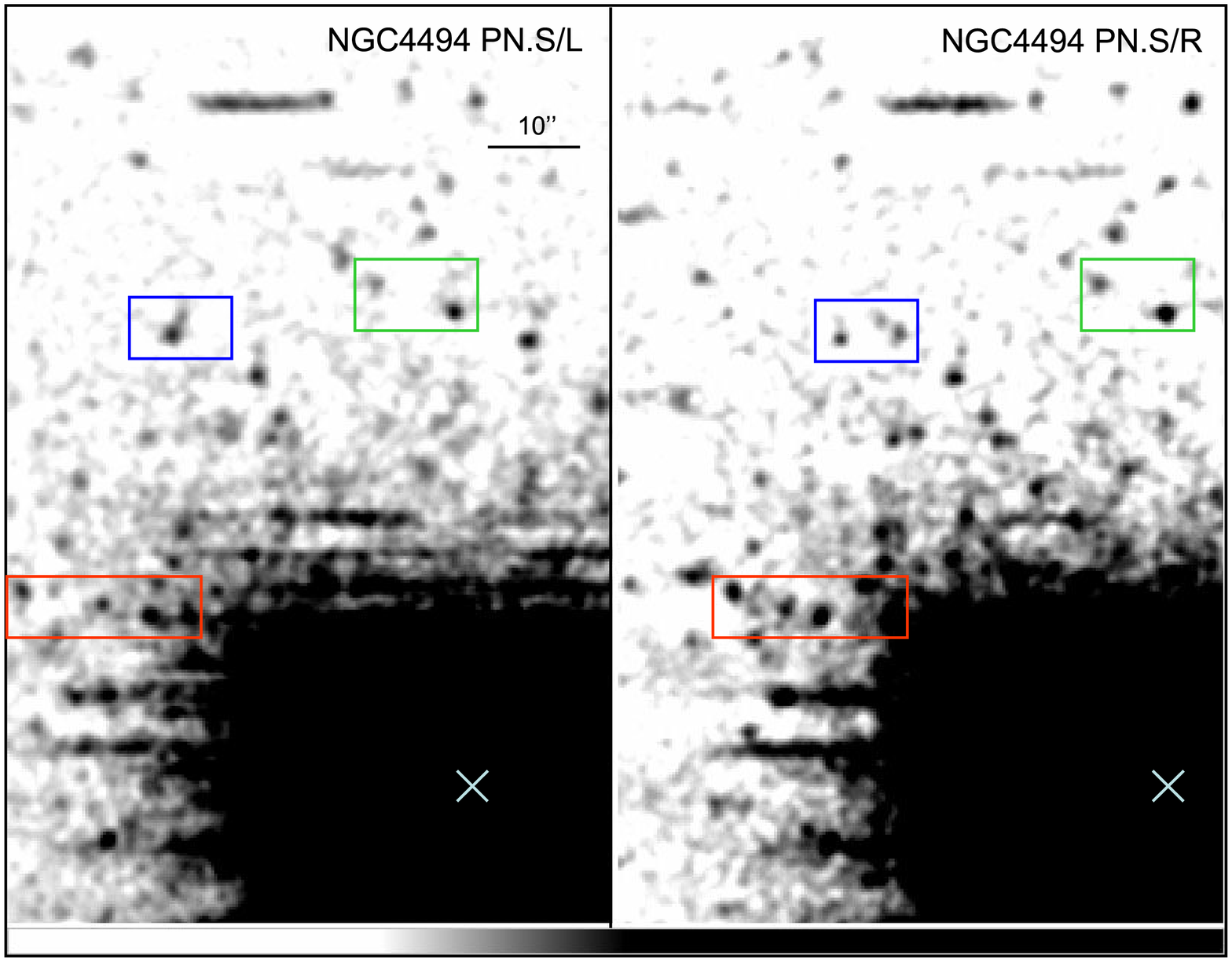,width=6.7cm,angle=-90}
\caption{Dispersed images of NGC~4494\ from the PN.S, with close-up of the
central regions; the ``Left'' and ``Right'' camera arm images are marked by
PN.S/L and PN.S/R.
The galaxy centre is marked by $\times$.
The images have been median subtracted and Gaussian smoothed in order to enhance the emission lines.
The stellar background precludes practically any detection for $R<20\arcsec$, and
the crowding is evident.
Some sample pair associations are framed:
in the green box two pairs clearly separated in the two arms;
in the blue box three emissions in the Right arm collapse into two in the Left arm;
in the red box there is a complicated set of 4 or 5 emissions.}
\label{fig:fig0}
\end{figure}
The observations were processed as detailed in Section 2 of D+07\footnote{
Data reduction is via a dedicated pipeline written in the {\sc iraf} script language
(distributed by NOAO, which is operated by AURA Inc., under contract
with the National Science Foundation), with some additional routines written in {\sc\ Fortran}.}.
An important step in the procedure is identifying PNe in the dispersed images
by their pointlike emission, since they are spatially and spectrally unresolved
objects.  Potential sources of contamination are noise, foreground and background
continuum objects, and background galaxies with line emission
(\otwo\ at redshift $z\sim0.35$ or Ly-$\alpha$ at $z\sim3.14$).
The latter two categories can often be identified by their extended shape in the
images, while  noise is rejected by requiring matching detections in both arms.
The point-like objects are extracted using the automated software
SExtractor \citep{1996A&AS..117..393B}, with extraction parameters optimized
by experimenting on simulated data sets.
The automated selection is complemented by an ``eyeball'' check on every PN
candidate by at least three observers.  This check also picks up some PNe missed
by SExtractor in certain regions of the image (near star-trails and the galaxy centre).

PN.S observations near galaxy centers are generally hampered by the
increasing dominance of the stellar background,
and by crowding as a consequence of the central concentration of the PNe themselves.
These effects lead to reduced detection completeness, as well as to confusion in
assigning PN pairs detected in the left and right arms,
and were more significant in NGC~4494 than in NGC~3379 because of
a poorer median seeing, higher stellar surface brightness,
and larger number of PNe in the case of the former (see Figure~\ref{fig:fig0}).
Fortunately, the central galactic regions are not of much concern to us,
since the power of the PN.S lies in probing the outer parts of galaxies.
We must simply keep in mind that the PN-based results inside $\sim$50$''$ should be
considered unreliable.

\begin{figure}
\hspace{-0.7cm}
\epsfig{file=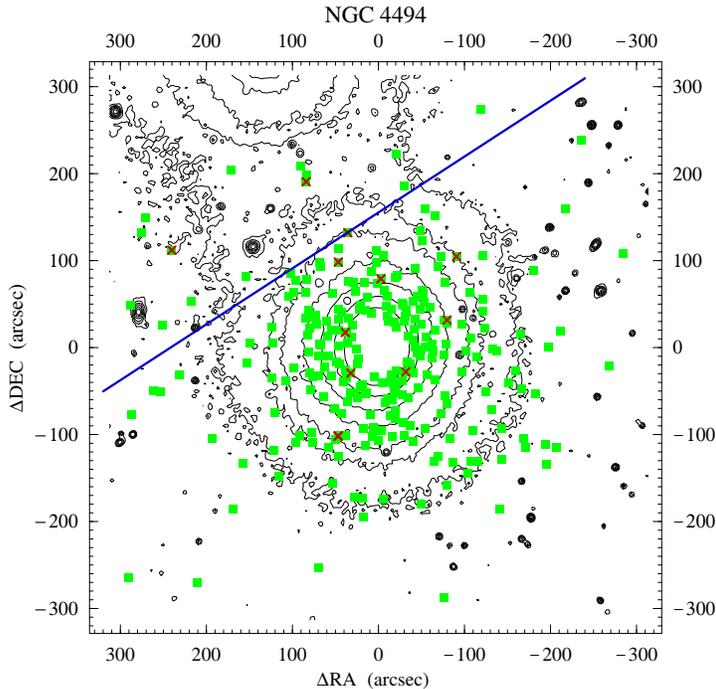,width=9.5cm}
\caption{Spatial distribution of objects around NGC~4494.
The PNe are marked as filled green boxes, with red $\times$ symbols showing
objects discarded as phase-space ``outliers'' (see text).
Surface brightness isophotes from (Megacam $g'$-band) are shown by contours,
and the blue solid line
marks the boundary where the surface photometry is unaffected by the bright
star to the Northeast.}
\label{fig:fig1}
\end{figure}

After rejecting $\sim$~30 objects for lack of unanimous approval from
the independent observers, we obtain a remarkable full sample of 267 PN candidates
(Table~\ref{tab:cat}).  These objects may still include a small number
of background line-emitter contaminants and pair mismatches
(see Section \ref{sec:compcont}).
Their spatial distribution is shown in Fig.~\ref{fig:fig1}.

\subsection{Completeness and contamination}
\label{sec:compcont}

It is important to understand the completeness of our PN sample as a function
of radius and magnitude (see \S\ref{sec:spatdist}).
Therefore, we followed the procedure used in D+07 for estimating the
fraction of sources missed by our PN search algorithm.
We added to our images a simulated sample of 400 point-like sources randomly placed,
with a luminosity function following that expected for the PNe of NGC~4494 and
covering a range up to 1~mag fainter than the bright-end cutoff (\citealt{1989ApJ...339...53C})
appropriate to this galaxy, $m^*\sim26.0\mag$ (\citealt{1996ApJ...462....1J}),
where the [O~III] magnitudes are defined as
$$
m_{5007}=-2.5 \log F_{5007}-13.74
$$
(\citealt{1989ApJ...339...39J}).
We ran the same SExtractor procedures used for real data, finding that
$\sim90\%$ of the sources were recovered outside $R\simeq3\Re$
($\simeq150''$).  The detectability decreased to $\sim75\%$ at $R\simeq\Re$,
falling sharply inside this radius.  Inside $R\simeq20\arcsec$, even the
brightest ($m^*$) PNe were not recovered.
In the real data, the completeness was higher because of the additional visual
checks, so we correct our completeness fraction upward by using the ratio of
automatically and visually recovered PNe in the real data, as a function of radius.
This correction is $\sim$15\% and $\sim$10\% in the inner and outer regions.
Overall, our 50\% completeness limit was $(m^*+1.2) \sim 27.2$~mag.

We next try to eliminate any residual false PN identifications, which could be due to
unresolved background emission-line galaxies, or to PN pair mismatches in crowded
regions.  As in D+07, we recognize such cases by their apparent ``velocities'' which
are outlying from the general distribution defined by the {\it bona fide} PNe.
To this end, we use the ``friendless'' algorithm introduced in \citet{2003MNRAS.346L..62M}
which flags objects deviating by more than $n\times\sigma$ from the velocity
distribution of their $N$ nearest neighbours.
With $n=2.5$ and $N=15$, we find 12 outliers among the NGC~4494 PN candidates
(see Fig.~\ref{fig:fig1}).

\begin{figure}
\centering
\epsfig{file=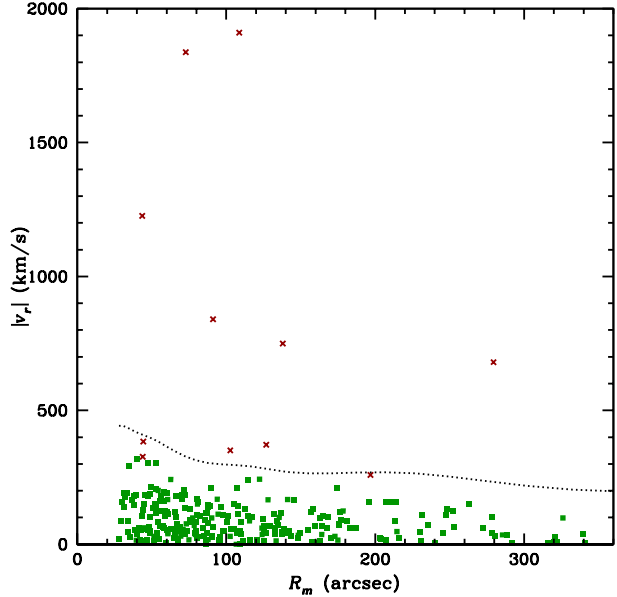,width=8.5cm}
\caption{Distribution of line-of-sight velocities of PN candidates around NGC~4494,
as a function of radius, and relative to the systemic velocity (1344~\kms).
Red $\times$ symbols mark objects designated as outliers by the friendless criterion
(one object is off the plot at 108$''$, 2518\kms{}),
and green boxes show the {\it bona fide} PNe.
The dotted line shows the 3~$\sigma$ velocity envelope.}
\label{fig:outli}
\end{figure}

Almost all of the friendless objects are in areas of the field that are either noisy or
prone to pair confusion; one or two remaining outliers are probably background
galaxies.  An alternative approach calculates an azimuthally-averaged projected
velocity dispersion profile using a moving-window technique, and rejects objects
outside the 3~$\sigma$ velocity envelope (see Fig.~\ref{fig:outli}).
This approach finds the same outliers as the friendless algorithm, except in
the very central regions which are not of importance to our study anyway.
The final PN catalogue for NGC~4494 therefore has 255 objects;
this is 3.5 times the 73 PNe reported in R+03, with most of the increase coming
from a greatly improved method of reducing and stacking the raw data.

We have compared the R+03 preliminary catalogue with our new one and found 70 PNe
in common between the two datasets, with
no velocity offset between the two samples (for comparison,
in D+07 we found an offset of $\sim30$~\kms{} for NGC~3379).
The distribution of velocity differences between the R+03 sample, $V_{\rm R+03}$,
and the present sample, $V_{\rm PN08}$,
$\Delta V=V_{\rm R+03}-V_{\rm PN08}$ is nearly Gaussian (skewness $\sim-0.4$ and kurtosis $\sim2.9$) with
a standard deviation of 18~\kms{}---which is smaller than the typical instrumental error of the PN.S
(20\kms{}, see D+07).
We find no systematic trend of velocity difference with spatial position.

We have also cross-checked the \citet[hereafter J+96]{1996ApJ...462....1J} photometric-only
catalogue of 183 PN candidates, all of them in our field of view.
We recover only 121 of their PNe; of the ones we miss, two-thirds are
{\it not} from the photometrically
complete part of their catalogue ($m_{5007} < 27.0$, $R > 1.0'$).
Photometric catalogues are known to be liable to false detections if the
off-band images are not deep enough (e.g. \citealt{2005AJ....129.2585A}), and
the fraction of the photometric PNe that we have confirmed is consistent with
typical recovery rate of photometric candidates in spectroscopic follow-up
(e.g. \citealt{1993ApJ...414..454C,1998ApJ...507..759A})---where the
losses can also be
due to different factors like astrometric problems, fiber positioning and losses etc.
On the other hand, the vast majority of the objects detected only by the PN.S
were below the completeness limit of J+96.

A comparison of spatial positions shows a large average difference in RA and DEC
($\sim-2''$ and $\sim$2$''$ respectively) between our positions and the ones from J+96.
However, the scatter is small (0$''$.6 and $0.''5$, respectively),
so there is simply a constant offset between the astrometric systems, which will
not compromise the kinematical and dynamical analyses of this paper.

As a final check we have compared the [OIII] magnitudes $m_{5007}$
between the two datasets, finding a scatter of $\sim 0.4$~mag for the whole
matched sample, which reduces to $\sim 0.3$~mag for the brightest objects
(within 0.3~mag of $m^*$).
Such scatter is consistent with the typical
photometric uncertainty of the PN.S,
of the order of 0.2--0.3 (see also D+07), once the photometric
errors quoted by J+96 are subtracted off in quadrature.

\section{PN distribution and kinematics}\label{sec:dist}

We now present the basic properties of the field stars and PNe in NGC~4494,
including their distributions in space and velocity.
Since an important assumption in strengthening our models is that the PNe
are a fair tracer population of the field stars, we compare throughout
the properties of the stars and PNe.
We examine the spatial distribution in \S\ref{sec:spatdist}, the rotation field
in \S\ref{sec:rot}, the velocity dispersion in \S\ref{sec:dispsec}, and
the kurtosis in \S\ref{sec:higher}.

\subsection{Surface photometry and PN spatial distribution}\label{sec:spatdist}

The surface photometry of NGC~4494\ available in the literature includes
{\it HST}-based observations in the $V$ and $I$ bands \citep{2005AJ....129.2138L},
and ground-based CCD observations in $BVI$ \citep[hereafter G+94]{1994A&AS..104..179G}.
As neither study extends far enough in radius to cover the region probed by the
PNe (to $\sim 6'$), new surface photometry was necessary.
To this end we have used imaging data taken on January 27, 2006, with Megacam
\citep{2000fdso.conf...11M} mounted on the MMT 6.5m telescope\footnote{The MMT
Observatory is a joint facility of the Smithsonian Institution and the University of Arizona.},
as part of a programme to image the globular cluster (GC) systems of nearby ETGs.
A mosaic of 36 2k$\times$4k backside-illuminated CCDs provide a 24$'$$\times$24$'$
field of view with a $0.08\arcsec$ native pixel scale.
Our data consist of $3\times$ 45~sec exposures of NGC~4494\ using the SDSS
$g'$ filter, read out using $2\times2$ binning, with a pixel scale of 0.16$\arcsec$.
The detailed data reduction and surface photometry analysis are beyond the
scope of this paper,
but we provide a brief summary here.

The initial data reduction (including overscan, trimming, flat-fielding, and
bad pixel identification) was performed using the standard routines in the
IRAF MSCRED package.
World Coordinate System solutions were then found for each exposure by comparison with the
2MASS catalogue (\citealt{2003AJ....125..525J}).
The images were projected to the tangent plane using SWarp\footnote{SWarp is software
developed by Emmanuel Bertin and it is publicly
available at http://terapix.iap.fr/rubrique.php?id\_rubrique=49/},
and stacked in IRAF to create
the final corrected images with a seeing-limited resolution of $\FWHM\sim1.4\arcsec$.
Isophotes from the reduced image are shown in Fig.~\ref{fig:fig1}, where it is evident
that contamination from a bright star becomes problematic beyond $\sim 2'$ in the
Northeast direction.  Therefore we have excluded this region as marked in the
Figure from our surface photometry analysis.

\begin{figure}
\centering
\hspace{-1cm}
\epsfig{file=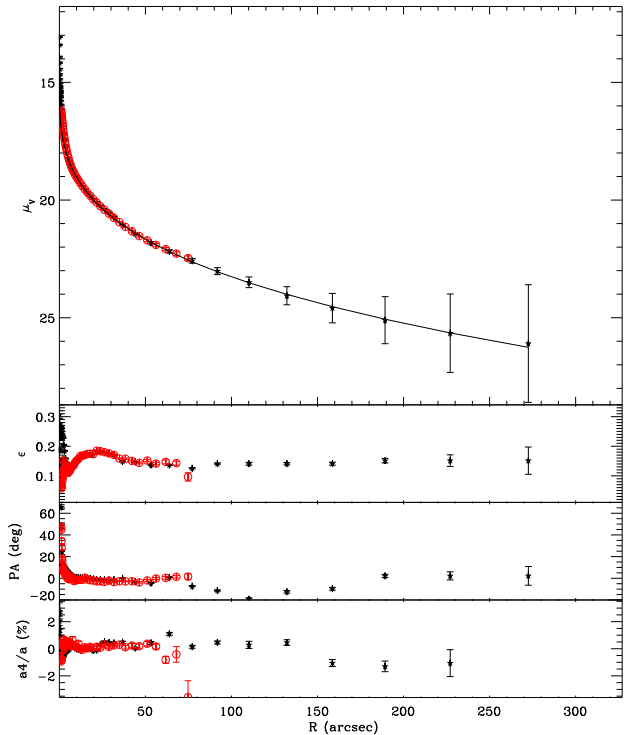,width=9cm}
\caption{Surface photometry of NGC~4494, as a function of the intermediate axis radius.
Literature data \citep{1994A&AS..104..179G,2005AJ....129.2138L} are shown by red circles,
and Megacam data by black stars (this paper), with the uncertainties shown by error bars.
{\it Top large panel:} composite $V$-band surface brightness profile.
The solid line is the best fitting S\'ersic (1968) formula for $R>5\arcsec$.
{\it Bottom small panels}: Profiles of ellipticity, position angle and
fourth-order isophote shape coefficient $a_4$, with G+94 data also shown.
}
\label{fig:phot}
\end{figure}

To convert the surface photometry into simple photometric profiles suitable for
modelling, we used the IRAF tasks IMSURFIT and ELLIPSE to subtract the background
light and fit elliptical isophotes.
The resulting profiles of surface brightness, ellipticity, and position angle are
shown in Fig.~\ref{fig:phot}, mapping from the $g'$-band to the $V$-band with
an arbitrary normalisation required to match the literature data over the
radial range of $R = 1.8''$--$60''$.
In this Figure, and in general in this paper, the radius used is from the projected
intermediate axis $R_m$, which is related to the major axis $R_a$ and ellipticity
$\epsilon$ by $R_m \equiv R_a (1-\epsilon)^{1/2}$.
The data sets are seen to be generally consistent, except notably inside $\sim 4\arcsec$, where
the Megacam data are affected by seeing.
We therefore combine the literature and Megacam data into a
single $V$-band surface brightness profile $\mu_V(R)$, listed in Table~\ref{tab:SB}
of Appendix \ref{app:photmega}, together with ellipticity and isophote
shape parameter $a_4$ (\citealt {1988A&AS...74..385B}) values.

In order to make use of the stellar luminosity profile in the dynamical models, we parametrize the
surface brightness profile by the S\'ersic law:
\begin{equation}\label{eq:sersic}
\mu(\Rm)-\mu(0) \propto (\Rm/a_S)^{1/m} ,
\end{equation}
where $a_S$ is a scale length and $m$ describes the ``curvature'' of the profile
\citep{1968adga.book.....S}.
Fitting the regions outside the inner disc ($R > 5\arcsec$), we find $a_S=0.115\arcsec$, $m=3.30$, and
$\mu_0 = 14.82$~mag~arcsec$^{-2}$.
These parameters translate into an effective radius of $\Re= 49.5\arcsec$;
a more detailed multi-component fit including the central regions yields
$\Re= 48.2\arcsec \pm 3.0\arcsec$, which is similar to other literature estimates
\citep{1989ApJS...69..763F,1991trcb.book.....D},
The total extinction-corrected luminosity in the $V$-band is
$2.64\times10^{10} L_{V, \odot}$, or $M_V = -21.26$;
the uncertainties in the outer surface brightness profile yield a (model-dependent) total
luminosity uncertainty of $\sim$5--10\%.
To estimate the same quantities in the $B$-band, we use the typical de-reddened
color of $(B-V)_0=0.81$ from G+94 to find $L_B=2.37\times10^{10} L_{B, \odot}$ and $M_B=-20.45$.
These and other global parameters for NGC~4494\ are listed in Table \ref{tab:galpar}.

\begin{figure}
\hspace{-0.5cm}
\epsfig{file=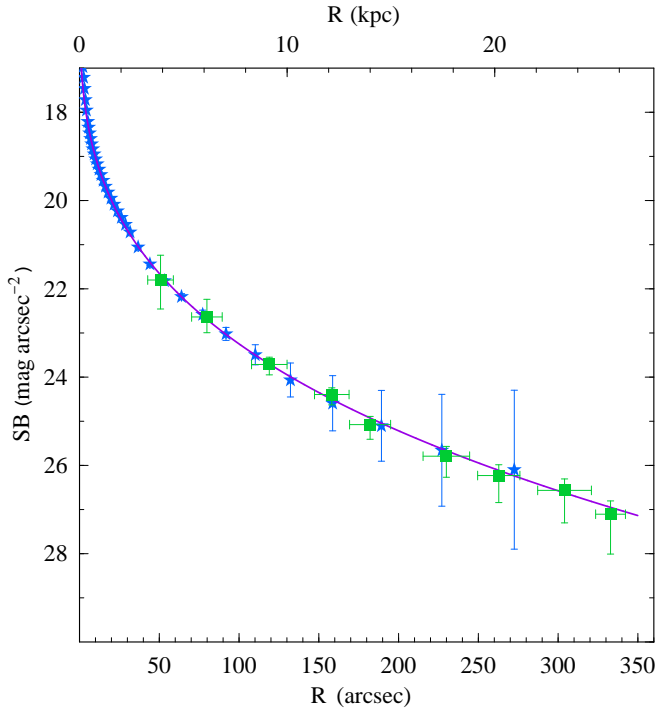,width=8.7cm}
\caption{Radial surface density profiles of the field stars ($V$-band; blue star symbols)
and of the PNe (green squares) in NGC~4494.
The PN number counts have been corrected for spatial incompleteness, and
arbitrarily normalized to match the stellar data.
The vertical error bars of the PN data in this and in the following figures represent the 1~$\sigma$
uncertainties (based in this case on counting statistics and completeness correction
uncertainties), while the horizontal error bars show 68\% of the radial range of
the PNe in each bin.
The purple solid curve is the S\'ersic model fit to the stellar photometry (see main text).}
\label{fig:spatcomp}
\end{figure}

We next compare the spatial density of the PNe with the field stars, using the
PN number density complete to $m^*+1$ (Section \ref{sec:compcont}).
Given an arbitrary normalisation, the PN profile matches the stellar photometry remarkably well (Fig.~\ref{fig:spatcomp})---as also generally found in a larger sample of
galaxies by \citet{Coccato08}.
Therefore we do not see a pattern emerging to support the scenario proposed by D+05,
wherein the observed PNe trace a young
stellar sub-population generated in a merger, with a steeper spatial density profile
than the bulk of the field stars.

The normalisation of the PN profile corresponds to a specific density parameter of PNe
per unit $V$-band galaxy luminosity of $\alpha_{V,1.0}= (11.5\pm3.0) (10^9 L_\odot)^{-1}$.
This translates to a $B$-band density of $\alpha_{B,1.0}= (10.3\pm2.6) (10^9 L_\odot)^{-1}$,
and over the standard magnitude interval to $m^*+2.5$,
$\alpha_{B,2.5}=(42\pm 11) (10^9 L_\odot)^{-1}$, which is consistent
with the previous result of \citet{1996ApJ...462....1J}:
$\alpha_{B,2.5}=(31\pm3) (10^9 L_\odot)^{-1}$ (assuming our
adopted extinction and distance).
The specific frequency is fairly high compared to other ETGs
of similar luminosity (cf. \citealt{2005ApJ...629..499C}, \citealt{2006MNRAS.368..877B}).
The 20\% discrepancy between the PNLF and SBF distances to NGC~4494
will be investigated in a future paper on PNLFs.

\begin{figure*}
\centering
\epsfig{file=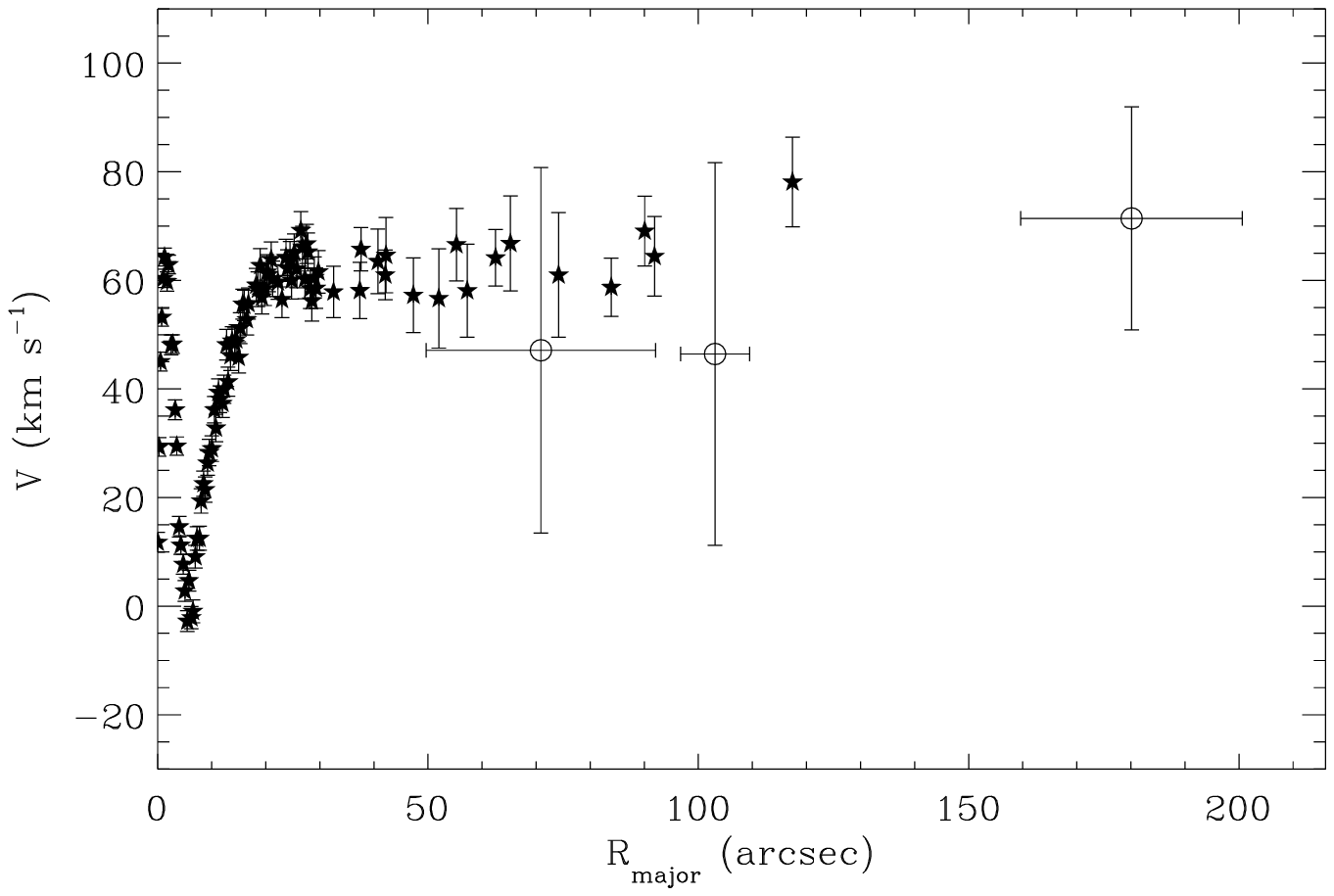,width=8.2cm}
\epsfig{file=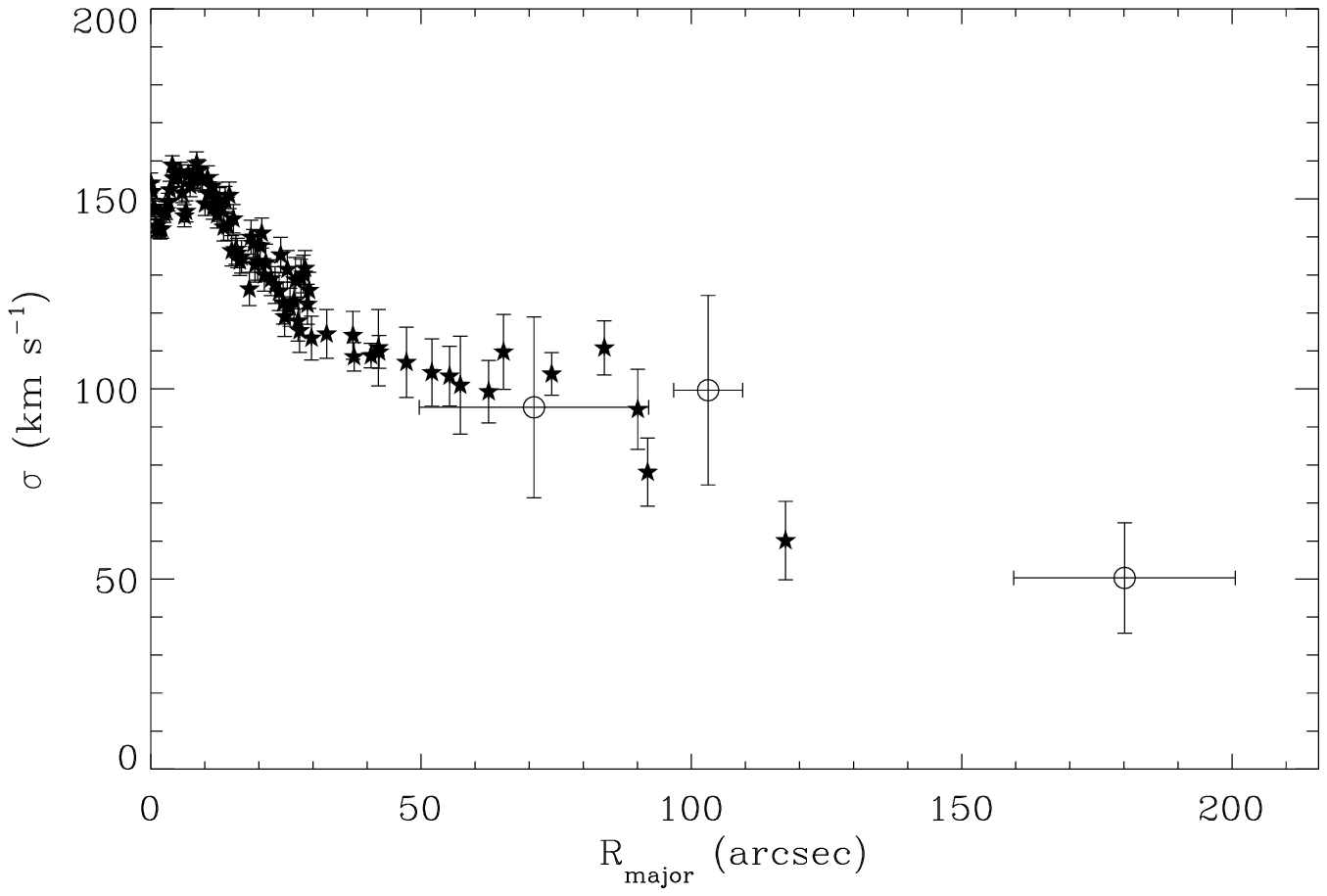,width=8.2cm}
\epsfig{file=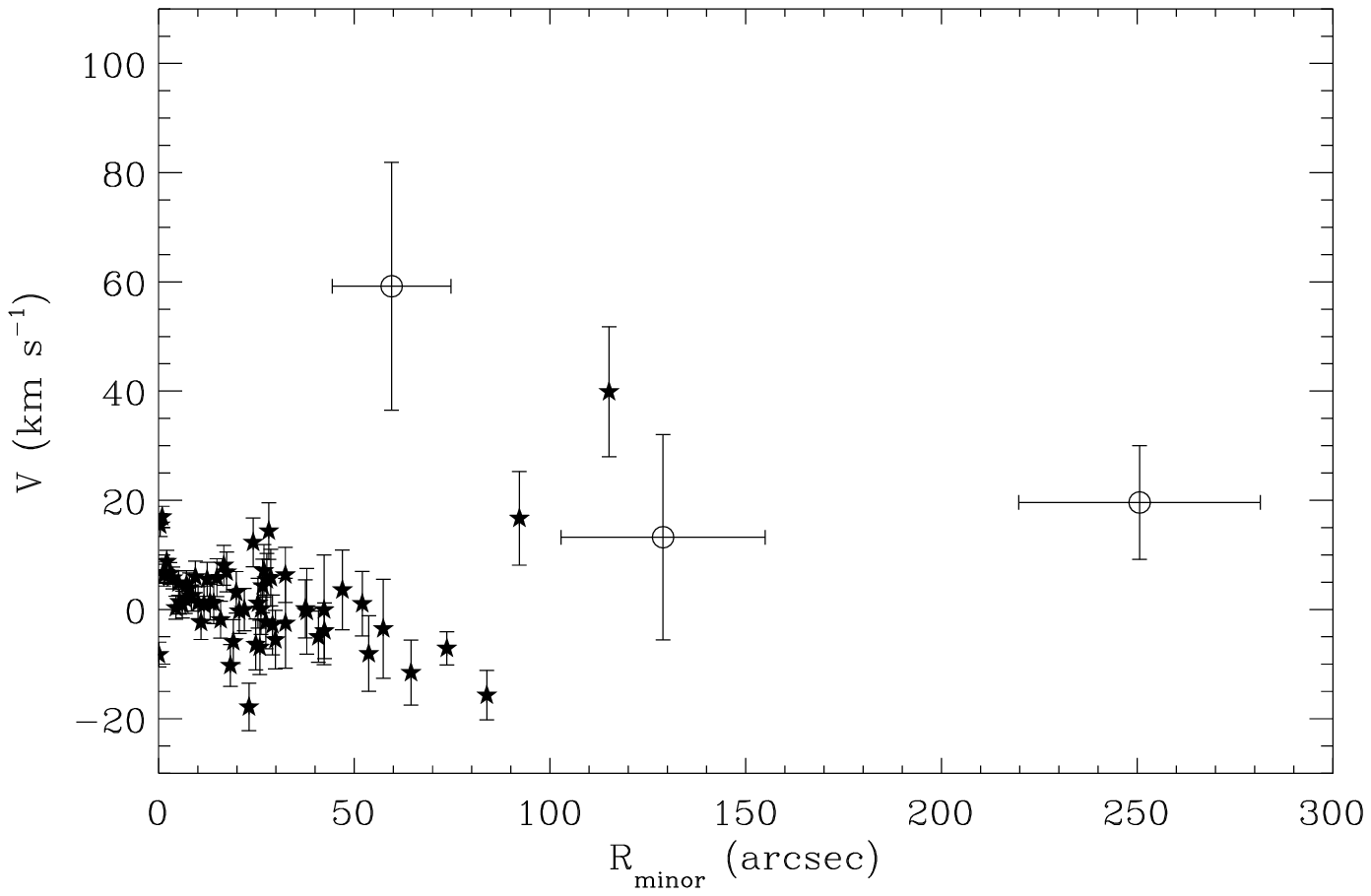,width=8.2cm}
\epsfig{file=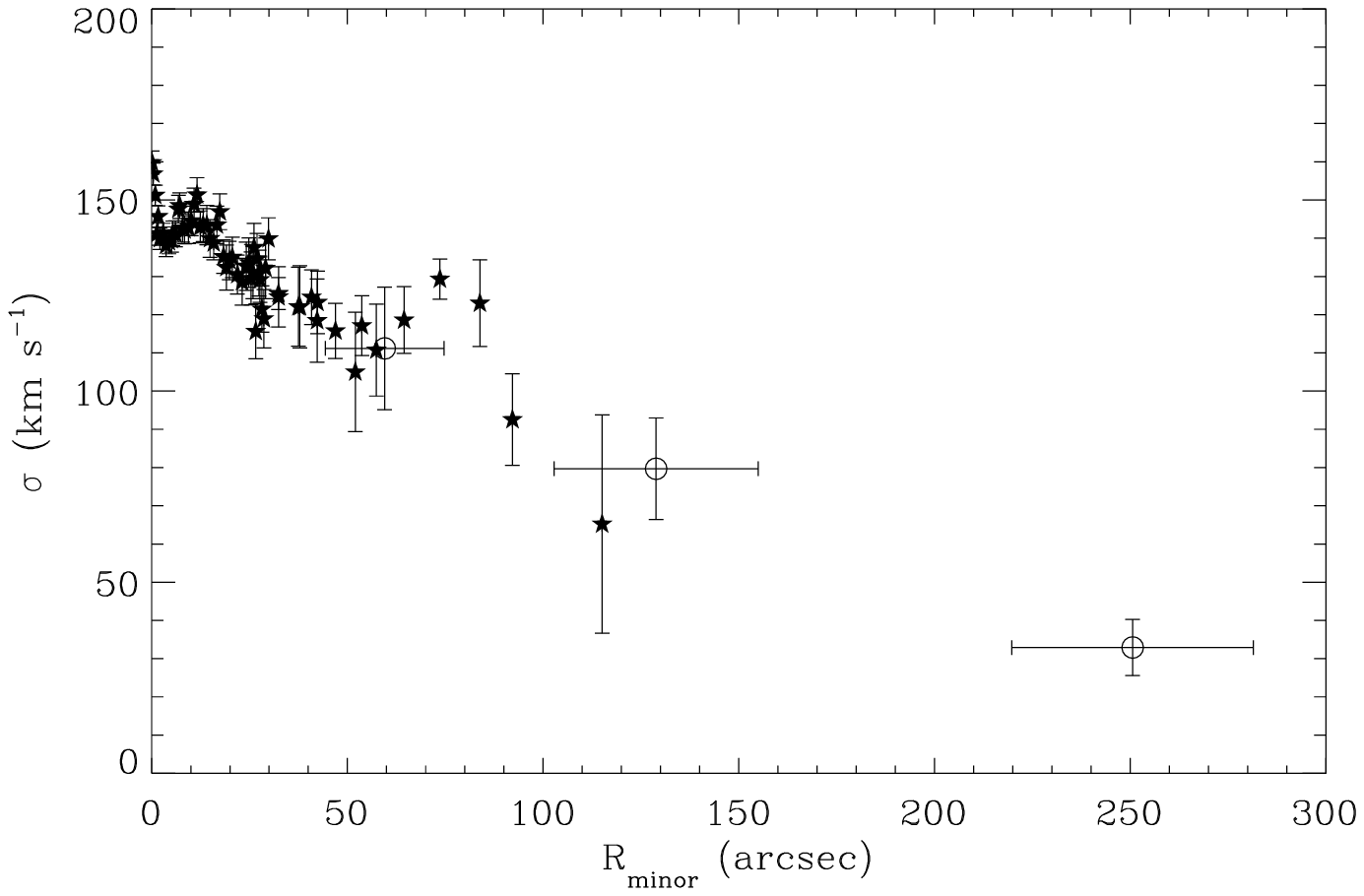,width=8.2cm}
\caption{Observed kinematics of stars (filled star symbols) and PNe (open circles)
in NGC~4494, with uncertainties shown by vertical error bars.
The data from opposite sides of the galaxy have been folded into a single radial axis.
The horizontal error bars show 68\% of the radial range of each PNe bin.
The left panels show the mean velocities relative to the systemic velocity,
and the right panels show the dispersions about the mean.
The top panels show the major axis profiles, and the bottom panels show the minor axis.}
\label{fig:majmin}
\end{figure*}

\subsection{Rotation}\label{sec:rot}

We now examine kinematical streaming motions in the data, i.e. rotation.
Although there are data available in the literature for the stellar kinematics
of NGC~4494 \citep[hereafter B+94]{1994MNRAS.269..785B}, we have obtained new long-slit data
as part of a programme to obtain extended stellar kinematics in ETGs.
These data will be presented in detail in a forthcoming paper \citep{Coccato08},
and we summarize the main features of the NGC~4494 data here.

Deep spectra were obtained with the ESO Very Large Telescope in service mode
[Programme ID 076.B-0788(A)]  using the FORS2 spectrograph, equipped with the 1400V grism,
through a 1\farcs0-wide slit.  The spectra were taken along the major
(PA=$0\degree$) and minor (PA=$90\degree$) axes of the galaxy, and were re-binned
in radius to reach $S/N\sim 30$ at $\Rm\sim 100\arcsec\sim 2\Re$.
Kinematical stellar templates were chosen from the Indo--U.S. Coud\'e Feed
Spectral Library \citep{2004ApJS..152..251V} and convolved with a Gaussian
function to match the instrumental FWHM.
Various moments of the velocity distribution were extracted using the Penalized
Pixel-Fitting method of \citet{2004PASP..116..138C}.
The fitted spectral range was $\sim$~4560--5860\AA{}.

The results for the stellar kinematics are shown in Fig.~\ref{fig:majmin}.
The rotation profile along the major axis
shows the kinematically decoupled core inside 5$''$, as
previously observed by B+94,
and coincident with the photometric stellar disc \citep{1997ApJ...481..710C}.
Since we are interested mostly in the outer kinematics,
this very central feature is of no concern for us.

Before examining the PN kinematics, we check their self-consistent
systemic velocity by taking the median of the individual PN velocities
in the azimuthally complete region of $40''$--$135''$.
This value is 1344~\kms, exactly the same as the NED
velocity for NGC~4494.

We then measure the rotation of the PN system by calculating the mean velocity in bins
within a wedge of 20 degrees opening angle, aligned with the directions
used for the long slit spectroscopy (PA=0\degree\ for the major axis and
PA=90\degree\ for the minor axis).
From Fig. \ref{fig:majmin} (left panels) it is apparent that the rotational
properties of the PNe and stars are generally consistent.
The discrepancy at small radii ($\sim50''$) on the minor axis may be caused by the
finite width of the PN measurement wedge (i.e., picking up off-axis rotation)
and by statistical fluctuations in the velocity sample
(see \citealt{2001A&A...377..784N}).

The constant major-axis rotation velocity of the stars outside the galaxy core of
$60\kms$ is found using the PNe to extend to at least 4~\Re{}.
The minor-axis rotation is generally small but does suggest some finite rotation
at large radii, i.e. a kinematic misalignment.
This feature could be caused by a transition in the orbit structure
or by projection effects in a smooth triaxial system
(e.g. \citealt{2008MNRAS.385..647V}), and may be related to the
isophote twisting between $\sim75''$ and $\sim200''$ (see Fig.~\ref{fig:phot}).

\begin{figure*}
\epsfig{file=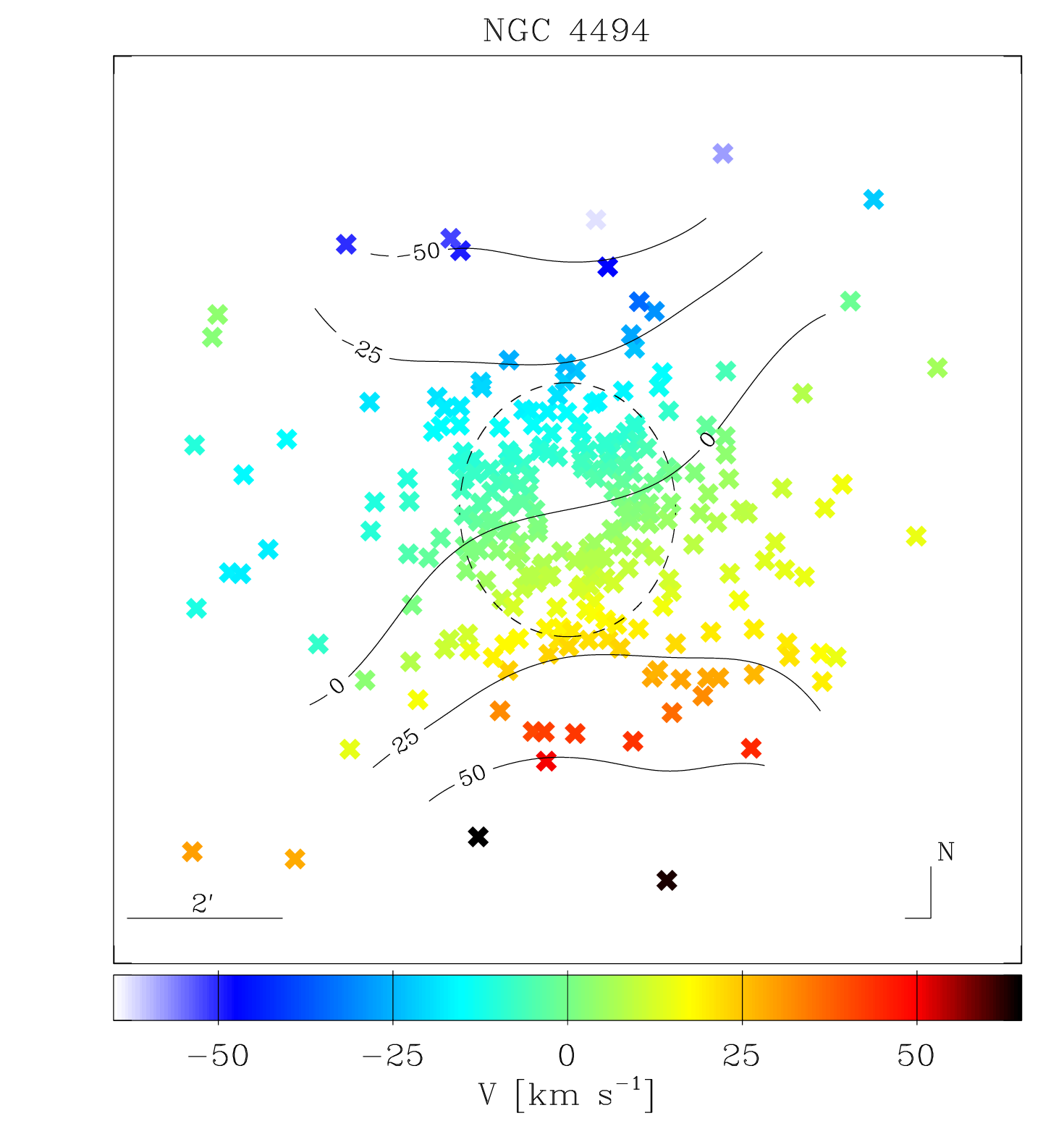,width=8cm}
\epsfig{file=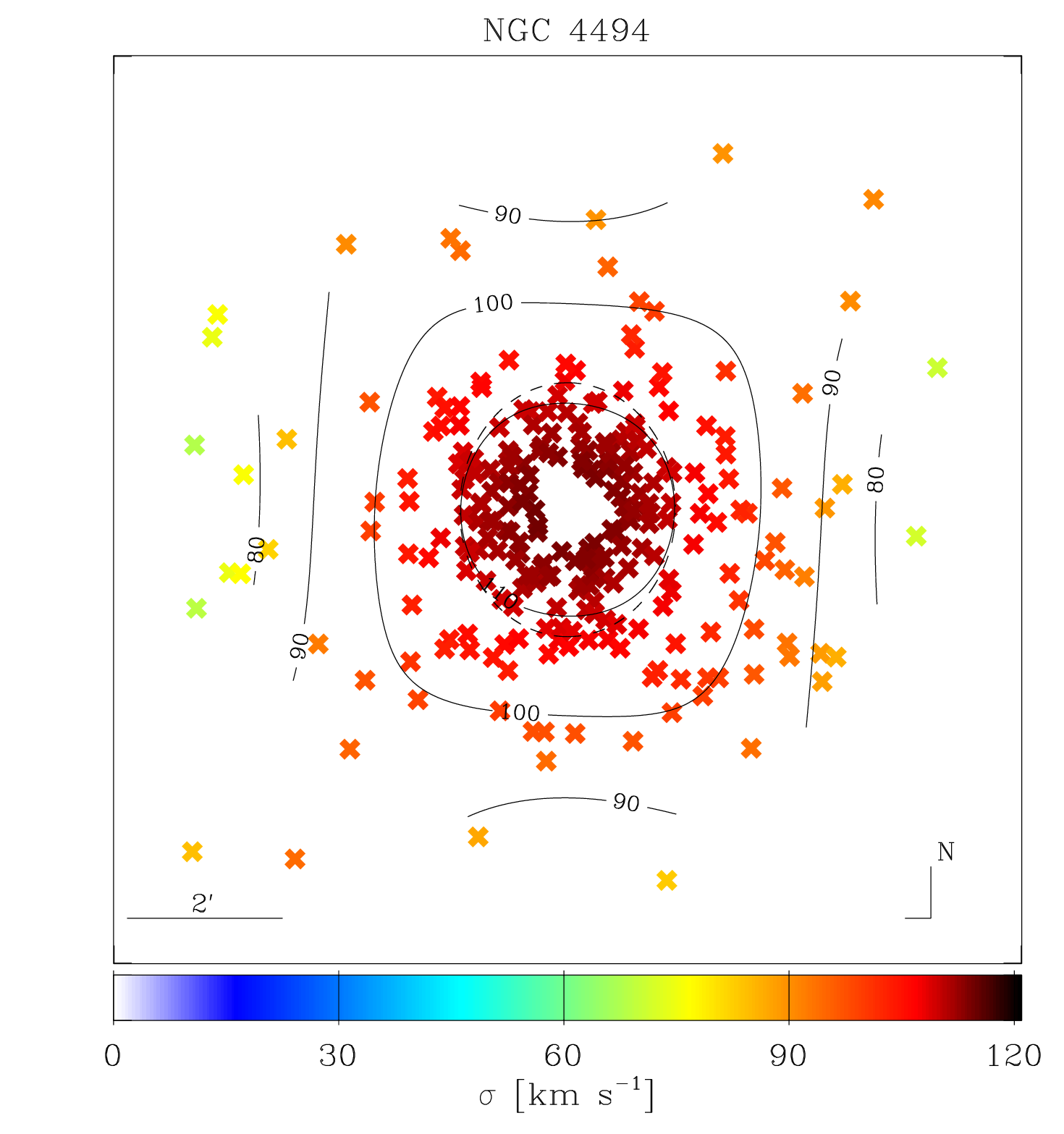,width=8cm}
\caption{Two-dimensional kinematics of PNe in NGC~4494.
Individual PN data points are shown as $\times$ symbols, with values
given by colors from the scale below.
The data set has been doubled by a reflection symmetry, and then smoothed
(see main text).
Solid curves illustrate iso-velocity contours.
The dashed ovals show the 2~\Re{} isophote.
{\it Left}: Mean velocity.
{\it Right}: Velocity dispersion about the mean (the squarish contour is partially
an artefact of the gridded smoothing technique).}
\label{fig:pnsmooth}
\end{figure*}

In order to make use of all the PN data rather than just those along the
principal axes, we construct 2-D maps of the velocity distribution, as in D+07.
Assuming at least triaxial symmetry, we produce higher-$S/N$ maps by mirroring
the PN data points in phase space from $(x,y,v)$ to $(-x,-y,-v)$.
We then smooth the velocity field using a median filter and a Gaussian kernel
of width $\sigma\sim60\arcsec$ ($\sim 4.6\kpc$) on a regularly spaced grid of
75\arcsec\ cell widths (see also D+07).
The resulting mean velocity field is shown in Fig.~\ref{fig:pnsmooth}
(left panel). Again we see evidence of rotation along the major axis (PA $\sim 0^\circ$)
in the galaxy's central parts, twisting to another axis ($\sim 20^\circ$) outside $\sim$~1~\Re.

We next try an alternative measure of the PN rotation that is
insensitive to azimuthal incompleteness.
It works by fitting the velocities to a cosinusoidal curve:
\begin{equation}
v(\phi)=v_0\times\cos(\phi-\phi_0),
\end{equation}
where $\phi$ is the position angle of the PNe relative to the major axis
(cf. \citealt{2001A&A...377..784N}).
We report the best-fit parameters for various radial bins in Table~\ref{tab:Z1}.
Using this technique, there is no significant evidence that the rotation is
misaligned with the major axis (PA=0\degree) at any radius. Although there is a
twist at intermediate radii (consistent with the smoothed velocity field in Fig.
\ref{fig:pnsmooth}), this is not statistically significant and
we will consider the whole sample as consistent with PA=0\degree
(as in the first row of Table \ref{tab:Z1}).

\begin{table}
\caption{Best-fit parameters of sinusoidal rotation model to PN velocity data, in radial bins.
$\phi_0$ is the direction of maximum velocity.}
\label{tab:Z1}
\begin{center}
\noindent{\smallskip}\\
\begin{tabular}{llll}
\hline
Radius [\arcsec] & N. PNe & $v_0$ [\kms] & $\phi_0$ [degrees] \\
\hline
0 -- 400 &255 & $-23\pm 9 $ & $-2\pm25$ \\
0 -- 75  & 98 & $-28\pm 18$ & $24\pm39$ \\
75 -- 133 & 79 & $-12\pm 16$ & $42\pm76$ \\
133 --400 & 78 & $-37\pm 14$ & $-8\pm19$ \\
\hline
\end{tabular}
\end{center}
\end{table}

Finally, we wish to quantify the dynamical significance of rotation
within 1~\Re.  This is best done with high-quality integral field stellar kinematics data
(e.g. SAURON; \citealt{2007MNRAS.379..418C}, hereafter C+07),
but we can make an estimate based on the available long-slit data.
C+07 derived an empirical calibration (their eqn. 23)
between the SAURON rotation dominance parameter
within 1~\Re, and the traditional measure of B+94:
\begin{equation}
\left(\frac{v}{\sigma}\right)_{\rm e} \approx 0.57 \times \left(\frac{v_{\rm max}}{\sigma_0}\right)_{\rm B+94},
\end{equation}
where $v_{\rm max}$ is the maximum observed rotation velocity along the major axis,
and $\sigma_0$ is the average velocity dispersion within \Re/2.
Given $v_{\rm max}=$~76\kms{} and $\sigma_0=163$\kms{} from B+94,
we find that $(v/\sigma)_{\rm e} \approx 0.26$.
Comparing this value to the isophote ellipticity from B+94 ($\epsilon = 0.16$),
we find that NGC~4494 is located firmly in the space occupied by ``fast rotators''
(C+07 Fig.~11).
Although the calibration above was done for the B+94 data only, very similar results would
be obtained using our new photometric and long-slit data.

\subsection{Velocity Dispersion Profile}\label{sec:dispsec}

We next examine the random motions in NGC~4494, represented by the
velocity dispersion profile.
Fig.~\ref{fig:majmin} shows that the PN and stellar dispersions are
consistent along the major and minor axes.
The dispersion declines steadily with radius, and random motions dominate
streaming motions out to $\sim 100''$ ($\sim 2 \Re{}$).
Fig.~\ref{fig:pnsmooth} (right) shows that the dispersion is roughly
constant along the galaxy's isophotes.

We reduce these data to a single velocity dispersion profile as a function of
the intermediate radius $R_m$.  The rotation and true dispersion profile are folded
into an root-mean-square velocity profile $v_{\rm RMS}=\sqrt{v^2 + \sigma^2}$,
where $v$ and $\sigma$ are the rotation and dispersion components respectively.
The RMS velocity is a measure of the total kinetic energy, and
we henceforth loosely refer to it as the velocity dispersion or \VD{}.
We combine the stellar data from the different axes by averaging, while folding
the (small) systematic differences into the final uncertainties\footnote{The uncertainties
in the PN dispersion use a classical analytic formula that assumes a Gaussian distribution,
i.e. $\Delta v_{\rm RMS} \sim \sqrt{\Sigma_i v_i^{2}/2N}$.
However, we expect this to produce accurate results in realistic systems
\citep{2001A&A...377..784N}, and we have carried out additional
Monte Carlo simulations of a simplified galaxy with radial orbits, finding
that the dispersion is very accurately recovered with our estimator, with a
possible bias to be $\sim5\%$ too high.
Also, the expression for stellar $v_{RMS}$ is valid for comparison with the
azimuthally-scattered PNe only because we are averaging the major and minor axis
long-slit data; if only major axis data were available, the expression would be
$v_{\rm RMS} = \sqrt{v^2/2+\sigma^2}$.}

\begin{figure}
\hspace{-0.5cm}
\epsfig{file=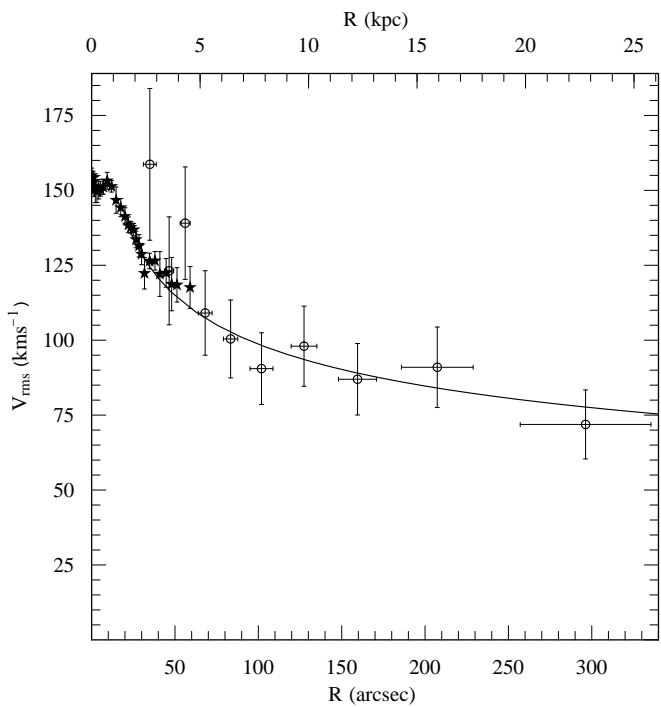,width=8.7cm}
\caption{Composite projected velocity dispersion profile of NGC~4494,
with data from stars (filled star symbols) and PNe (open circles).
The solid curve shows a power-law fit to the PN data outside 60$''$.}
\label{fig:dispprof}
\end{figure}

The resulting dispersion data are plotted in Fig.~\ref{fig:dispprof}.
The PN dispersions are similar to those reported in R+03, but
are based on 3.5 times as many velocities, and extend to twice the radius.
In our subsequent dynamical models, the stellar dispersion will be fitted out
to 60$''$ ($\sim 1 \Re{}$), where it is more accurate,
and the PN dispersion outside this radius.
To characterize the decline of the dispersion with radius, we fit a
power-law to the PN data outside 60$''$, using a maximum-likelihood method that
fits the discrete data without binning, though it does assume an underlying
Gaussian velocity distribution \citep{2006A&A...448..155B}.
We find a power-law exponent of $-0.2\pm0.1$ (see Fig.~\ref{fig:dispprof}),
which is significantly shallower than $-0.6\pm0.2$ found for NGC~3379 (D+07),
suggesting a dynamical difference between the two galaxies
(discussed further in \S\ref{sec:dynamics}).

We also test the sensitivity of the derived velocity dispersion profile to the
outliers selection (Sec.~\ref{sec:compcont}).
Restoring all the objects to the PN sample which were rejected at $n=2.5$ but not $n=5$,
we find that the
only affected feature is a peak in the dispersion between $100\arcsec$ and
$150\arcsec$, which does not change the broad trend.

\subsection{Higher-Order Velocity Moments}\label{sec:higher}

The velocity dispersion is a crude measure of the particle speeds in a
collisionless system, whose finer orbital properties can be probed using higher-order
moments of the line-of-sight velocity distribution (LOSVD).
Stars on isotropic orbits in a galaxy halo with a flat rotation curve have a
Gaussian LOSVD \citep{1993MNRAS.265..213G}.
In a tangentially-anisotropic system where the particles follow nearly circular orbits, the observed velocities
tend to pile up at the circular speed, producing an LOSVD which is flatter than
a Gaussian, or even double-peaked.
Radial orbits produce centrally peaked LOSVDs with long tails.
In the case of integrated-light stellar velocities with fairly high $S/N$,
these LOSVD shapes are described via the Gauss-Hermite moments such as
$h_4$ \citep{1993ApJ...407..525V,1993MNRAS.265..213G}.
In the case of discrete velocities, it is more straightforward to compute the
classical dimensionless kurtosis,
$\kappa\equiv\overline{v^4}/(\overline{v^2})^2-3$
(see \citealt{Joanes98} for exact expression and uncertainties\footnote{Again,
the uncertainty calculation assumes a Gaussian distribution, while for a more
general distribution the errors are formally infinite.
However, the Monte Carlo simulations mentioned in \S\ref{sec:dispsec}
demonstrate accurate recovery of the kurtosis using our estimator, with a
systematic deviation of no more than $\sim 0.1$.}).
Broadly speaking, $\kappa \simeq 0$ for isotropic orbits, $\kappa < 0$ for
tangential orbits, and $\kappa > 0$ for radial orbits.

\begin{figure}
\hspace{-0.5cm}
\epsfig{file=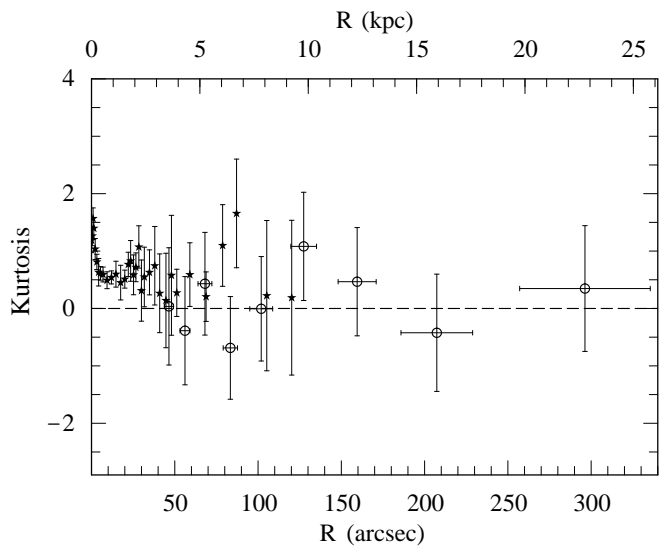,width=8.7cm}
\caption{Projected kurtosis profile of NGC~4494, with symbols as in Fig.~\ref{fig:dispprof}.}.
\label{fig:kurt}
\end{figure}

We quantify the shapes of the stellar and PN LOSVDs in NGC~4494
in Fig.~\ref{fig:kurt}, using the approximate relation $\kappa\simeq 8\sqrt{6} h_4$
for the stars \citep{1993ApJ...407..525V}.
The PN kurtosis is fairly noisy, but is consistent with the stellar properties
in the region of overlap.
However, there is a systematic offset between our stellar $h_4$ values and those
reported by B+94 inside $20''$,
where we believe our improved
template library procedures yield more reliable results.
In the halo, the PN kurtosis is consistent with zero.

\section{Dynamical Models}\label{sec:dynamics}

We now combine the photometric and kinematical data for the stars and PNe in NGC~4494
into integrated dynamical models in order to derive the
mass profile of the galaxy, and test whether or not it hosts a DM halo
as expected from \LCDM.
In \S\ref{sec:massinv} we begin with the simple
pseudo-inversion mass model\footnote{This is a phenomenological mass invertion
method where we reconstruct the mass distribution from the velocity dispersion data.
As shown in \S\ref{sec:massinv}, since we make use of a parametrized model for the
$\sigma_r$ and the anisotropy parameter in the radial Jeans equation, this is not a full inversion method.}
used in R+03 and D+07, and introduce a new series of physically-motivated
kurtosis-based models in \S\ref{sec:massmod}.
We address systematic issues in \S\ref{sec:comp}.

In both of our main modelling exercises (\S\ref{sec:massinv} and \S\ref{sec:massmod}),
we will apply the spherical Jeans equations to NGC~4494, since
the observed stellar isophotes are very round (see \S\ref{sec:spatdist} for photometry, and
D+07 \S8.1 for discussion of the spherical approximation).
We will examine a non-spherical model in \S\ref{sec:comp} and Appendix~\ref{app:flatt}.
Given the consistent agreement seen in \S\ref{sec:dist} between the stellar and PN
properties, we will assume that all of these data are drawn from
the same underlying dynamical tracer population.
We will also in general omit the stellar kinematics data inside 10$''$
from our model fits, since there appears to be a strong dynamical change in this
nuclear disc region which our smooth Jeans models are not designed to reproduce.

\begin{figure}
\hspace{-0.5cm}
\epsfig{file=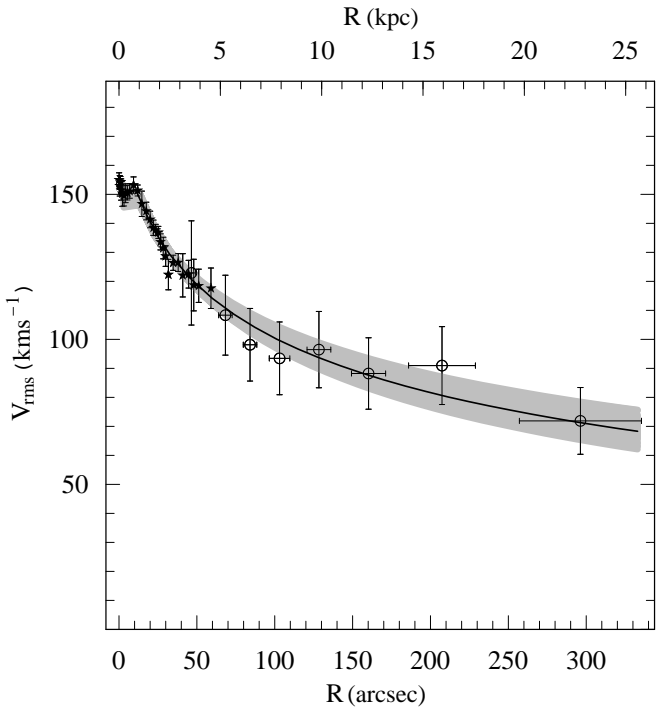,width=8.7cm}
\caption{Model fit to the NGC~4494 velocity dispersion data using
the pseudo-inversion mass model
and assuming isotropy. The best fit and
1~$\sigma$ range of uncertainty are shown by the solid curve and
shaded region, respectively.}
\label{fig:modelfitsone}
\end{figure}

\subsection{Pseudo-inversion mass model}
\label{sec:massinv}

Our first series of models uses simple assumptions to convert the observed
kinematics into a mass profile $M(r)$.  This phenomenological approach,
as introduced in R+03 and D+07, is computationally light, need not involve
Abel inversion integrals, and has no prejudice about the form that
$M(r)$ should take, nor about the stellar $M/L$ value
(which will be discussed later in this section).
On the other hand, it is not possible to directly test any theoretical
predictions for the DM distribution, and the resulting mass profile may not even
be physical.

The mass modelling procedure consists of five steps:
\begin{enumerate}
\item Adopt a simple smooth parametric function for the intrinsic radial velocity dispersion profile:
\begin{equation}
\sigma_r(r) = \sigma_0 \left[1+\left(\frac{r}{r_0}\right)^\eta \right]^{-1},
\label{eq:v0eqn}
\end{equation}
where ${\sigma_0,r_0,\eta}$ are a minimalistic set of free parameters.
\item Assume a fixed anisotropy profile:
\begin{equation}\label{eq:beta}
\beta(r) \equiv 1-\sigma^2_{\theta}/\sigma^2_r ,
\end{equation}
where $\sigma_\theta$ and $\sigma_r$ are the tangential and radial components of
the velocity dispersion ellipsoid, expressed in spherical
coordinates\footnote{Since we are folding the observed rotational and pressure support into a single parameter $v_{\rm RMS}$,
we use a similar approach in the Jeans equations, so that technically
$\beta(r) \equiv 1 - (\sigma_\theta^2+\overline{v_\phi^2})/(2 \sigma_r^2)$.
We thus assume that
$\overline{v_\theta^2} = \overline{v_\phi^2}$
everywhere, in contrast to an alternative assumption that
$\sigma_\theta^2 = \sigma_\phi^2$, with the rotation support modelled
separately (e.g. \citealt{1995ApJ...449..592H}).
Strictly speaking, a spherical model is not warranted for a rotating system anyway,
but in the limit of weak rotation, it is a matter of adopting a convenient
dynamical approximation.  The rotational contribution does rise in the outer parts
of NGC~4494 to as high as $v/\sigma \sim 1.4$, potentially causing systematic issues
in our mass analysis which would require a more generalized study to quantify.}.
\item Project the radial component of the spatial velocity dispersion $\sigma_r$ (see Eq. \ref{eq4})
for comparison with the line-of-sight velocity dispersion data $\sigma_{\rm los}(R)$.
\item Iteratively adjust the free parameters in equation~\ref{eq:v0eqn},
to best fit the model to the observed dispersion profile.
\item Use the best-fit model (Eq.~\ref{eq:v0eqn}) in the Jeans equation 4-55 of
\citet{1987gady.book.....B} to calculate $M(r)$:
\begin{equation}
M(r) = -\frac{\sigma_r^2~r}{G}\left( \frac{d\ln j_*}{d\ln r}+\frac{d\ln\sigma_r^2}{d\ln r} +2\beta \right) ,
\label{eq:masseq}
\end{equation}
where $j_*(r)$ is the spatial density of the PNe, and corresponds
to an Abel deprojection of a smoothed density law fitted to the stellar data.
Additional quantities may then be computed, such as the cumulative $M/L$.
\end{enumerate}

\begin{figure}
\hspace{-0.5cm}
\epsfig{file=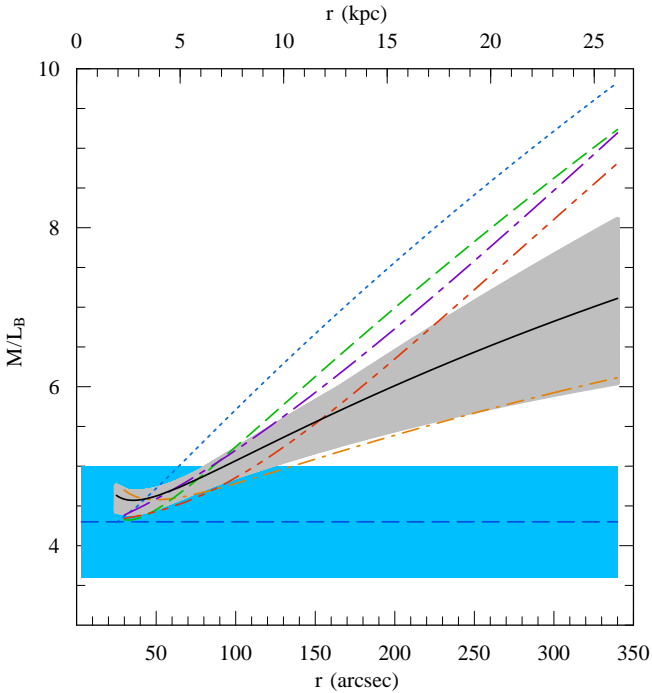,width=8.7cm}
\caption{Cumulative $B$-band mass-to-light ratio ($M/L$) of NGC~4494.
Most of the curves show results from the
mass-inversion method:
the black solid curve with grey shaded region shows the best-fit $\beta=0$ model with
its statistical uncertainties;
the orange dot-dashed and blue dotted curves show the best fit models with
$\beta=-0.5$ and $\beta=+0.5$, respectively;
the green short dashed curve shows the adopted $\beta(r)$ model of Eq.~\ref{eq:MLbeta}.
The purple dot-long-dashed line shows the best cosmologically-based model
[``NFW + $\beta(r)$''] and the red double-dot-dashed curve is the logarithmic potential model
with $\beta(r)$
(see \S\ref{sec:resNFW} and \S\ref{sec:resLOG}).
The horizontal blue long-dashed line with surrounding shaded region
shows the stellar $M/L$ and its uncertainty.}
\label{fig:modelfitstwo}
\end{figure}

Starting with the minimalist assumption of an isotropic galaxy ($\beta=0$),
we find that the simple model~(\ref{eq:v0eqn}) is able to fit the dispersion
data very well (Fig.~\ref{fig:modelfitsone}).
The resulting $M/L$ profile increases clearly with the radius
(Fig.~\ref{fig:modelfitstwo}),
providing a strong indication for the presence of an extended DM halo.
Using 5\Re{} ($241''$) as a benchmark value,
we find that the $B$-band $M/L$ within this radius is
$\Upsilon_{{\rm 5}, B} = 6.3^{+0.6}_{-0.7}$~$\Upsilon_{\odot,B}$,
where quoted errors
account for the 1-$\sigma$ confidence region in the parameter space
(${\sigma_0,r_0,\eta}$) of the
dynamical model.
As we will see in \S\ref{sec:massmod}, the derived value for $\Upsilon_5$ is
sensitive to assumptions about $\beta(r)$ but not about the detailed
form for $M(r)$, and so is useful as a robust quantity for comparison to theory.

The shaded regions in Figs.~\ref{fig:modelfitsone} and \ref{fig:modelfitstwo} show
the statistical modelling uncertainty, but there remains the {\it systematic}
uncertainty implied by the anisotropy in step (ii) of the procedure.
Adopting constant values of $\beta=\pm$~0.5 for a plausible (though not exhaustive)
range of the stellar anisotropy, we can find fits to the data just as
good as in Fig.~\ref{fig:modelfitsone}.
However, the famous mass-anisotropy degeneracy implies very different $M(r)$ profiles,
as shown in Fig.~\ref{fig:modelfitstwo}.  Assuming $\beta=+0.5$ implies more DM
than with $\beta=0$, while $\beta=-0.5$ implies a smaller but {\it non-zero} amount
of DM (i.e. a constant $M/L$ is excluded at the 1~$\sigma$ level).
Note though that the kurtosis values found in \S\ref{sec:higher} suggest
$\beta \gsim 0$ and thus that the higher-mass solutions are to be preferred
(as will be explored in the next section).

As a final refinement, we wish to test
an anisotropy profile based on theoretical expectations.
In general, collisionless systems such as galaxy haloes and clusters that were
formed through mergers should have some degree of radial anisotropy
(e.g., \citealt{1999MNRAS.309..610D,2000ApJ...539..561C,2004MNRAS.351..237R,2005MNRAS.361L...1W,2006MNRAS.365..747A})---a
prediction that has found empirical support from
the Galactic halo and from elliptical galaxies
(\citealt{2000AJ....119.2843C,2001AJ....121.1936G}; DL+08a,b).
In particular, M{\L}05 have introduced a radially-varying anisotropy
profile,
\begin{equation}\label{eq:MLbeta}
\beta(r)= \beta_0 {r\over r+r_{\rm a}} ,
\end{equation}
finding $\beta_0 \simeq 0.5$ and
$r_{\rm a}\simeq1.4\Re$ when applied to the merger simulations of D+05.
Adopting this profile with $r_{\rm a}=65''$,
we find a $M/L$ profile that is slightly shallower than the pure $\beta=+0.5$ model
(Fig.~\ref{fig:modelfitstwo}).

The overall plausible range for the benchmark-radius $M/L$ of NGC~4494 is
$\Upsilon_{{\rm 5},B}=5.0$--9.1~$\Upsilon_{\odot,B}$
(including both statistical and assumed systematic $\beta$ uncertainties).
This result is very similar to the
pseudo-inversion mass approach
from R+03,
which was based on a different photometric model and a much more limited data set:
$\Upsilon_{{\rm 5},B}=5.6$--8.0~$\Upsilon_{\odot,B}$
(distance assumption revised, and only the $\beta$ uncertainty included).

To further contextualise our result, we calculate the $M/L$ gradient
parameter introduced by N+05:
\begin{equation}
\dML \equiv \frac{\Re \Delta \Upsilon}{\Upsilon_{\rm in} \Delta R} ,
\end{equation}
where $\Upsilon_{\rm in}$ is the central dynamical $M/L$.
This parameter is independent of the bandpass and the distance, and is
fairly insensitive to the modelling details within the spherical Jeans formalism
(with the exception of anisotropy).
For simple interpretations we can assume that $\Upsilon_{\rm in}$ is dominated
by the stars as in the case of NGC~4494, but independently of this assumption,
the \dML{} parameter has been found to show strong observational correlations
with other galaxy properties (N+05).
For NGC~4494 we find $\dML =$~0.02--0.23.
This result is nearly identical to our previous finding, placing
the galaxy among the population of low-$\dML$ (i.e. low-DM) systems
(see N+05 and \citealt{2005ApJ...623L...5F}).

For NGC~3379, the
mass-inversion results of D+07 imply $\dML =$~-0.03--0.14.
The two galaxies thus appear to have consistent amounts of DM given the total
uncertainties, but assuming the {\it same} anisotropy profile for both,
the two galaxies are just barely consistent at the 1~$\sigma$ level.
The reason for this difference is that the dispersion profile of NGC~4494
is shallower than that of NGC~3379 (see \S\ref{sec:dispsec}),
while its stellar luminosity falls off more {\it steeply} (which is a factor
that should {\it steepen} the dispersion profile).

We can also independently estimate the amount of mass contained in the stars
(i.e. the stellar $M/L$, $\Upsilon_*$)
via independent modelling of the stellar population based on line strengths.
Unfortunately, there is not a radially-extended population map of NGC~4494 available
as for NGC~3379,
and we are forced to resort to constraints on its very central regions.
Based on luminosity-weighted line strengths reported inside $\sim 1.2''$
\citep{1998ApJS..116....1T}, we use the on-line \citet{1994ApJS...95..107W}
models\footnote{http://astro.wsu.edu/worthey/dial/dial\_a\_model.html}
to estimate an age of $\simeq$~10.6~Gyr and a metallicity of [Fe/H] $\simeq +0.21$,
implying $\Upsilon_{*,B} \simeq$~$4.3\pm0.7$~$\Upsilon_{\odot,B}$
(Kroupa IMF assumed, and $\sim20\%$ typical age errors adopted from
\citealt{2000AJ....120..165T})\footnote{Somewhat lower $\Upsilon_*$ values
are implied by the results of \citet{2001AJ....121.1936G}, \citet{2005MNRAS.358..813D}
and \citet{2008A&A...487..177Z}},
while higher values would be obtained using a Salpeter IMF.
A caveat here is that
central line indices in general are very likely to be contaminated by young
stellar subpopulations confined to the central regions (e.g. in the nuclear
disc of NGC~4494) and thus to produce lower $\Upsilon_*$ values than in
the halo regions that concern us.

We plot the $\Upsilon_*$ value in Fig.~\ref{fig:modelfitstwo}, and see that
it is similar to the dynamically-inferred $M/L$ in the galaxy's centre.
Although there are large uncertainties in this comparison, the coincidence
of the best-guess $M/L$ values suggests that
the stars contribute the large majority of the mass in the central regions
(similar conclusions were found in the case of NGC~3379 by D+07;
see also \citealt{2001AJ....121.1936G}).
It is only outside $\sim$~2~\Re{} where the effects of DM become noticeable.
In the next section we will attempt to constrain $\Upsilon_*$ by dynamical
means, and to reconstruct the DM profile in more detail.

\subsection{Multi-component models}\label{sec:massmod}

We now move beyond the
pseudo-inversion mass
method used in R+03, D+07
and \S\ref{sec:massinv}, and construct physically-based dynamical
models of NGC~4494, allowing for more direct comparisons of observation and theory.
Along with these models, we
include a
kurtosis-based
application for alleviating the mass-anisotropy degeneracy.
This approach has similarities to the orbit models used for NGC~3379
in R+03, as we will discuss in \S\ref{sec:dynmeth}.
It is outside the scope of this paper to construct fully rigorous and general dynamical models.
Rather, our aim is to generate plausible ``best-guess'' models that reproduce the broad features
of the data.

Our two-component mass model is simple but realistic, consisting of a luminous field star
distribution and (optionally) a DM halo.
The total gravitational potential may thus be expressed as
$\Phi=\Phi_*+\Phi_{\rm d}$.
The stellar gravitational potential $\Phi_*(r)$ is derived from the
stellar luminosity $j_*(r)$ (according to the S\'ersic model of \S\ref{sec:spatdist}),
combined with some assumed constant $\Upsilon_*$.

We first describe our suite of mass models, where the DM distribution follows either the
NFW profile (\S\ref{sec:lcdm}) or a pseudo-isothermal form (\S\ref{sec:LOG}).
Our dynamical methods are outlined in \S\ref{sec:dynmeth}, and our results given
in \S\ref{sec:dynres}--\ref{sec:resLOG}.

\subsubsection{NFW model}\label{sec:lcdm}

Our first suite of mass models is based on simulations of collisionless DM
halo formation in a $\Lambda$CDM cosmology.
In this case the DM density takes the approximate form of an NFW profile:
\begin{equation}
\rho_{\rm d}(r)=\frac{\rho_s}{(r/r_s)(1+r/r_s)^2} ,
\label{rhoNFW}
\end{equation}
where $\rho_s$ and $r_s$ are the characteristic density and scale radius of the halo.
The cumulative dark halo mass is
\begin{equation}
M_{\rm d}(r)=4 \pi \rho_s r_s^3 A(r/r_s) ,
\end{equation}
where
\begin{equation}
A(x) \equiv \ln (1+x)-\frac{x}{1+x}.
\end{equation}
The potential is:
\begin{equation}
\Phi_{\rm d}(r) = \frac{4\pi G \rho_s r_s^3}{r} \ln \left(\frac{r_s}{r+r_s}\right) ,
\end{equation}
where $G$ is the gravitational constant.

The three free parameters describing the NFW mass model are thus $\Upsilon_*$,
$\rho_s$ and $r_s$.
The halo can alternatively be parametrized by the virial mass and concentration,
$M_{\rm vir}\equiv 4\pi\Delta_{\rm vir}\rho_{\rm crit}r_{\rm vir}^3/3$ and
$c_{\rm vir}\equiv r_{\rm vir}/r_s$, where the critical density is
$\rho_{\rm crit}=1.37\times10^{-7} M_\odot$~pc$^{-3}$ and
the virial overdensity value is $\Delta_{\rm vir}=101$.
The expected values for these model parameters are not arbitrary in $\Lambda$CDM---a theme
to which we will return in \S\ref{sec:impl}.

\subsubsection{LOG model}\label{sec:LOG}

Our second mass model consists of a logarithmic potential
(\citealt{1987gady.book.....B} \S2.2.2)
which was motivated by observations of spiral galaxy rotation curves.
The potential is:
\begin{equation}
\Phi_{\rm d}(r)=\frac{v_0^2}{2}\ln(r_0^2+r^2) ,
\label{potLOG}
\end{equation}
where $v_0$ and $r_0$ are the asymptotic circular velocity and core radius of the halo.
The corresponding DM density and cumulative mass profiles are respectively:
\begin{equation}
\rho_{\rm d}(r)=\frac{ v_0^2 (3 r_0^2+r^2)}{4\pi G (r_0^2+r^2)^2} ,
\label{rhoLOG}
\end{equation}
and
\begin{equation}
M_{\rm d}(r)=\frac{1}{G}\frac{v_0^2 r^3}{r_0^2+r^2} .
\label{massLOG}
\end{equation}
The three free parameters of this ``LOG'' model are thus
$\Upsilon_*$, $v_0$ and $r_0$.
We define a virial mass according to the same definition as in \S\ref{sec:lcdm}.

Unlike the cuspy $r^{-1}$ density core of the NFW halo, the DM density of the
LOG halo is constant in the centre.
Outside of the core, the density decreases as $r^{-2}$ to produce a constant
circular velocity profile $v_{\rm c}\equiv GM(r)/r^2$,
similar to the NFW model near $r=r_s$.
Such a model allows us to maximize the stellar contribution to the mass
in the central regions, and thus test a ``minimal DM halo'' scenario.
Similar models have been used to successfully fit galaxies of all types
(e.g. \citealt{1980MNRAS.193..189F,1991MNRAS.249..523B,1995ApJ...447L..25B,2000A&AS..144...53K,2007MNRAS.382..657T,2008MNRAS.383.1343W}; \citealt{2008MNRAS.385.1729D}, hereafter DL+08a; DL+08b).

\subsubsection{Dynamical methods}\label{sec:dynmeth}

The Jeans modelling approach employed here is more traditional than that of \S\ref{sec:massinv}:
one starts with a trial mass model, solves equilibrium equations for the internal
velocity moments, and projects those moments onto the sky for comparison with the data.
This approach typically means solving for the second-order velocity
moments (i.e., the velocity dispersions) and comparing to projected velocity dispersion data.
However, such data cannot constrain the mass model without additional strong assumptions;
the entire LOSVD in a spherical system is necessary to have any hope of uniquely determining
its phase-space distribution (e.g. \citealt{1992ApJ...391..531D}).
In practice,
the systematic uncertainties can be strongly reduced by using fourth-order moments of
the LOSVD to diagnose orbit anisotropies (e.g., \citealt{1993MNRAS.265..213G}).

One such method is to construct fourth-order Jeans equations
in addition to the usual second-order, and to constrain them with kurtosis data
(e.g. \citealt{2001MNRAS.322..702M,lok02,lokmam03}).
Although the higher-order Jeans equations are not closed in general,
one can adopt a simple choice for the distribution function which makes the
problem tractable (see Appendix~\ref{app:eqs}).
This simplification is arbitrary (e.g. $\beta$ is assumed to be constant
with radius) and does restrict the generality of our
results, but the model is still more general than an assumption of isotropy.
Given these restrictions, we can now constrain $\beta$ using the data.

The basic steps of our analysis are as follows, with details for steps
(ii) and (iii) provided in Appendix~\ref{app:eqs}:
\begin{enumerate}
\item Set up a multi-dimensional grid of model parameter space to explore, including $\beta$ and the mass profile parameters ($\Upsilon_*, \rho_s, r_s$) or ($\Upsilon_*, v_0, r_0$).\\
\item For each model grid-point, solve the 2$^{\rm nd}$- and 4$^{\rm th}$-order Jeans equations.\\
\item Project the internal velocity moments to $\sigma_{\rm los}$ and $\kappa_{\rm los}$.\\
\item Compute the $\chi^2$ statistics, defined as
\begin{equation}
\chi^2=\sum_{i=1}^{N_{\rm data}} \left[\frac{p^{\rm obs}_i-p^{\rm mod}_i}{\delta p^{\rm obs}_i}\right]^2 ,
\label{chi2}
\end{equation}
where $p^{\rm obs}_i$ are the observed data points ($\sigma_{\rm los}$ and $\kappa_{\rm los}$),
$p^{\rm mod}_i$ the model values, and
$\delta p^{\rm obs}_i$ the uncertainties on the observed values,
all at the radial position $R_i$.
We fit the PN data outside 40$''$ and the stellar dispersion between 20$''$ and 60$''$.
The stellar kurtosis is used between 30$''$ and $60''$.
\\
\item Find the best fit parameters minimising the $\chi^2$.
In practice, we find that the VD is affected by both the mass and anisotropy profiles,
while the kurtosis is almost entirely driven by the anisotropy.
\end{enumerate}

For comparison, R+03 modelled NGC~3379 with a similar suite of spherical mass profiles in
non-parametric orbit models (A.K.A. ``Schwarschild's method'') that included direct fits to
the stellar $h_4$ moments as well as to the full PN LOSVDs.
Such methods (see also \citealt{2008ApJ...682..841C} and DL+08b) may be more powerful than our Jeans
approach, but the latter is computationally faster and somewhat more intuitive.

\begin{table*}
\caption{Summary of best-fit multi-component model parameters.}
\label{tab:jeanssumm}
\scriptsize
\noindent{\smallskip}\\
\hspace{-1.5cm}
\begin{tabular}{lcccccccccccc}\hline\hline
Model  & $\beta_5$$^1$ & \Ystar$^2$& log ${M_*}^3$ & $c_{\rm vir}$$^4$ & log ${M\vir}^5$& $f\vir$$^6$ & $f_{\rm DM,5}$$^7$ & $\Upsilon(\Re)$$^8$ & $\Upsilon_{B5}$$^9$ & $\Upsilon(R\vir)$$^{10}$ &$\nabla_{\ell} \Upsilon$$^{11}$ & $\chi^2$/d.o.f.$^{12}$ \\
 & & (\Ysol)  & (\Msun) &  & (\Msun) &  &  & (\Ysol)  & (\Ysol)  & (\Ysol)  & &  \\
\hline\hline
\multicolumn{12}{c}{No-DM model} \\
\hline
star iso & 0 & $4.6$ & 11.04$\pm$0.06 & -- & 11.04$\pm$0.06 & 0 & 0 & 4.6 & 4.6 & 4.6 & 0 & 25/37\\
star tang & -1.3 & $4.4$ & 11.02$\pm$0.06 & -- & 11.02$\pm$0.06 & 0 & 0 & 4.4 & 4.4 & 4.4 & 0 & 27/36\\
\hline
\multicolumn{12}{c}{NFW model} \\
\hline
NFW iso & 0 & $4.1$ & 10.99$^{+0.05}_{-0.07}$ & 6.9$^{+0.5}_{-0.4}$ & 12.03$\pm$0.12& 10$\pm$3 & 0.41$\pm$0.08 &4.5$^{+0.2}_{-0.1}$ & 7$^{+1}_{-1}$ & 45$^{+10}_{-8}$ &0.14 & 21/35\\
NFW+$\beta_0$& 0.46$\pm$0.15 & $4.3$ & 11.01$^{+0.05}_{-0.06}$ & 6.5$^{+1.6}_{-0.6}$ & 12.17$^{+0.16}_{-0.20}$ & 13$\pm$5 &  0.42$\pm$0.07 & 4.7$^{+0.3}_{-0.2}$ & 7$^{+2}_{-1}$ & 62$^{+21}_{-19}$ &0.15 & 10/34 \\
NFW+$\beta(r)$ & 0.43$\pm$0.05 & $4.1$ & 10.99$^{+0.05}_{-0.07}$& 8.3 $^{+0.9}_{-0.7}$ & 12.05$\pm$0.10& 11$\pm$3 & 0.44$\pm$0.08 & 4.6$^{+0.2}_{-0.2}$ & 8$^{+1}_{-1}$ & 47$^{+8}_{-6}$ & 0.14 & 5/22\\
\hline
\hline
\multicolumn{12}{c}{LOG model}\\
\hline
Model &  $\beta_5$$^1$ & \Ystar$^2$ & log ${M_*}$$^3$ & $v_0^{13}$  & log $M\vir^{5}$ & $r_0^{14}$ & $f_{\rm DM,5}$$^7$ & $\Upsilon(\Re)$$^8$ & $\Upsilon_{B5}$$^9$ & $\Upsilon(R\vir)$$^{10}$ &$\nabla_{\ell} \Upsilon$$^{11}$ & $\chi^2$/d.o.f.$^{12}$ \\
 & & (\Ysol) & (\Msun) & (kms$^{-1}$)  & ($M_\odot$) & (arcsec) &  & (\Ysol) & (\Ysol) & (\Ysol) & &  \\
\hline
LOG iso & 0 & $4.3$ & 11.01$^{+0.07}_{-0.06}$ & 147 & 12.19$\pm$0.18& 244 & 0.34$\pm$0.14 & 4.4$^{+0.1}_{-0.1}$ & 7$^{+2}_{-1}$ & 65$^{+23}_{-20}$ &0.11 & 12/35\\
LOG+$\beta_0$& 0.44$\pm$0.10 & $4.3$ & 11.01$^{+0.05}_{-0.06}$ & 172 & 12.39$\pm$0.15& $205$ & 0.45$\pm0.13$ & 4.4$^{+0.2}_{-0.1}$ & 8$^{+2}_{-2}$ & 100$^{+30}_{-25}$ &0.18 & 10/34 \\
LOG+$\beta(r)$ & 0.43$\pm$0.07 & $4.3$ & 11.01$^{+0.05}_{-0.06}$& 146 & 12.19$\pm$0.18& $190$ & 0.39$\pm0.16$ & 4.4$^{+0.2}_{-0.1}$ & 7$^{+2}_{-1}$ & 65$^{+25}_{-20}$ & 0.15 & 4/22\\
\hline
\hline

\end{tabular}
\noindent{\smallskip}\\
\begin{minipage}{16.5cm}

NOTES:
$1$) Anisotropy at the benchmark radius of 5\Re;
$2$) dynamical stellar mass-to-light ratio $M/L$, in $B$-band Solar units:
typical uncertainty is $\pm0.2$\Ysol;
$3$) log of stellar mass in solar units;
$4$) concentration parameter (see \S\ref{sec:lcdm});
$5$) log of virial mass;
$6$) ratio of total dark and luminous matter within the virial radius, $f\vir=M_{\rm d}/M_*$ at $r_{\vir}$;
$7$) dark matter fraction, $f_{\rm DM}=M_{\rm d}/(M_{\rm d}+M_*)$ at  5\Re;
$8$) dynamical $M/L$ at \Re;
$9$) dynamical $M/L$ at 5\Re;
$10$) dynamical $M/L$ at the virial radius;
$11$) $M/L$ logarithmic gradient;
$12$) $\chi^2$ statistic (see text for details of data included);
$13$) asymptotic circular velocity: typical uncertainty is $\pm30\kms$;
$14$) halo core radius: typical uncertainty is $\pm50''$.
\end{minipage}
\noindent{\smallskip}\\
\end{table*}

\subsubsection{Results: no-DM case}\label{sec:dynres}

\begin{figure*}
\centering
\epsfig{file=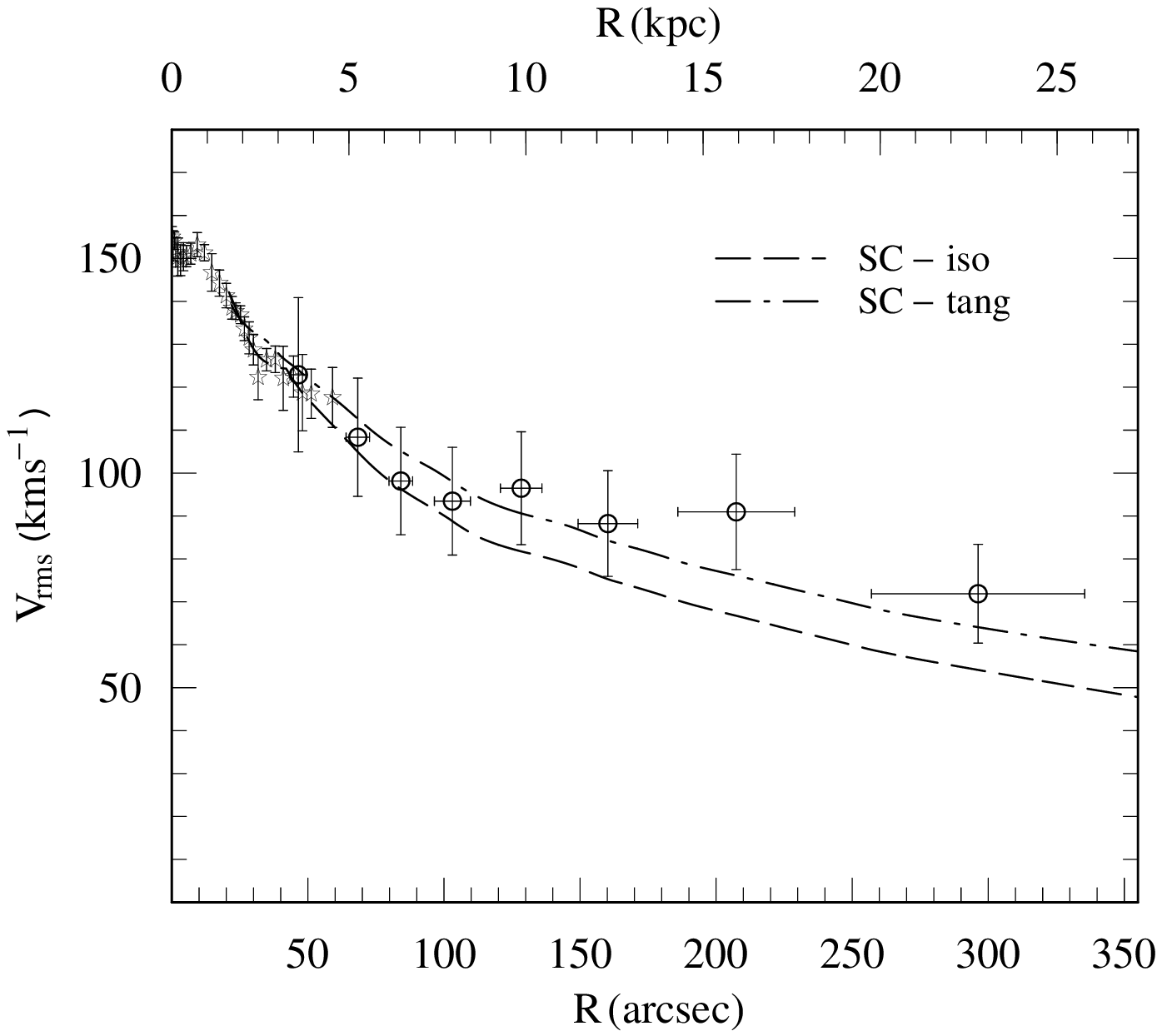,width=8.cm}
\epsfig{file=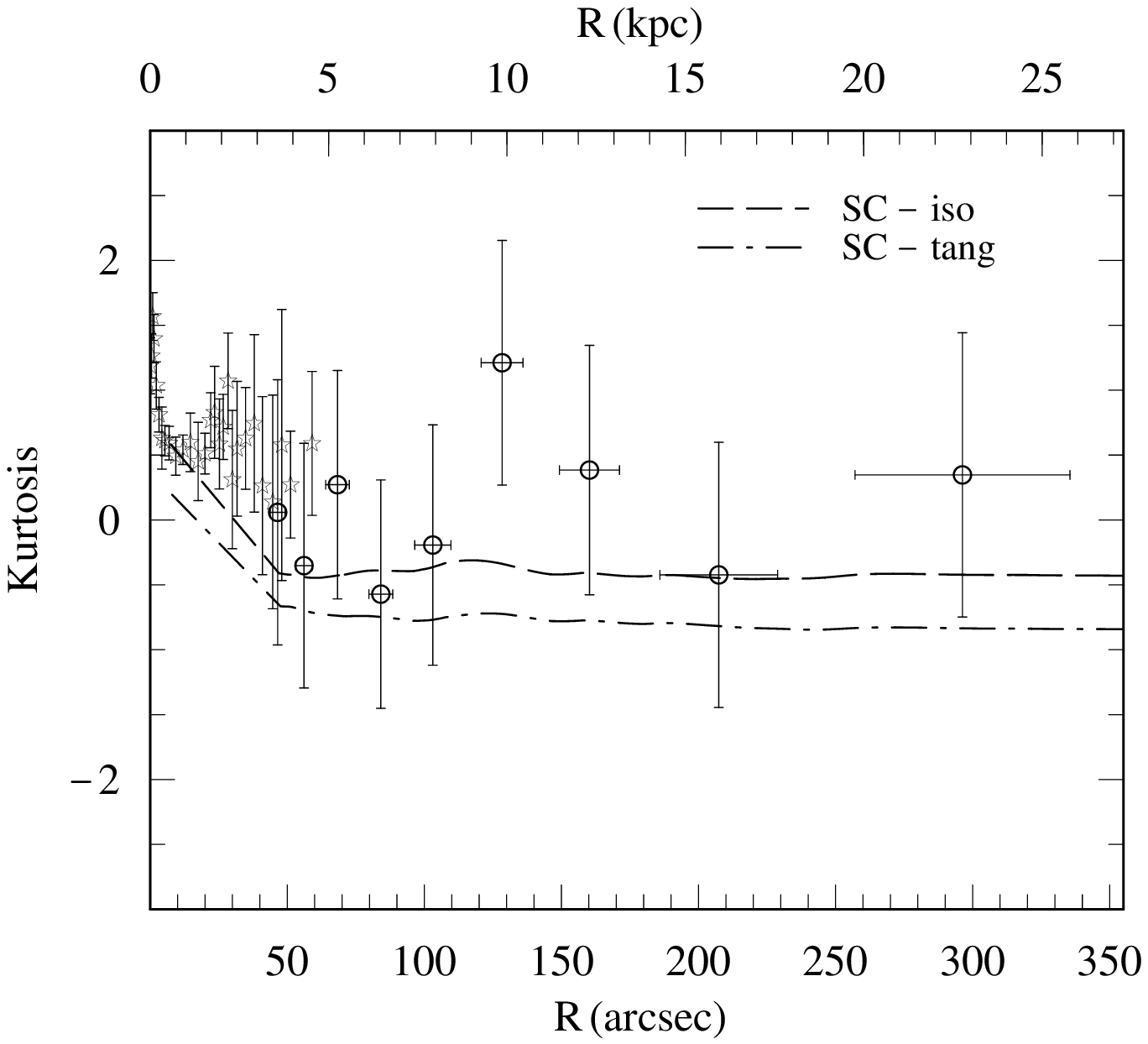,width=8.cm}
\epsfig{file=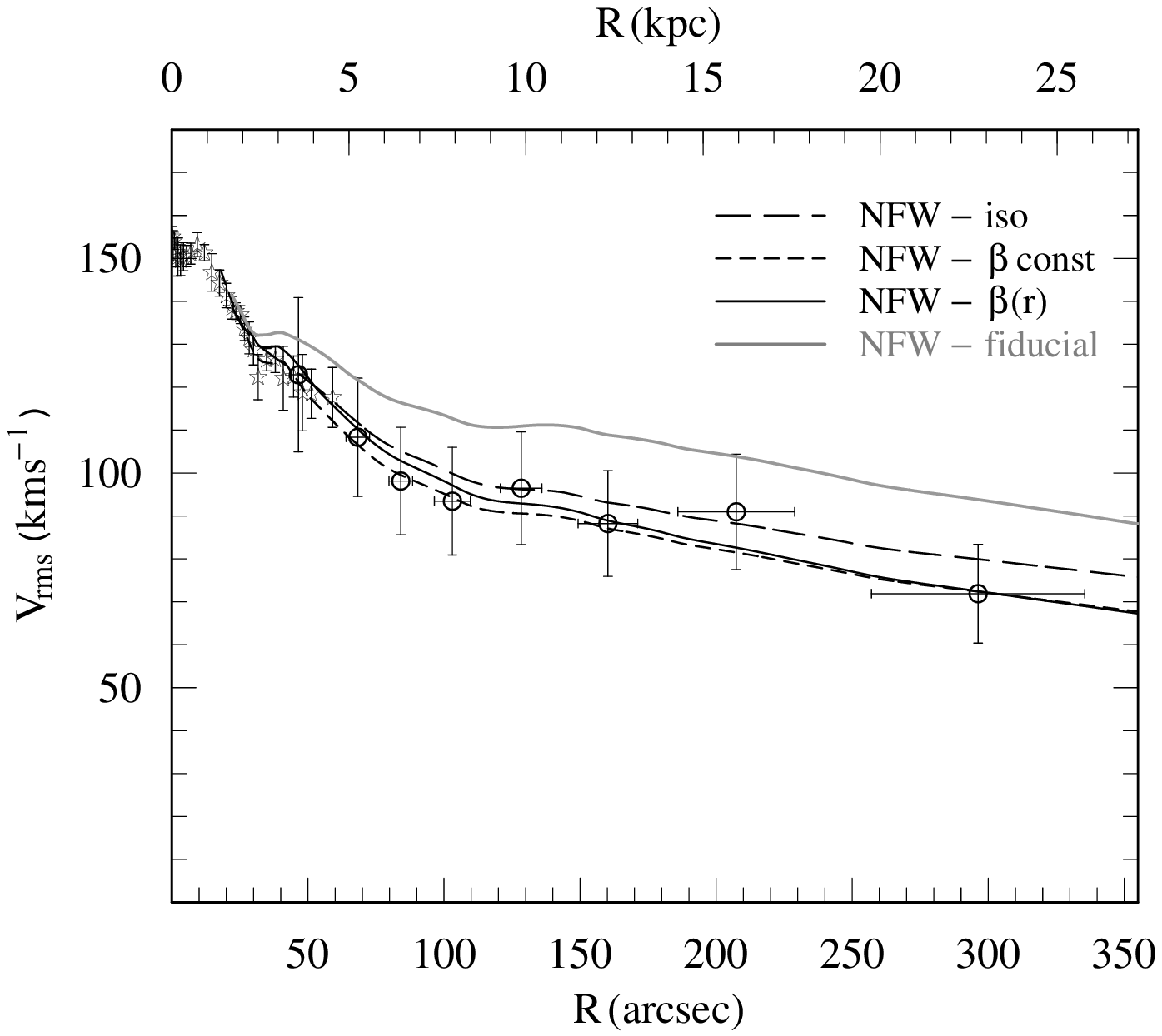,width=8.cm}
\epsfig{file=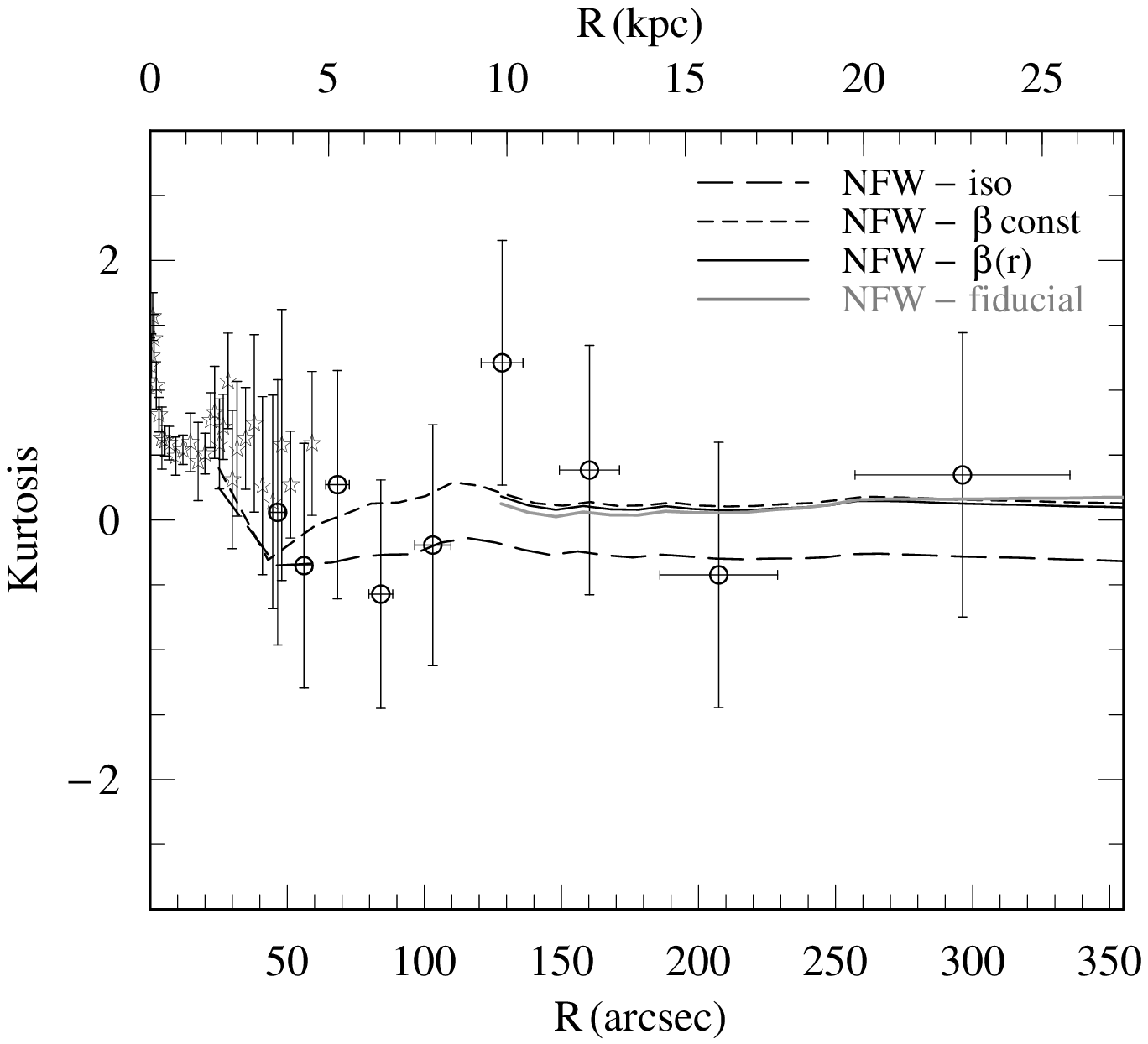,width=8.cm}
\epsfig{file=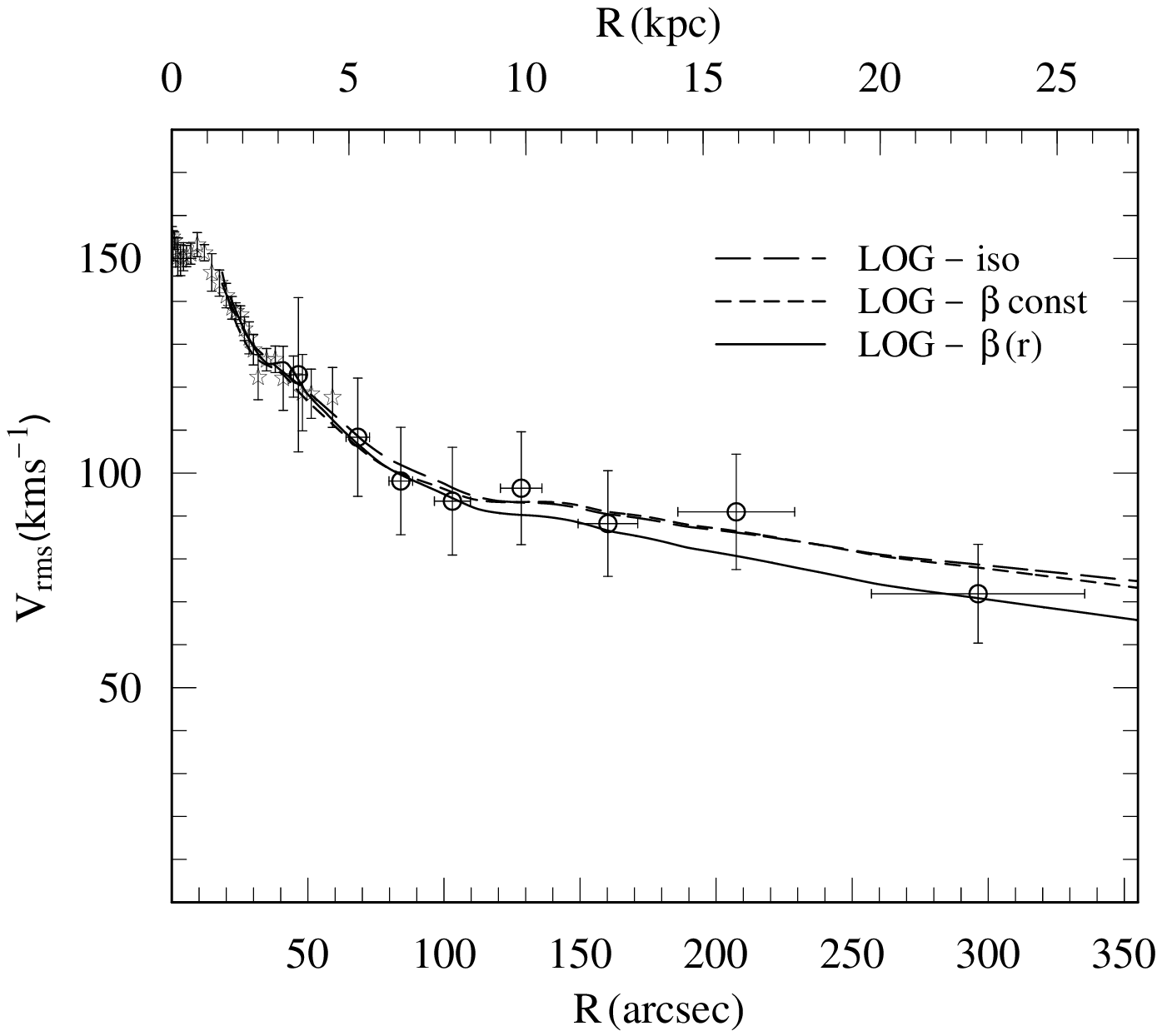,width=8.cm}
\epsfig{file=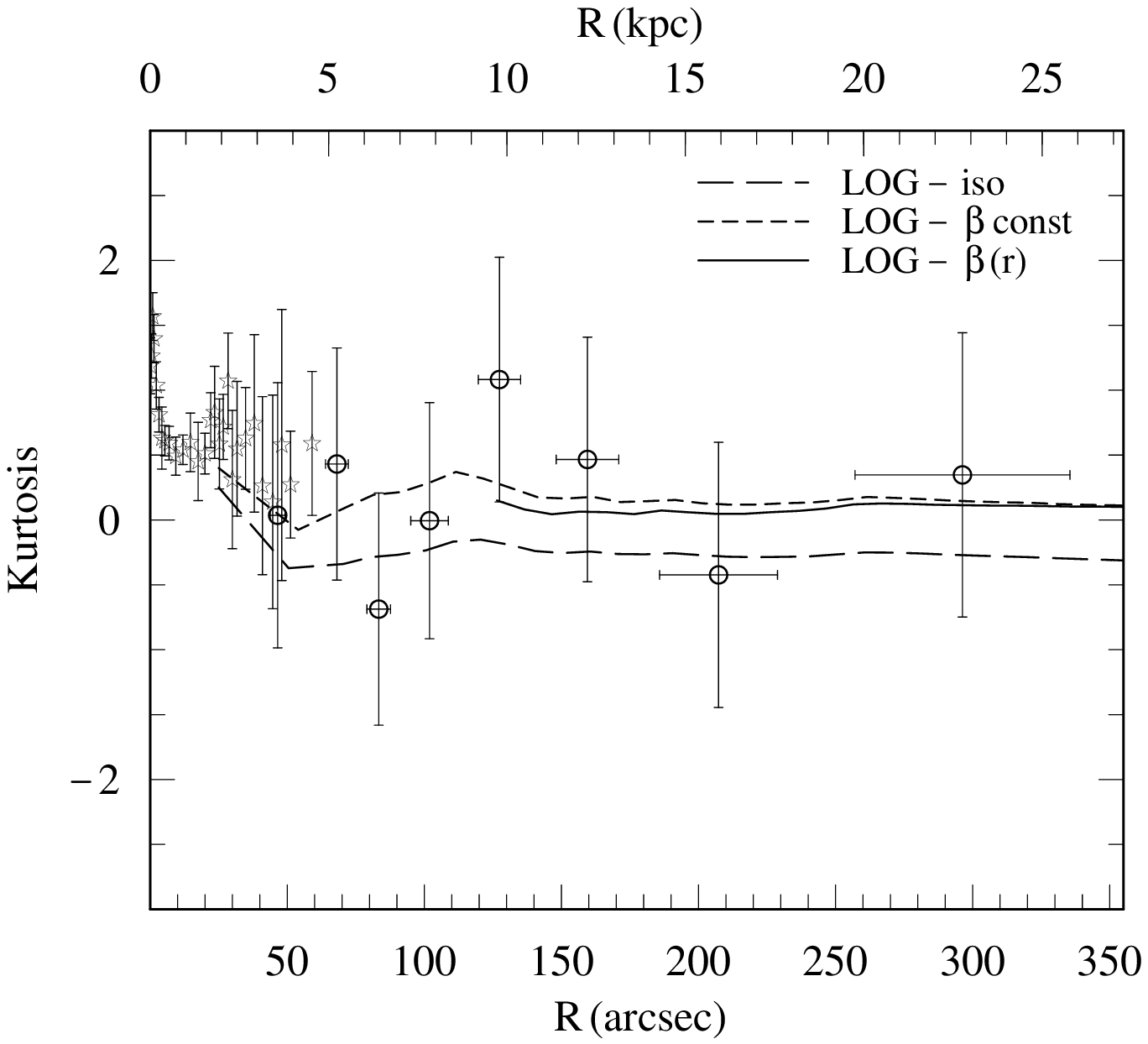,width=8.cm}
\caption{Multi-component Jeans model fits to the NGC~4494 kinematics data.
The stellar data are shown by star symbols, while the PN data are open circles.
The left panels show the projected RMS velocity profiles, while the right panels
show the projected kurtosis.
The top panels show fits from self-consistent stars-only models, the central panels
show stars+NFW halo models and the bottom panels show
stars+LOG halo models.
The curves correspond to models as in the panel legends.
See text for details.}
\label{fig:SCall}
\end{figure*}

We start with the simplest model, assuming a pure-stellar potential ($\rho_s=0$
or $v_0=0$) and isotropy for the velocity ellipsoid ($\beta=0$).
The best-fit parameters of the model are given in Table~\ref{tab:jeanssumm} together with
their $\chi^2$ values.
Given the freedom to adjust $\Upsilon_*$, this model can fit the VD in the
central regions $(\lsim 2 \Re)$, but falls off too quickly in the outer regions
(Fig.~\ref{fig:SCall}, upper left).
Although the kurtosis is included in the fit
(Fig.~\ref{fig:SCall}, upper right), omitting these constraints would not
provide an appreciably better fit to the VD, since the $M(r)$ and $\beta$ assumptions
already specify fully the shape of the VD.
Note that in R+03 we {\it were} able to fit a pure-stellar model to the VD data,
with the difference caused by a slightly steeper new stellar luminosity profile,
by a $\sim$~10\% lower VD normalisation from the new stellar kinematics data,
and by a shallower outer VD slope as derived from our larger, more extended new PN data set.

In order to increase the VD at large radius, we experiment with introducing a
degree of tangential anisotropy ($\beta < 0$).
As seen in Fig.~\ref{fig:SCall}, this strategy does improve the VD fit
(best-fit $\beta=-1.3$),
but requires a negative kurtosis in the halo that is ruled out by the PN data.
Given the observational uncertainties in the stellar luminosity profile $j_*(r)$
(\S\ref{sec:spatdist}), we test out a shallower outer profile
(S\'ersic index $m=5.2)$\footnote{Our
surface photometry is originally $g$-band transformed to the
$B$- and $V$-bands, so the true stellar mass density profile ($\sim K$-band)
is likely to be {\it steeper}.}.
We find that we can indeed improve the fit to the VD, but the kurtosis is
still problematic for this model.
The use of higher-order velocity moments therefore leads us to conclude that there is
a non-zero amount of DM in NGC~4494 (as also found by R+03, D+07 and DL+08b for NGC~3379).

Given the presence of DM in this system, we next determine what halo parameters
are best consistent with the data for the two assumed DM profiles.\\

\subsubsection{Results: NFW model}\label{sec:resNFW}
Once again we start by assuming isotropy, and find a best fit as shown in
the central panels of Fig.~\ref{fig:SCall},
with parameters again reported in Table~\ref{tab:jeanssumm}.
Note that inside $\sim 50'' \sim \Re$, both the stars-only and the stars+DM models
can recover the VD profile with simple assumptions, suggesting that in
these regions it is the stellar mass that determines the main kinematical features.

\begin{figure*}
\centering
\epsfig{file=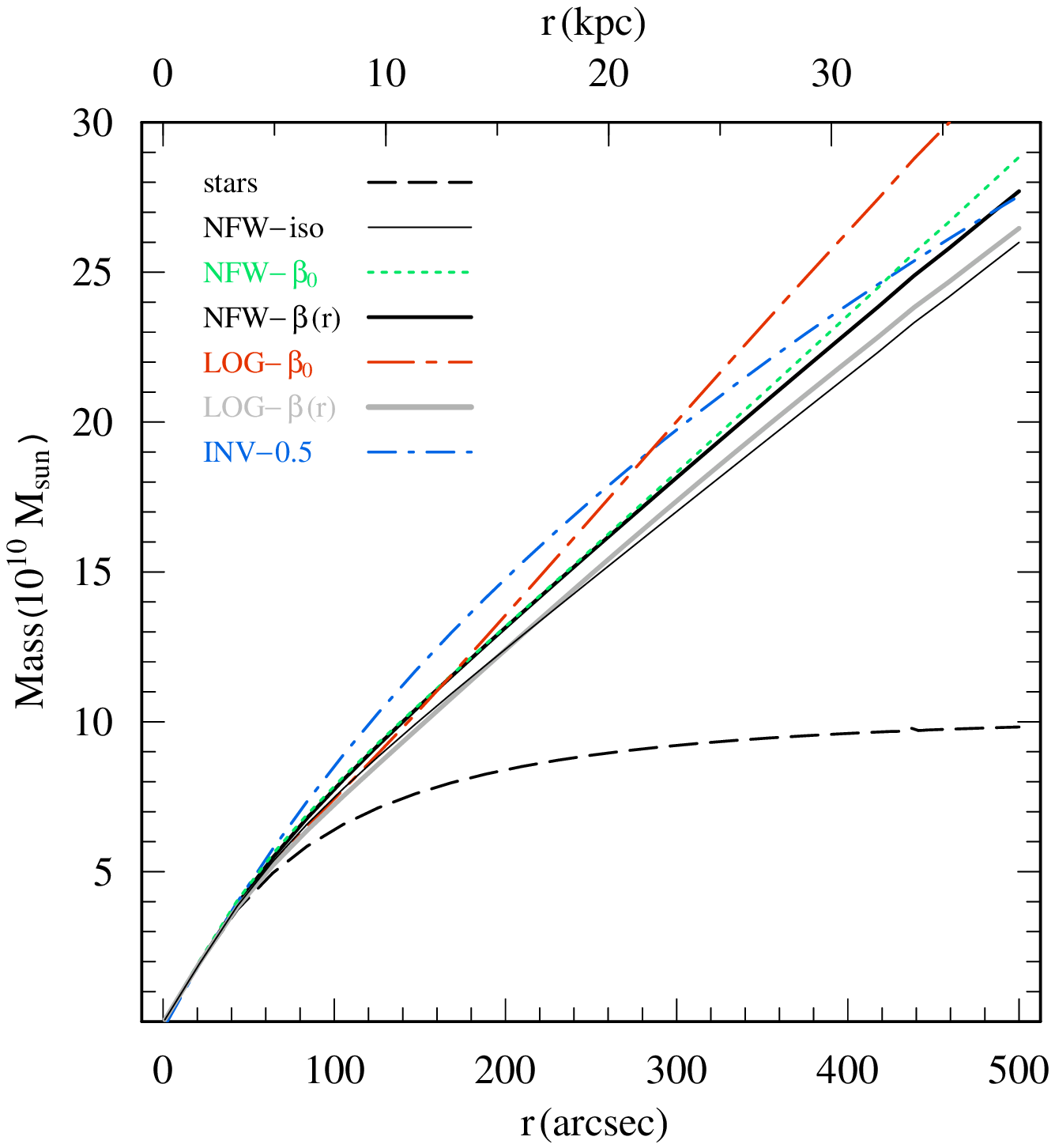,width=8.cm}
\epsfig{file=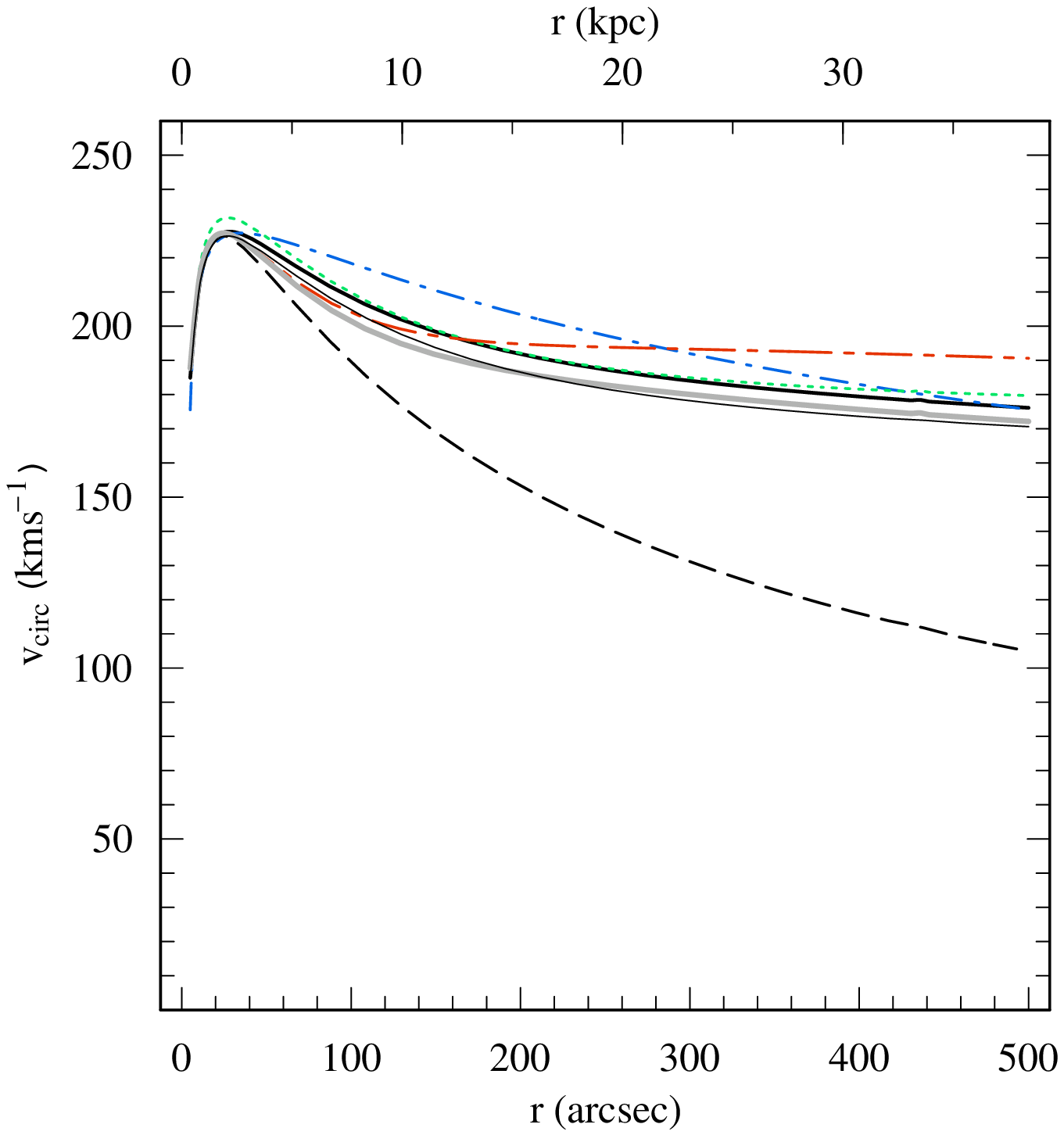,width=8.15cm}
\caption{Radial mass distribution of NGC~4494.
The left panel shows the cumulative mass, and the right panel shows the circular velocity profile.
The heavy solid curves are the best-fit stars+DM model with varying anisotropy (black is NFW, gray LOG) while the light black solid line is the isotropic NFW case;
the dashed curve is the stellar distribution for the LOG model (\Ystar=4.3\Ysol).
The green dotted curves show the $\beta=0.5$ stars+NFW model and the red dot-long-dashed curves the
$\beta=0.5$ stars+LOG model while the blue dot-dashed curves show the $\beta=0.5$
pseudo-inversion mass result.
A legend for the model symbols is in the left panel.}
\label{fig:masscomp}
\end{figure*}

The isotropic NFW solution is a reasonably good match to the data, but it
can be seen that the predicted outer VD is somewhat too high,
and the predicted kurtosis somewhat low\footnote{Although a galaxy halo with $\beta=0$ and
constant $v_{\rm c}$ should result in $\kappa_{\rm los}=0$, our mass model deviates
enough from this picture to produce non-zero kurtosis.}.
Therefore we next allow for the anisotropy $\beta$ to be a free parameter,
though constant with the radius.
The best-fit solution has a $\chi^2$ value decreased by 9, although
there is only one less degree of freedom.
This implies that the anisotropic model is strongly preferred over the isotropic one,
as also visible in Fig.~\ref{fig:SCall}, where both the VD and the kurtosis are
at last reproduced well at all fitted radii.
The solution has a moderate degree of radial anisotropy ($\beta\sim0.5$), and
is formally preferred at the 3~$\sigma$ level over the best
anisotropic stars-only solution;
the inferred mass profiles are dramatically different, with the NFW model
$v_{\rm c}$ remaining much flatter with radius than the stellar model
(see Fig.~\ref{fig:masscomp}).
The radial anisotropy decreases the VD at $\sim\Re$ and requires a slightly higher
$\Upsilon_*$ value than in the isotropic case.

\begin{figure}
\centering
\epsfig{file=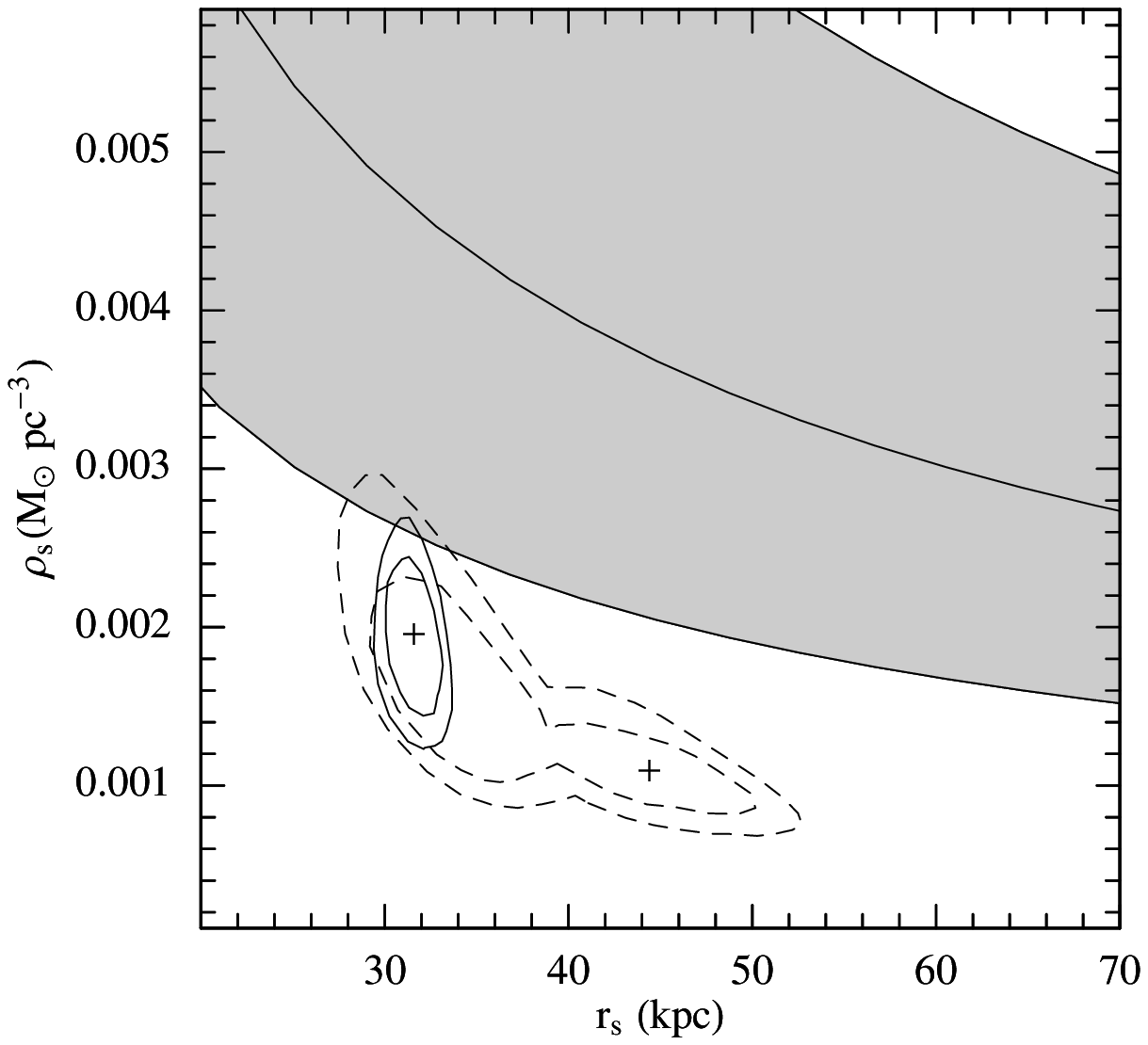,width=8.5cm}
\caption{Halo parameter solutions for NGC~4494 NFW-based model,
plotted as halo scale radius $r_s$ vs central density $\rho_s$.
The contours show the two-parameter $\chi^2$ confidence levels (1 and $2\sigma$).
Dashed contours show the constant-$\beta$ model, and solid contours the $\beta(r)$ model.
The best-fit solutions are marked with crosses.
The solid line with shaded region shows the theoretical expectations for the
halo parameters (see \S\ref{sec:halo}) and the 1--$\sigma$
scatter.
}
\label{fig:parcont}
\end{figure}

The halo parameters in this anisotropic analysis are illustrated by Fig.~\ref{fig:parcont},
which shows the joint region of permitted values for $r_s$ and $\rho_s$, marginalized
over the other free parameters, $\Upsilon_*$ and $\beta_0$.
It can be seen that there
is a limited degeneracy between the parameters, with alternatively low-$\rho_s$ and
high-$r_s$ models (corresponding to high $M_{\rm vir}$ and low $c_{\rm vir}$), or
high-$\rho_s$ and low-$r_s$ (low $M_{\rm vir}$ and high $c_{\rm vir}$).

The best-fit anisotropy value in our modelling of $\beta=0.46$
is remarkably similar to theoretical expectations
for galaxy haloes (see \S\ref{sec:massinv}).
Therefore in order to make an improved estimate of a plausible $\Lambda$CDM halo
that would reproduce our data, we adopt Eq.~\ref{eq:MLbeta} for the
anisotropy profile $\beta(r)$.
This model is isotropic in the centre and radially anistropic in the outer parts,
where
the transition radius $r_{\rm a}$ is free to vary (within the range 30$''$ to 110$''$).
Because our Jeans modelling procedures are not entirely designed to cope with
a radially-varying $\beta$ (see Appendix~\ref{app:eqs}),
we adopt a slightly different procedure than the preceding.
We fit the VD data as before, but fit the kurtosis data only at large radii ($R > 100''$),
where the anisotropy is fairly constant, and considered fixed to its value at 200$''$ ($\sim 4 \Re$); see eqn.~(\ref{eq7}).

This final best-fit theoretically-motivated model is again a good match to the data
(see Fig.~\ref{fig:SCall} and $\chi^2$ in Table~\ref{tab:jeanssumm}).
The anisotropy transition radius is $r_{\rm a}=45''\pm20''$, i.e.
radial anisotropy may set in at smaller radii in NGC~4494 than in the simulations of D+05.
The stellar $M/L$ of this solution of $\Upsilon_{*,B}=4.1\pm0.2$~$\Ysol$ agrees
remarkably well with the independent estimate from stellar populations of
$\Upsilon_{*,B} \simeq 4.3$~$\Ysol$ (\S\ref{sec:massinv}).
The central NFW halo parameters of
$\rho_s=0.0019\pm0.0006$~$ M_\odot$~pc$^{-3}$ and $r_s=410''\pm30''=32\pm2$~kpc
correspond to a virial radius, mass and concentration of
$r_{\rm vir}=261\pm30$~kpc,
$M_{\rm vir}=(1.12^{+0.33}_{-0.24})\times 10^{12} M_\odot$ and
$c_{\rm vir}\sim8\pm1$.

As shown in Fig~\ref{fig:parcont}, this $\beta(r)$ solution is better constrained than the
constant-$\beta$ one, and coincides with its secondary $\chi^2$ minimum---showing that the
results were not sensitive to the potentially worrisome stellar kurtosis
points between 30$''$ and 60$''$.
The data in both models are
inconsistent at the $\sim1$~$\sigma$ level with the predicted relation from $\Lambda$CDM
(shown in the same Figure, see Eq.~\ref{eq:rho_rs} in \S\ref{sec:halo}).
As a visual demonstration of the significance of this result,
in Fig.~\ref{fig:SCall} we show also a ``typical'' expected $\Lambda$CDM halo,
with $c_{\rm vir}=12.3$ and $M_{\rm vir}=2\times10^{12}M_\odot$,
and a $\beta(r)$ profile with $r_{\rm a}=67''$. This model is a relatively poor fit
to the dispersion profile, and required a smaller \Ystar (=3.4\Ysol) in order to match the central
data points. On the other hand, the kurtosis is well matched, as the anisotropy profile is similar
to our best-fit NFW+$beta(r)$ model.

\subsubsection{Results: LOG model}\label{sec:resLOG}

We next carry out an equivalent model sequence for the LOG halo model,
with results shown in Fig.~\ref{fig:SCall} (bottom) and Table~\ref{tab:jeanssumm}.
For the cases of $\beta=0$ or $\beta=$const,
the LOG model can fit the data significantly better than the NFW model,
owing to the former's slower transition with radius from stellar to DM dominance.
In the case of variable anisotropy, the LOG and NFW models have comparably good fits
and similar $\beta(r)$ profiles, but in the context of the LOG model,
we cannot distinguish between the different $\beta$ models.
However, we will adopt the $\beta(r)$ solution as the best LOG model because of
its theoretical motivation.

The $M/L$ at 5~\Re{} is very similar for the $\beta(r)$ NFW and LOG models,
demonstrating that this quantity is well constrained by the data,
independently of the details of the mass models.
The mass profiles are also similar given the modelling uncertainties
(Fig.~\ref{fig:masscomp}), with the NFW model
assuming a relatively low concentration which mimics the constant-density core of
the LOG model, and with the NFW scale radius value of $r_s \sim 9 \Re$ producing a
roughly flat $v_{\rm c}$ profile in the halo regions as in a LOG model.
The remaining differences in the model profile shapes result in slightly
different values for \Ystar, which is lower for the NFW model because of the
stronger central DM contribution from its cuspy core.

Although we cannot at this stage distinguish between the NFW and LOG models,
there are possibilities for doing so in the future.
New data extending to larger radii could verify our current results and
provide stronger discriminants between
the mass models surveyed here and alternative ones---but would not be able
to discriminate between the NFW and LOG models, which begin to diverge only
near the virial radius (note the differing extrapolated $M_{\rm vir}$ values in
Table~\ref{tab:jeanssumm}).
More crucial would be generalized modelling
which fits the fine details of the stellar kinematics in order
to ease out the differences in the total mass profile shapes in the central regions.
In this context, additional data over a wide baseline in radius would also help
to build up the $S/N$ of the constraints.

\subsection{Systematics and mass comparisons}\label{sec:comp}

We next investigate the sensitivity of our mass results to some systematic issues and
to different modelling methods and mass tracers.
We can first examine the results from the different modelling methods and assumptions
we have used thus far.  As seen in Figs.~\ref{fig:modelfitstwo} and \ref{fig:masscomp},
the total mass profile at the outer extent of the data ($\sim 300''$) is fairly well
determined (within $\sim\pm30\%$) regardless of the anisotropy and mass model assumptions,
and of the Jeans method used (cf. also the consistency of the $\Upsilon_{B5}$
benchmark parameter in Table~\ref{tab:jeanssumm}).
Similarly, the inferred value for \Ystar\ is the same at the $\sim \pm 5\%$ level,
and is reassuringly consistent with the independent estimate from stellar populations analysis.
Using the more detailed models of \S\ref{sec:massmod} with their anisotropy constraints,
we update our estimate of the $M/L$ gradient (\S\ref{sec:massinv}) to
$\dML\sim0.1$--$0.2$.

One caveat about all of these Jeans models is that a somewhat arbitrary radial region is
chosen for fitting the central stellar kinematics (thereby normalising the
entire solution), since the models are not sufficiently flexible to reproduce
the high-$S/N$ bumps and wiggles representing fine structure in the stellar data.
The only such model to-date of NGC~4494 was by \citet{2000A&AS..144...53K},
based on spherical non-parametric distribution function fitting.
Using the B+94 stellar kinematics data-set, they found
a circular velocity profile at $\sim 0.5 \Re$ which is slightly different from ours
(perhaps owing to the data differences: see \S\ref{sec:higher}).
Similar results without the benefit of $h_4$ constraints were found by
\citet{2001MNRAS.322..702M}.

It is of critical importance for more fully understanding this system to employ
a more general dynamical model that can fit the stellar data in detail,
while not requiring the {\L}okas-type simplification (see Appendix \ref{app:eqs})
we have imposed on the distribution function.
There are already some indications from our Jeans modelling that other regions of
model space may be permitted (with lower \Ystar\ and higher central DM content).
While more general modelling is outside the scope of this paper,
we can estimate the possible systematic effects from the Jeans smoothing
simplifications.  In the generalized NMAGIC modelling of NGC~3379 (DL+08b),
the halo anisotropy could not be constrained to better than $\Delta \beta \sim \pm 0.3$.
Taking this as a guide, we re-run our NFW-based Jeans models with $\beta_0$ fixed at
either $+0.2$ or $+0.8$, finding that the corresponding systematic uncertainty in $c_{\rm vir}$ is
then $\pm 1.8$, and in $\log M_{\rm vir}$ is $\pm 0.26$.

Our models make powerful use of the assumption that the
PNe and field stars are drawn from the same underlying dynamical population.
This premise was questioned by D+05, who posited that the PNe could
reflect a recent burst of star formation with distinct dynamics from the observed field stars---a
hypothesis with some support from observations
\citep{2006AJ....131..837S,2008arXiv0803.3626F}.
We discussed this issue in more detail in D+07 and \citet{Coccato08},
and here note that NGC~4494,
although a reasonable candidate for a young stellar subpopulation
\citep{2005MNRAS.358..813D}, in fact shows
good
agreement between
its PN and stellar properties (see e.g. \S\ref{sec:spatdist}).

Our $\Lambda$CDM mass models assume a fairly basic collisionless NFW form for the DM distribution.
This profile is bound to be altered by interaction with the collisional baryons;
although the amount and even the sign of the effect are not clear,
the standard view is that the change can be approximated by
adiabatic contraction (e.g. \citealt{2004ApJ...616...16G}).
The effect of this contraction would be to draw more DM into the central regions,
flattening the rotation curve, and thereby requiring a lower stellar $M/L$
(possibly losing the consistency of our solution with the independent value) and a
lower concentration.
A more recent revision of the collisionless DM distribution
(the Einasto model: \citealt{2008MNRAS.387..536G})
would tend to push things in the opposite direction, but more weakly.
Of course, these conventional mass-model pictures could be completely wrong,
as we have seen that a cored-halo model fits our data equally well
(see also e.g.
\citealt{2001AJ....122.2396D,2003MNRAS.341.1109B,2007MNRAS.375..199G}),
and alternative gravity models should also be explored
(e.g. \citealt{2007A&A...476L...1T}).

A final concern, and a perennial problem in modelling any galactic
system, is the symmetry assumption.
The spherical symmetry adopted here has in particular been called repeatedly
into question, and certainly more robust results should eventually be derived
using less restrictive symmetries and more detailed modelling.
NGC~3379 is a similar case to NGC~4494, where its round apparent shape could be
produced by a very flattened system seen near face-on.
DL+08b used flattened particle-based ``NMAGIC'' models of NGC~3379
to explore the simultaneous effects of shape and anisotropy on the mass inferences from PNe.
Intriguingly, they found that relaxing the spherical assumption hardly affects
the inferred halo mass:
the intrinsic flattening and inclination are
strongly limited by the observed rotation field,
and the main source of systematic uncertainty remains the halo anisotropy
(which the available data do not constrain).
Thus the effects of asphericity on mass modelling may not be as important as
previously thought
(cf. \citealt{2006MNRAS.370.1737B}),
although the effects of triaxiality have yet to be fully explored.

It is beyond the scope of the current paper to construct similar advanced models for NGC~4494,
but we do explore a flattened scenario
using similar Jeans machinery to \S\ref{sec:massmod},
as described in Appendix~\ref{app:flatt}.
We find that the inferred DM halo differs little from our spherical conclusion.
It will be the object of a future paper to fully explore
multiple modelling methods with NGC~3379, NGC~4494 and other galaxies,
to get a handle on the robustness of the results to the methods.
But so far, although very general models allow a wider range of solutions,
we have seen no evidence that these models imply a systematic shift in the
preferred mass solutions.

Fully independent mass constraints from another observational tracer such as
GCs or X-ray emitting gas are desirable.
\citet{2004MNRAS.349..535O} analysed {\it XMM-Newton} observations of
NGC~4494, finding a weak component of very soft
($k_{\rm B}T_{\rm X}\sim$~0.2 keV) X-ray emitting gas
in a compact arrangement (within $\sim \Re$).
They proposed that the X-ray underluminosity of this and similar ETGs
could be due to either a weak or diffuse DM halo, or to a young field star population
which has had little time to build up an X-ray halo through winds (though there are
problems with this scenario).
More recently, it was suggested that the apparent emission from hot gas in
such galaxies may in fact be dominated by cataclysmic variables and coronally active
binaries \citep{2006A&A...450..117S,2008arXiv0804.0319R}.

Whether or not it is feasible to measure the halo mass of NGC~4494 using hot gas
is thus a very open question.  Assuming that there is any such gas which could be
studied with deeper observations, it is more likely to be in an outflowing wind state
than in hydrostatic equilibrium, which would greatly complicate the mass analysis
\citep{2006MNRAS.370.1797P}.
Given these caveats, we note that
\citet{2006ApJ...636..698F} did derive a basic mass estimate for NGC~4494 from {\it Chandra}
observations of apparent compact X-ray gas emission.
They found an $M/L$ of $6.2\pm1.9$~$\Upsilon_{B,\odot}$
inside 1~\Re\ (converted to our distance, with minimal uncertainties
estimated from the uncertainty in $T_{\rm X}$),
which is consistent with our finding of $\sim 4.5$~$\Upsilon_{B,\odot}$ at this
radius (see Fig.~\ref{fig:modelfitstwo}).
Note though that a different analysis by \citet{2007ApJ...668..150D} found no trace of X-ray gas in this system.

\section{Implications: ordinary ellipticals revisited}\label{sec:impl}

Given the dynamical solutions found for NGC~4494 in \S\ref{sec:massmod},
we compare them to other galaxies, and
consider the possible implications for DM and for galaxy formation.
We start with a small sample of galaxies consisting of all four ``ordinary''
ellipticals discussed in R+03, whose PN dynamics originally suggested
unexpectedly low DM content: NGC~821, NGC~3379, NGC~4494 and NGC~4697
(with data from the latter presented in \citealt{2001ApJ...563..135M}).
These galaxies have similar central velocity dispersions $\sigma_0 \sim 200 \kms$
and near-$L^*$ luminosities ($M_B \sim -20.3$;
see Table~\ref{tab:galmodels}),
and are of the discy/cuspy ETG sub-type found by N+05 to have lower
DM content on $\sim$~5\Re{} scales than the boxy/cored ETGs.
These galaxies (dubbed the ``R+03 sample'')
are also of the ``fast rotator'' variety of ETG
(\S\ref{sec:rot} and \citealt{2007MNRAS.379..401E}),
which as a class shows hints of less central DM content
than the slow rotators \citep{2006MNRAS.366.1126C}.

\begin{table}
\caption{Mass results for fast-rotator early-type galaxies, based
on extended kinematics data and higher-order LOSVD modelling. The
absolute magnitudes $M_B$ and central velocity dispersions $M_B$
and $\sigma_0$ are adapted from HyperLeda, D+07, and this work.}
\scriptsize
\begin{center} \setlength\tabcolsep{5pt}
\begin{tabular}{lcccccc}
\hline \hline \noalign{\smallskip}

Name & $M_B$ & $\sigma_0(\kms)$ & $\log M_*$ & $\log M_{\rm vir}$ & $c_{\rm vir}$ & $\Upsilon_{\rm vir}$\\
 & & & ($M_{\odot}$) & ($M_{\odot}$) & & ($\Upsilon_{\odot,B}$) \\
\hline\\
NGC 821  & -20.5 & 210 & 11.24 & 14.46 & 2.5 & 1700 \\
NGC 3379 & -19.8 & 207 & 11.01 & 12.17 & 6.8 & 110 \\
NGC 4494 & -20.5 & 150 & 10.99 & 12.06 & 7.6 & 47 \\
NGC 4697 & -20.2 & 174 & 11.14 & 12.80 & 6.6 & 300 \\
\hline
\hline\\
\end{tabular}
\label{tab:galmodels}
\end{center}
\end{table}

Each galaxy has now been modelled using constraints
on higher-order LOSVD moments to handle the mass-anisotropy degeneracy,
and to infer the \Ystar\ component by dynamical means:
NGC~821 using three-integral axisymmetric orbit models
(stellar kinematics only; \citealt{2008arXiv0803.3626F}, but see
also preliminary analysis of PN kinematics in \citealt{2007IAUS..244..289N});
NGC~3379 using spherical orbit models (R+03) and spherical, axisymmetric and quasi-triaxial
particle modelling (DL+08b);
NGC~4494 using kurtosis-based Jeans models;
and NGC~4697 using axisymmetric particle modelling (DL+08a).
One caveat is that these results are tied
to particular assumptions about the form of the DM density profile (NFW or LOG),
which affects the decomposition of the total mass into stars and DM.
E.g. dynamics on its own does not completely rule out a scenario
(though unrealistic) where the stars have
zero mass and the central DM profile is similar to the stellar luminosity profile.

We examine the DM fraction for these galaxies, as well as for theoretical models,
in  \S\ref{sec:frac}.
In \S\ref{sec:halo}, we consider the global DM halo parameters,
and in \S\ref{sec:ext} we place them in the wider context of galaxy formation.

\subsection{The dark matter fraction}\label{sec:frac}

The first basic parameter we consider is the DM fraction within a given radius $r$:
\begin{equation}
f_{\rm DM} (r) \equiv \frac{M_{\rm DM}(r)}{M_{\rm tot}(r)},
\end{equation}
where we find for NGC~4494 from the best-fit NFW and LOG models that at 5~\Re{},
$f_{\rm DM,5}=1-\Upsilon_*/\Upsilon_5 \sim $~0.2--0.5.
For NGC~3379,
the comparable calculation from the R+03 spherical model yields
$f_{\rm DM,5} \sim $~0.2--0.4 (cf. D+07).
The flattened models for NGC~3379 from DL+08b imply
$f_{\rm DM,5} \sim $~0.2--0.6,
while similar models for NGC~4697 give
$f_{\rm DM,5} \sim $~0.3--0.6.
These galaxies are therefore all consistent with a typical value of
$f_{\rm DM} \sim $~0.4 at 5~\Re;
this value is notably lower than one would expect from strong gravitational lensing
studies which typically find $f_{\rm DM} \sim 0.3$ at \Re\
(e.g. \citealt{2007ApJ...667..176G}), i.e. at much smaller distances from the centre.
However, lensing studies so far include very few of the fainter
ordinary galaxies which would be equivalent to this fast rotator sample,
although there are emerging indications of lower $f_{\rm DM}$ systems
\citep[]{2005ApJ...623L...5F,covone08}.
The constraints on NGC~821 (based on stellar kinematics only) do not extend to large
enough radii for a direct comparison, and the models do suggest that this galaxy is different,
with $f_{\rm DM} \sim$~0.5 at 1.5--2~\Re\ already. However, our preliminary analysis
based on PN kinematics (\citealt{2007IAUS..244..289N}) seems to favor a lower $f_{\rm DM}
\sim$~0.4 at 5~\Re, as found for the other three observed galaxies.

On the theoretical side,
N+05 found for a simplified $\Lambda$CDM-based suite of models that one
should expect $f_{\rm DM,5}=$~0.5--0.8, depending on various assumptions about
the galaxy and DM masses.
A similar framework from M{\L}05 predicted $f_{\rm DM,5}\sim0.5$.
Simulations of ETG formation including baryonic processes are now beginning
to produce plausible results, with the schematic pair-merger simulations of D+05
yielding $f_{\rm DM,5} \sim $~0.5--0.7.
The full cosmological simulations of \citet{2007ApJ...658..710N}
and \citet{2007MNRAS.376...39O}
suggest $f_{\rm DM,5} \sim $~0.4 and $\sim$~0.6--0.7, respectively
(though it is not clear if these simulated galaxies are better
analogues of the fast or slow rotators).
This variety of predictions are all roughly consistent with a typical
$f_{\rm DM,5} \sim $~0.5 (a quantity that is presumably not very sensitive
to baryonic effects)---which appears
similar to the empirically-obtained values.

\begin{figure}
\hspace{-0.5cm}
\epsfig{file=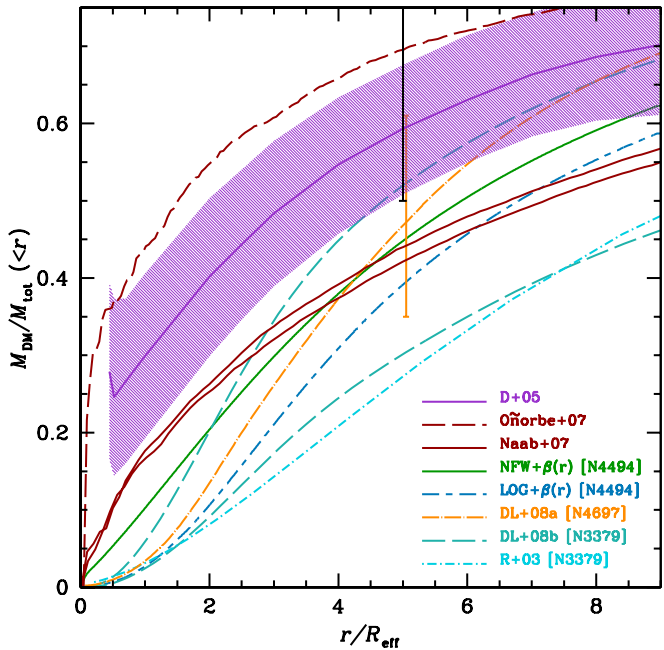,width=9.2cm}
\caption{Cumulative dark matter fraction as a function of radius.
The green solid line shows the best-fit $\beta(r)$ stars+NFW model while
the short-long-dashed blue line is the $\beta(r)$ stars+LOG model for NGC~4494:
typical uncertainties are 20--30\% at 5\Re\ (see Table \ref{tab:jeanssumm}) and drop
to $\sim10\%$ at 1 \Re.
Light green long-dashed and cyan dot-short-dashed lines are the best-fit
results of NGC 3379: DL+08b (model C90 - lower line - and D90 - higher line -)
and R+03, respectively.
Orange dot-long-dashed is the dark matter fraction of NGC 4697 (DL+08a model G with
errorbar showing the DM fraction at 5 \Re\ for models D to J).
The purple line with surrounding shaded region shows the mean results
and scatter from the galaxy merger simulations of D+05.
The red lines show simulations from \citet[solid curves]{2007ApJ...658..710N}
and \citet[dashed curve]{2007MNRAS.376...39O}. The black
error-bar shows the theoretical prediction from N+05 based on $\Lambda$CDM
cosmology and the full range of allowed star formation efficiency
($\epsilon_{\rm SF}=$0.07--1; see Sec.~\ref{sec:halo}).
}
\label{fig:fDM}
\end{figure}

For additional insight, as in D+07 we show the DM fraction as a function of effective radius
in Fig.~\ref{fig:fDM}, with various of the aforementioned empirical and
theoretical results displayed.
The plot is somewhat tricky to interpret because the stellar and DM mass profiles
probably do not scale with luminosity in a homologous way.
E.g. NGC~3379 and the \citet{2007ApJ...658..710N} simulated ETGs have similarly
small values for \Re\ and hence should be compared;
the larger galaxies NGC~4494 and NGC~4697 should presumably be compared to
the D+05 and \citet{2007ApJ...658..710N} simulations.
It thus becomes clear that simulations which include baryonic effects predict
systematically more DM within 5~\Re\ than is inferred observationally, and that
this mismatch becomes even stronger at smaller radii\footnote{Note that the
uncertainty bounds on the observational results are not shown in the Figure for the
sake of clarity.  On a case-by-case basis, few of the galaxies may be formally
inconsistent with the theoretical predictions, but the overall mean pattern
is a systematic offset between observations and theory---while there
is no indication of a systematic shift in the observational results
(see also \S\ref{sec:halo}).}.

While the dark matter fraction $f_{\rm DM,5}$ is a directly-constrained observational
quantity, its meaning may be muddled by the whims of the baryonic processes in
individual galaxies.
For example, NGC~4494 appears to be more DM dominated in its central regions than
NGC~3379 (Fig.~\ref{fig:fDM}), but
the global halo properties of the two galaxies turn out to be very similar---even
though the PN velocity dispersion in NGC~3379 declines much more
steeply than in NGC~4494 (\S\ref{sec:dispsec}).
This suggests that halo VD gradients are
strongly affected by the individual galaxy scale-lengths,
and careful dynamical modelling is necessary to interpret the implications for DM.

\subsection{Halo parameters: the R+03 sample}\label{sec:halo}

For less ambiguous comparisons to theory, one can instead consider directly
the properties of the DM haloes.
A halo's virial properties are not directly constrained by PN kinematics data which
extend to only $\sim 0.1 r_{\rm vir}$,
but we can at least hazard some guesses by assuming {\it a priori} a functional form
for the mass profile to be extrapolated outwards (e.g. the NFW model).
Dynamical modelling results based on NFW haloes were presented for NGC~821,
NGC~3379 and NGC~4494 in \citet{2008arXiv0803.3626F}, R+03 and this paper, respectively.
For NGC~4697, DL+08a used a LOG potential (model H), to which we have fitted
an NFW model {\it post hoc}.
The results are summarized in Table~\ref{tab:galmodels}, where it should be
noted that a cluster mass halo is not credible for the isolated galaxy NGC~821,
and is a product of extrapolating from data extending to only 2 \Re.
However, the joint mass-concentration result for this galaxy is relevant,
as discussed below.

Disregarding NGC~821, we see that the other three galaxies have virial $M/L$
values of $\sim$~50--150~$\Upsilon_{\odot,B}$, and virial DM fractions of
$\sim$91\%--95\%.
Assuming a primordial baryon fraction of 0.17 \citep{2008arXiv0803.0732H},
this implies that the net star formation efficiency was $\sim$~30\%--50\% for
these galaxies.
These results are broadly consistent with the current consensus that star formation
was most efficient in galaxies near the $L^*$ luminosity
(e.g., \citealt{2002ApJ...569..101M,2006MNRAS.368..715M,2007MNRAS.376..841V}).

Rather than measuring the virial mass, the kinematics data are best suited to measuring
the amount of DM within $\sim$~5\Re, and to some extent the detailed profile of DM
with radius in the same region.  In the context of an NFW model, we can determine
jointly the mass and concentration parameters $M_{\rm vir}$ and $c_{\rm vir}$.
Given perfect observational constraints from small to large radius,
we could uniquely determine these parameters as well as $\Upsilon_*$
based on the detailed shape of the $v_{\rm c}(r)$ profile.
With real-world uncertainties in the observations and the dynamical modelling,
there is in practice a strong degeneracy between these parameters, within certain limits
(cf. Fig.~\ref{fig:parcont} and M{\L}05 Fig.~1).
Roughly speaking, a larger value of $M_{\rm vir}$ will require a smaller
value of $c_{\rm vir}$ (in order to decrease the fraction of DM inside 5~\Re)
and a higher value of $\Upsilon_*$ (compensating for the reduced
amount of DM in the very central regions).

We show the mass-concentration results (Table~\ref{tab:galmodels})
for the R+03 sample in Fig.~\ref{fig:confb1} ({\it filled triangles}).
Except for the case of NGC~4494,
the uncertainties are generally not shown because at this juncture
they are not uniformly and robustly determined.
As previously discussed, the uncertainty regions can be expected to follow
a diagonal degeneracy track from low-mass and high-concentration, to
high-mass and low concentration.
This suggests that the NGC~821 result at $M_{\rm vir} \sim 10^{14} M_\odot$ and
$c_{\rm vir} \sim 3$ could be consistent with the other three galaxies
at $M_{\rm vir} \sim 10^{12} M_\odot$ and $c_{\rm vir} \sim 6$.

Next we can compare these empirical results to theoretical expectations.
In a collisionless $\Lambda$CDM universe, a mean relation is expected between
mass and concentration such as the following:
\begin{equation}
c\vir(M\vir)\simeq18\left(\frac{M\vir}{10^{11}\Msun}\right)^{-0.125} ,
\label{cMvir}
\end{equation}
where the relation has a 1~$\sigma$ scatter of  0.14 dex, and is valid for
$z=0$, $\Omega_m=0.3$, $\Omega_{\Lambda}=0.7$, $h=0.7$, $\Delta_{\rm vir}=101$
and $\sigma_8=0.9$ (\citealt{2001MNRAS.321..559B}; N+05).
Although more recent work has revised this relation with higher-resolution
simulations and updated cosmologies
(e.g. \citealt{2007MNRAS.381.1450N,2008arXiv0804.2486D}), the differences below $M_{\rm vir} \sim 10^{14} M_{\odot}$
are only at the level of $\sim$~10\%.
The prediction is thus that higher-mass haloes have lower concentrations, which
conveniently coincides roughly with the direction of the uncertainties in the
observational results, and means that we can make useful comparisons to
theory despite the modelling degeneracies.

As shown in Fig.~\ref{fig:confb1}, the R+03 sample suggests a mass-concentration
relation that is parallel to the theoretical expectation, but offset to lower
concentration values by a factor of $\sim$2.
This discrepancy was also manifested in \S\ref{sec:frac} as a difference in DM fractions.
One could perhaps systematically increase the observed concentrations into
agreement with theory, by postulating that all of the galaxies systematically have
very radial halo anisotropy (e.g. $\beta\gsim0.8$ at $\sim$~7\Re; DL+08b)
which may be technically allowed by the data.
However, this anisotropy is not consistent with the $\Lambda$CDM-based
simulations of D+05, nor perhaps with any reasonable theory of galaxy formation:
for this ETG sample, it appears difficult to
simultaneously fit both the mass and anisotropy profiles of $\Lambda$CDM galaxies.
However, since the anisotropy is not yet well constrained by observations or
by theory, we will continue to focus here on the better-understood
issue of mass profiles.

For comparing with models parameterized by the scale radius $r_s$
and density $\rho_s$ (e.g. Fig.~\ref{fig:parcont}), we find that eqn.~\ref{cMvir}
is equivalent to the following relation:
\begin{equation}
\rho_s \simeq \left(\frac{r_s}{10 {\rm pc}}\right)^{-2/3} \Msun {\rm pc}^{-3} ,
\label{eq:rho_rs}
\end{equation}
where the scatter in $\rho_s$ at fixed $r_s$ is a factor of 1.8.

\subsection{Halo parameters: wider context}\label{sec:ext}

\begin{figure}
\centering
\hspace{-1cm}
\epsfig{file=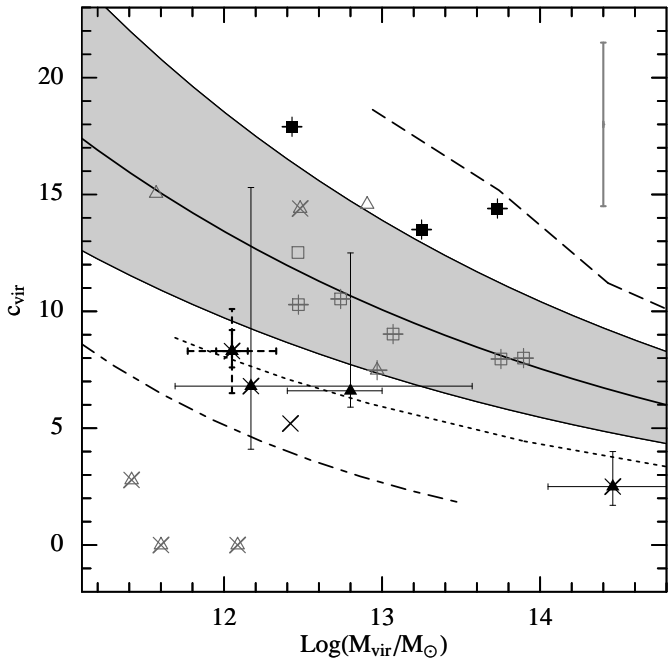,width=8.9cm}
\caption{
Dark matter halo virial mass and concentration parameters.
{\it Triangles} and {\it boxes} mark fast-rotator and slow-rotator ETGs, respectively.
{\it ``$\times$''} and {\it ``$+$''} symbols mark discy/cusped and boxy/cored galaxies respectively, classified
following \citet{1997AJ....114.1771F} and N+05.
The {\it filled symbols} mark detailed ETG dynamical results using PNe and GCs (light error bars, where
available), with {\it solid} and {\it dashed} heavy error bars showing the
statistical and systematic (\S\ref{sec:comp}) uncertainties for NGC~4494,
The {\it open symbols} show the dynamics-based ETG results from N+05,
with error bars in the upper right corner showing the typical uncertainties.
The {\it dashed line} shows the mean result for X-ray bright groups and clusters,
the {\it dot-dashed line} is an inference for late-type galaxies,
and the {\it dotted line} is the trend from weak lensing of all types of galaxies and groups.
The solid curve with shaded region
shows a mean relation expected from $\Lambda$CDM, with its 1~$\sigma$ scatter.
}
\label{fig:confb1}
\end{figure}

The mass-concentration comparison will obviously need much more scrutiny using a large observational
sample along with rigorous dynamical modelling---which is the goal of our ongoing
PN.S survey of ETGs.
But based on the preliminary trend that we see, we may start putting the results in context, starting with
some comparisons to previous results for ETGs.
For the sake of uniformity, we initially include only studies of ETGs using extended
kinematics data fitted to basic NFW-based models as in this paper.
These galaxies are NGC~1399, M87, and NGC4636---in all cases modelled with a
combination of stellar and GC dynamics, though not in all cases directly
constraining the anisotropy
\citep{2001ApJ...553..722R,2006A&A...459..391S,2008A&A...478L..23R}.
Intriguingly, these galaxies' haloes ({\it filled boxes in Fig.~\ref{fig:confb1}})
all have much higher concentrations than the R+03 sample (by a factor of $\sim$3--4),
and are slightly higher than the theoretical expectation,
though consistent within typical modelling uncertainties.

We do not believe this difference is a systematic modelling effect from using PN vs GC dynamics,
as our preliminary PN.S work also suggests a population of high-concentration galaxies
(\citealt{2007IAUS..244..289N}).
The ETG stellar dynamics analysis of \citet{2001AJ....121.1936G} found a main
sequence of high DM halo densities consistent with the three high-concentration galaxies,
along with a few lower-density galaxies similar to the four low-concentration galaxies.
A related analysis from \citet{2007MNRAS.382..657T}
also found a large scatter in DM halo densities for cluster ETGs.

In order to increase the sample size, we next plot the results from N+05, who used
dynamical results on ETGs from the literature to infer NFW-based mass-concentration
parameters.  We classify the galaxies as fast- or slow-rotators based on C+07 where
possible, and otherwise using literature results as described in \S\ref{sec:rot}.
While these mass inferences are not as individually reliable as direct NFW-based
modelling, we do not detect any systematic difference between the two techniques
when comparing overlapping cases.
Interestingly, the main effect of this supplemental sample is to fill in the
``missing'' population of normal-concentration haloes (Fig.~\ref{fig:confb1}).
Examples of such galaxies include NGC~4406, NGC~5128, NGC~5846 and NGC~6703.

When compared to the theoretical prediction,
the overall ETG dynamics-based results suggest a normal population of haloes,
with a systematic bias for fast-rotators to reside in low-concentration haloes,
and slow-rotators in high-concentrations.
Although it is outside the scope of this paper to determine if this
difference is more strongly linked to rotation, luminosity, isophote
shape, environment, etc., Fig. \ref{fig:confb1} suggests that isophote shape is
at least as important as rotation.
The concentration gap seen in the initial sample may be due to selection effects,
e.g. the systems with the most GCs amenable to dynamical study
would be systematically the most massive systems
\citep{2008MNRAS.385..361S}.
But we also note that given no theoretical prejudice,
the data themselves would suggest a shotgun
pattern of masses and concentrations,
with no hint of a systematic decrease of concentration with mass
(if anything, the data suggest an {\it increase}).

Since the initial three high-concentration systems are galaxies at the
centres of massive X-ray-bright groups and clusters,
we compare the results from a large sample
of low-redshift groups and clusters \citep{2008arXiv0804.2486D}, where
we have transformed the halo parameters from $\Delta_{\rm vir}=200$ to
$\Delta_{\rm vir}=101$,
and from $z=0.1$ to $z=0$ by the trend $c_{\rm vir} \sim (1+z)^{-1}$.
As shown in Fig.~\ref{fig:confb1}, this relation is similar to ($\sim$15\% higher than)
the high-concentration dynamical sample.
Other studies using gravitational lensing have also found relatively high
concentrations for clusters
\citep{2007ApJ...654..714H,2007MNRAS.379..190C,2008arXiv0801.1875B,2008arXiv0805.2617B}.

We next broaden the picture to include {\it late-type} galaxies.
There is an emerging consensus that such galaxies near the $L^*$ luminosity
(including the Milky Way) have NFW-based concentrations that are unexpectedly low
(\citealt{2006ApJ...643..804K,2007ApJ...654...27D,2007ApJ...659..149M,2007ApJ...671.1115G,2008arXiv0801.1232X}; see \citealt{2008arXiv0805.1926M} for a differing view).
We take the mass-velocity relation from Fig.~10 of \citet{2007ApJ...659..149M},
and adapt it for Fig.~\ref{fig:confb1},
where we see that it is similar to the low-concentration ETGs ($\sim$25\% lower).

The overall impression we gain from these combined detailed studies of
individual systems is of an ``S''-shaped relation, where the galaxy-scale haloes
follow a trend of relatively low concentrations, and there is a transition at
$M_{\rm vir} \sim 5\times10^{12} M_{\odot}$ to group haloes with higher concentrations.
This transition is comparable to the DM ``dichotomy'' between low- and high-stellar-mass
galaxies found by N+05, occurring at $M_* \sim 2\times10^{11} M_{\odot}$.

As a final piece to the puzzle,
we consider a recent weak-gravitational-lensing study of
both late and early-type galaxies, from galaxy to cluster masses
\citep{2008arXiv0805.2552M}.
The mean trend is shown in Fig.~\ref{fig:confb1} (after conversions
for $\Delta_{\rm vir}$ and $z$) and is seen to roughly coincide with the
apparent trends for late-types and fast-rotator ETGs,
with no indication of the high-concentration trend found by dynamics and by X-rays.

Although there are potential sources of systematic error with
any of these mass-measurement techniques,
and some of the sample sizes are small,
the observational evidence so far agrees on a main population of relatively
low-concentration galaxy-scale DM haloes.
At higher masses, the observational inconsistency might be resolved if there were a
strong selection effect in the X-ray and dynamical studies for high-concentration
systems.
Alternatively, keeping in mind that weak lensing probes the outer regions of
haloes while the other techniques probe the central regions,
the inner haloes of groups and clusters might deviate strongly
from the assumed NFW profiles.

We next consider some theoretical scenarios that might explain these observational
constraints.
The impact of baryons on their host DM haloes could be a significant factor;
e.g. the observed concentration transition occurs at roughly the same mass
scale where a change in the physics of gas cooling and heating is expected
\citep{2006MNRAS.370.1651C}.
As discussed in \S\ref{sec:comp}, the standard model for baryonic modification
of a collisionless DM halo involves adiabatic contraction, which
draws additional DM into the central regions.
The implication for inferring unmodified NFW concentrations would be to
{\it decrease} them, i.e. the observational data points in Fig.~\ref{fig:confb1}
would be shifted slightly down and to the left.
This effect {\it might} bring the slow-rotator group-central ETGs
into agreement with theory, but would make matters worse for the fast-rotator ETGs
and the late-types.
For those, a strong net halo {\it expansion} would be required, perhaps
as a consequence of baryonic mass expulsion
(e.g. \citealt{2004MNRAS.353..829M,2008Sci...319..174M}).
Alternatively, other effects such as dynamical friction might flatten
the very central DM cusp (e.g. \citealt{2008arXiv0808.0195R}), which would not affect
the overall concentration
but could conceivably bias the observational inferences when fitting
NFW mass models.
However,
the low concentrations found by weak lensing could not be ascribed to
baryonic effects, so another explanation is needed.

On a related note, it should be remembered that the NFW profile is not a
prediction for a real halo even in the absence of baryons, as it represents
a composite average of many haloes.
Individual haloes would have departures from the NFW density, which
might systematically connect to galaxy or group type in a way that creates the
appearance of systematic concentration differences.

As already mentioned,
another possibility is that Fig.~\ref{fig:confb1} reflects selection
effects, i.e. the X-ray and GC studies are preferentially
picking up systems that have high central DM concentrations (producing unusually
high X-ray luminosities and rich GC systems; e.g. \citealt{2007A&A...473..715F}).
Because higher DM densities should arise from halo collapse at earlier times
(when the background densities were higher),
the implication is that the high-concentration groups collapsed earlier.
In order to not contradict the generic expectation of hierarchical structure
formation that higher-mass haloes collapse later {\it on average}
(producing trends such as Eq.~\ref{cMvir}),
the small observed subset of high-concentration systems would come from a high-$\sigma$
tail of earlier-collapsing high-mass haloes.
The different collapse times might also couple to baryonic effects,
e.g. adiabatic contraction was stronger at earlier times when the gas content was higher.
Such a scenario could be related to suggestions that
the slow rotators as a class formed in rapid multiple mergers at higher redshifts,
while fast rotators (and presumably late-types) formed at relatively low redshifts,
experiencing a small number of isolated major mergers \citep{2007arXiv0710.0663B}.

Thus, there may be a combination of effects in operation:
a mass dependency of baryonic, collapse, and merger processes in galaxy halo
formation, coupled with an observational bias.
Whether such a combination could quantitatively explain the observed concentration trends
is a question far beyond the scope of this paper,
but we do highlight some potential problems.

One issue is the sheer magnitude of the difference between the high- and low-concentration
populations.
The collapse redshift $z_{\rm c}$ of a DM halo scales roughly with its central
density $\rho_s$ as $(1+z_{\rm c}) \propto \rho_s^{1/3}$
(e.g., \citealt{2001MNRAS.321..559B}).
The difference between the halo densities on the upper- and lower-concentration
sequences is a factor of $\sim 20$, which implies markedly different formation
redshifts at the same halo mass.
E.g. if a typical $10^{13} M_\odot$ halo assembled at
$z_{\rm c} \sim 0.9$ (7~Gyr look-back time),
then its high-concentration counterpart formed at
$z_{\rm c} \sim 4$ (12~Gyr halo age).
These haloes would comprise a very small fraction of the
total population of the same mass
(e.g. \citealt{2006MNRAS.367.1039H}), whose presence should be
difficult to miss by the dynamical and X-ray observations.

The other problem is that the observed concentrations for the main halo population
are too low.  Unless there is a vast population of higher-concentration
``dark'' galaxy-scale haloes that are still missed by observational surveys, the remaining
avenue is alteration of the cosmological parameters.
The obvious candidate, $\sigma_8$, would have to be lowered
substantially to produce the low-concentration trend,
perhaps to implausible values \citep{2007ApJ...659..149M}.
An alternative is to fine-tune the mass dependence of the DM power spectrum
(e.g. \citealt{2002ApJ...572...34A,2003ApJ...584..566M,2008A&A...486...35F,2008arXiv0805.1926M}).

The foregoing discussion is a speculative attempt to explain the entire observational
picture in the context of the $\Lambda$CDM paradigm.
However, there is still little direct evidence for $\Lambda$CDM halo profiles
among all the observational studies of galactic and super-galactic mass distributions;
e.g. in NGC~4494, the NFW and LOG potentials fit the data equally well.
The possibility remains that the universe is dominated by an alternative form
of DM or of gravity, which might more naturally explain the observations
summarized in Fig.~\ref{fig:confb1}.

\section{Conclusions}\label{sec:concl}

We have presented observations of 255 PNe with line-of-sight velocities in NGC~4494,
out to $\sim$~7\Re.
The basic spatial and kinematical properties of these PNe (including rotation,
dispersion and kurtosis) agree well with the field stars, and do not support
a distinct origin for the PN progenitors.
The mean rotation is low but the system may become rotationally dominated in its
outer parts, and we also see evidence for a {\it kinematically-decoupled halo} which
may be a merger signature akin to kinematically-decoupled cores.
The projected velocity dispersion profile declines with the radius, although not
as steeply as in NGC~3379.

We have constructed spherical dynamical models of the system, including
a pseudo-inversion Jeans mass model
as well as a more standard Jeans analysis using
multi-component mass models that include NFW or logarithmic DM haloes,
with fourth-order moments being used to constrain the orbital anisotropy.
Both approaches give similar results for the mass profile within the radial range
constrained by the data.
Some DM is required by the data; our best-fit solution has a radially anisotropic
stellar halo, a plausible stellar $M/L$,
and a DM halo with a fairly low concentration.

We review the halo parameters determined observationally for other ordinary
early-type galaxies such as NGC~4494, as well as
for galaxies and groups from the literature.
There are preliminary indications that most haloes follow
a similar mass-concentration sequence which implies rather low DM densities,
accompanied by a population of bright slow-rotator ETGs
in group-scale haloes with much higher DM densities.
We discuss some possible theoretical scenarios for tying these observations together.
Clarifying the situation will require a broad observational survey of galaxies
of different types and environments, combined with
rigorous dynamical models and improved simulations of galaxy formation.
It is the focus of our ongoing PN.S survey
(e.g. \citealt{Coccato08}) to fill in the critical observational gap
of the halo properties of ordinary early-type galaxies.
The primary programme includes 12 ellipticals representing a range of sub-types and
environments, with an extended survey of 40 early-types in progress,
limited only by volume and magnitude.

\section*{Acknowledgments}
We would like to thank the Isaac Newton Group staff on La Palma for supporting the \PNS\ over the years.
We thank Gary Mamon for many stimulating discussions on data modelling and comparison with simulations,
Ralf Bender, Stacy McGaugh, Thorsten Naab and Jos\'e O\~norbe
for providing their results in tabular form,
Mark Wilkinson for further conversations,
and the referee for constructive comments.
NRN has been funded by CORDIS within FP6 with a Marie Curie European Reintegration Grant, contr. n. MERG-FP6-CT-2005-014774, co-funded by INAF.
AJR was supported by the National Science Foundation Grant AST-0507729 and by
the FONDAP Center for Astrophysics CONICYT 15010003.
EOS acknowledges support from NASA through Chandra Award Number AR5-6012X,
and thanks Brian McLeod for his assistance with Megacam data reduction.
This research has made use of the NASA/IPAC Extragalactic Database (NED) which is operated by the Jet Propulsion Laboratory, California Institute of Technology, under contract with the National Aeronautics and Space Administration.
We acknowledge the usage of the HyperLeda database (http://leda.univ-lyon1.fr).

\appendix

\section[]{Photometry of NGC~4494}\label{app:photmega}
We have obtained a composite photometric profile of NGC~4494 by combining new Megacam
observations, detailed in \S\ref{sec:spatdist}, with literature data which include
{\it HST}-based observations in the $V$ and $I$ bands \citep{2005AJ....129.2138L},
and ground-based CCD observations in $BVI$ \citep[hereafter G+94]{1994A&AS..104..179G}.
The {\it HST} data are used inside 4.3$''$, then G+94 to 32$''$, and Megacam
outside 32$''$. The $a_4$ isophote shape parameter (\citealt{1988A&AS...74..385B})
is not available for the HST data.
A small color
gradient is found between the G+94 and the Megacam data at the outer data points,
probably due to the background subtraction.

We derive some global photometric parameters for the galaxy by averaging the
radially-binned values outside 5$''$ (the central disc region), and weighted
by bin luminosities.  Thus we find $\overline{\epsilon} = 0.162 \pm 0.001$,
$\overline{a_4} = 0.16 \pm 0.07$ and $\overline{PA} = -0.9^\circ \pm 0.3^\circ$.

\begin{table}
\caption{Combined $V$-band surface photometry of NGC~4494}
\begin{center}
\setlength\tabcolsep{5pt}
\begin{tabular}{cccccc rccccc}
\hline \hline
\noalign{\smallskip}

$R_m$ & $\mu_V \pm \Delta \mu_V$ & $\epsilon$  & $a_4\pm \Delta a_4$& PA\\
(arcsec) & (mag~arcsec$^{-2}$) & & (degree)\\
\hline\\
0.020 & 13.077 $\pm$ 0.000 & 0.187 &   - &  -   \\
0.041 & 13.406 $\pm$ 0.028 & 0.187 &   - &  -    \\
0.076 & 13.920 $\pm$ 0.005 & 0.306 &   - &  -    \\
0.103 & 14.168 $\pm$ 0.025 & 0.436 &   - &  -     \\
0.138 & 14.440 $\pm$ 0.036 & 0.429 &   - &  -     \\
0.173 & 14.666 $\pm$ 0.034 & 0.424 &   - &  -     \\
0.233 & 14.874 $\pm$ 0.024 & 0.273 &   - &  -     \\
0.295 & 15.054 $\pm$ 0.012 & 0.144 &   - &  -     \\
0.349 & 15.170 $\pm$ 0.007 & 0.082 &   - &  -     \\
0.393 & 15.244 $\pm$ 0.008 & 0.078 &   - &  -     \\
0.437 & 15.344 $\pm$ 0.007 & 0.076 &   - &  -     \\
0.481 & 15.385 $\pm$ 0.006 & 0.076 &   - &  -    \\
0.523 & 15.427 $\pm$ 0.005 & 0.082 &   - &  -    \\
0.533 & 15.470 $\pm$ 0.014 & 0.113 &   - &  -    \\
0.612 & 15.577 $\pm$ 0.011 & 0.152 &   - &  -    \\
0.688 & 15.649 $\pm$ 0.007 & 0.229 &   - &  -     \\
0.796 & 15.790 $\pm$ 0.006 & 0.254 &   - &  -     \\
0.929 & 15.952 $\pm$ 0.006 & 0.265 &   - &  -     \\
1.092 & 16.127 $\pm$ 0.005 & 0.266 &   - &  -     \\
1.291 & 16.330 $\pm$ 0.005 & 0.259 &   - &  -     \\
1.531 & 16.537 $\pm$ 0.005 & 0.247 &   - &  -     \\
1.817 & 16.753 $\pm$ 0.004 & 0.234 &   - &  -    \\
2.146 & 16.969 $\pm$ 0.004 & 0.228 &   - &  -     \\
2.563 & 17.210 $\pm$ 0.004 & 0.204 &   - &  -    \\
3.053 & 17.462 $\pm$ 0.003 & 0.184 &   - &  -    \\
3.649 & 17.716 $\pm$ 0.003 & 0.158 &   - &  -   \\
4.331 & 17.953 $\pm$ 0.003 & 0.143 &      0.54 $\pm$    0.17 &     7.1 $\pm$ 0.3 \\
5.184 & 18.210 $\pm$ 0.003 & 0.110 &      0.49 $\pm$    0.26 &     5.2 $\pm$ 0.3 \\
5.703 & 18.340 $\pm$ 0.003 & 0.129 &      0.45 $\pm$    0.19 &     4.8 $\pm$ 0.4 \\
6.273 & 18.470 $\pm$ 0.004 & 0.118 &      0.24 $\pm$    0.26 &     4.1 $\pm$ 0.3 \\
6.900 & 18.597 $\pm$ 0.004 & 0.128 &      0.24 $\pm$    0.29 &     2.9 $\pm$ 0.3 \\
7.590 & 18.718 $\pm$ 0.004 & 0.133 &      0.63 $\pm$    0.26 &     1.9 $\pm$ 0.2 \\
8.349 & 18.830 $\pm$ 0.005 & 0.135 &      0.24 $\pm$    0.30 &    0.9 $\pm$ 0.2 \\
9.184 & 18.946 $\pm$ 0.004 & 0.146 &      0.17 $\pm$    0.17 &    0.7 $\pm$ 0.2 \\
10.103 & 19.059 $\pm$ 0.005 & 0.152 &    0.09 $\pm$    0.18  &    0.8 $\pm$ 0.2 \\
11.113 & 19.170 $\pm$ 0.005 & 0.157 &     0.37 $\pm$    0.18 &    0.8 $\pm$ 0.2 \\
12.224 & 19.291 $\pm$ 0.006 & 0.167 &    0.013 $\pm$    0.14 &    0.7 $\pm$ 0.2 \\
13.447 & 19.413 $\pm$ 0.007 & 0.168 &   -0.07 $\pm$    0.13  &    0.5 $\pm$ 0.2 \\
14.791 & 19.544 $\pm$ 0.007 & 0.170 &     0.11 $\pm$    0.12 &    0.2 $\pm$ 0.2 \\
16.270 & 19.678 $\pm$ 0.008 & 0.174 &    0.03 $\pm$    0.15  &   0.1 $\pm$ 0.1 \\
17.897 & 19.810 $\pm$ 0.009 & 0.175 &     0.14 $\pm$    0.13 &   -0.2 $\pm$ 0.1 \\
19.687 & 19.949 $\pm$ 0.011 & 0.173 &    0.06 $\pm$    0.10  &   -0.6 $\pm$ 0.1 \\
21.656 & 20.085 $\pm$ 0.012 & 0.185 &     0.15 $\pm$    0.11 &   -1.0 $\pm$ 0.1 \\
23.821 & 20.236 $\pm$ 0.014 & 0.183 &    0.08 $\pm$   0.09  &    -1.3 $\pm$ 0.2 \\
26.204 & 20.384 $\pm$ 0.016 & 0.180 &     0.33 $\pm$   0.09  &    -1.5 $\pm$ 0.2 \\
28.824 & 20.541 $\pm$ 0.018 & 0.176 &     0.14 $\pm$    0.11 &    -1.6 $\pm$ 0.2 \\
31.706 & 20.715 $\pm$ 0.021 & 0.171 &     0.25 $\pm$   0.10  &    -1.3 $\pm$ 0.2 \\
36.729 & 21.055 $\pm$ 0.024 & 0.149 &     0.48 $\pm$  0.08   &  -0.01 $\pm$ 0.3 \\
44.130 & 21.44 $\pm$ 0.03 & 0.147   &    0.06 $\pm$   0.06   &    -3.2 $\pm$ 0.2 \\
53.266 & 21.82 $\pm$ 0.05 & 0.137   &     0.43 $\pm$  0.11   &    -4.3 $\pm$ 0.4 \\
63.918 & 22.18 $\pm$ 0.07 & 0.137   &      1.1 $\pm$  0.11   &    0.8 $\pm$ 0.4 \\
77.203 & 22.58 $\pm$ 0.10 & 0.126   &     0.15 $\pm$  0.11   &    -7.3 $\pm$ 0.5 \\
91.818 & 23.02 $\pm$ 0.15 & 0.141   &     0.47 $\pm$  0.11   &    -11.4 $\pm$ 0.5 \\
110.18 & 23.49 $\pm$ 0.23 & 0.141   &     0.29 $\pm$  0.26   &    -19.4 $\pm$ 1.5 \\
132.22 & 24.1 $\pm$ 0.4 & 0.141     &     0.46 $\pm$  0.21   &    -12.4 $\pm$ 1.1 \\
158.66 & 24.6 $\pm$ 0.6 & 0.141     &     -1.1 $\pm$  0.3    &    -9.6 $\pm$ 1.2 \\
189.26 & 25.1 $\pm$ 1.0 & 0.151     &     -1.3 $\pm$  0.4    &     2.2 $\pm$ 1.5 \\
227.11 & 25.7 $\pm$ 1.7 & 0.151     &     -1.1 $\pm$  1.0    &     2.2 $\pm$ 3.7 \\
272.53 & 26.1 $\pm$ 2.5 & 0.151     &      -           &     2.2 $\pm$ 8.6 \\
\hline\\
\end{tabular}\label{tab:SB}
\end{center}
\end{table}

\section[]{Jeans equations} \label{app:eqs}

Here we present the second- and fourth-moment Jeans equations and projections
to velocity dispersion $\sigma_{\rm los}$ and kurtosis $\kappa_{\rm los}$,
as needed in \S\ref{sec:dynmeth},
and closely following \citet{lok02}.
It is the first time this particular method has been applied to an elliptical galaxy.

We assume that the system is spherically symmetric and that there are no net streaming
motions (e.g. no rotation) so that the odd velocity moments vanish and the different
component of the dispersion velocity tensor are $\sigma_r^2=\overline{v_r^2}$,
$\sigma_\theta^2=\overline{v_\theta^2}$ and
$\sigma_\phi^2=\overline{v_\phi^2}$, with $\sigma_\theta^2=\sigma_\phi^2$ by symmetry.
Under these assumption the Jeans equations reduce to the simple
radial equation (e.g. \citealt{1982MNRAS.200..361B}):
\begin{equation}
\label{eq1}
\frac{\rm d}{{\rm d} r} (j \sigma_r^2) + \frac{2 \beta}{r} j \sigma_r^2 = - j \frac{{\rm d} \Phi}{{\rm d} r} \ ,
\end{equation}
where $j_*(r)$ is the 3D number density distribution of the tracer population
(e.g. the stars or PNe)
and $\Phi(r)$ is the gravitational potential $G M(r)/r^2$ with $G$ the gravitational constant.
In equation (\ref{eq1}), the anisotropy parameter,
$\beta \equiv 1 - \sigma_\theta^2/\sigma_r^2$,
can be either constant or variable with the
radius $r$. In the first case the equation is simply written as

\begin{equation}
\label{eq2}
j_* \sigma_r^2 (\beta={\rm const})= r^{-2 \beta}
    \int_r^\infty r'^{2 \beta} j_* \frac{{\rm d} \Phi}{{\rm d} r'} \ {\rm d}r'
\end{equation}
under the boundary condition
$\sigma_r \rightarrow 0$ at $r \rightarrow \infty$.
In the second case the general solution is
\begin{equation}
\label{eq3}
j_* \sigma_r^2 = \int_r^\infty j_* \frac{{\rm d} \Phi}{{\rm d} r'} \int_{r'}^\infty 2 \frac{\beta(r'')}{r''} {\rm d}r'' {\rm d}r'
\end{equation}
under the same conditions.

The line-of-sight velocity dispersion is obtained from the 3D velocity dispersion
by integrating along the line of sight \citep{1982MNRAS.200..361B}
\begin{equation}
\label{eq4}
  \sigma_{\rm los}^2 (R) = \frac{2}{I(R)} \int_{R}^{\infty}
  \left( 1-\beta \frac{R^2}{r^2} \right) \frac{j_* \,
  \sigma_r^2 \,r}{\sqrt{r^2 - R^2}} \,{\rm d} r \ ,
\end{equation}
where $I(R)$ is the surface density of the tracer and $R$ is the projected radius.
For various simple functional forms for $\beta(r)$ (e.g. $\beta$ constant),
the calculation of $\sigma_{\rm los}$ can be reduced to a one-dimensional numerical
integration:
\begin{equation}
  \sigma_{\rm los}^2(R) = \frac{2 G}{I(R)} \int_R^\infty \frac{K j_* M dr}{r} ,
\end{equation}
where $K(r)$ is a special function depending on $\beta(r)$ (see Appendix A of M{\L}05).

To address the well-known mass-anisotropy degeneracy,
we consider higher-order moments of the velocity distribution.
For the fourth-order moments, the three distinct components $\overline{v_r^4}$, $\overline{v_\theta^4}=\overline{v_\phi^4}$
and $\overline{v_r^2 v_\theta^2}=\overline{v_r^2 v_\phi^2}$ are related by two higher
order Jeans equations \citep{1990AJ.....99.1548M}.

In order to solve these equations we need additional information about the distribution function.
We therefore restrict ourselves here to functions which can be constructed from the
energy-dependent distribution function by multiplying it by a function of
angular momentum $f(E, L) = f_0 (E) L^{-2 \beta}$ with $\beta =\rm const$ .
This is an ansatz widely used (\citealt{DF1}, \citealt{DF2}, \citealt{DF3}, \citealt{DF4}),
which has the advantage of being easy to integrate even though it does not allow to generalize
to the case of $\beta=\beta(r)$ for the 4$^{\rm th}$-order moment.\\
The two Jeans equations then reduce to one:
\begin{equation}
\label{eq5}
\frac{\rm d}{{\rm d}r}(j_*\overline{v_r^4})+\frac{2 \beta}{r}
j_* \overline{v_r^4} + 3 j_* \sigma_r^2 \frac{{\rm d}\Phi}{{\rm d} r}=0 ,
\end{equation}
whose solution is:
\begin{equation}
\label{eq6}
j_* \overline{v_r^4} = 3 r^{-2 \beta}
\int_r^\infty r'^{2 \beta} j_* \sigma_r^2
\frac{{\rm d} \Phi}{{\rm d} r'} \ {\rm d} r'
\end{equation}
(\citealt{lok02}).
By projection, we obtain the line-of-sight fourth moment
\begin{equation}
\label{eq7}
\overline{v_{\rm los}^4} (R) = \frac{2}{I(R)} \int_{R}^{\infty} \left(1 - 2 \beta \frac{R^2}{r^2} + \frac{\beta(1+\beta)}{2}
  \frac{R^4}{r^4}\right)
\frac{j_* \, \overline{v_r^4} \,r}{\sqrt{r^2 - R^2}}
\,{\rm d}r ,
\end{equation}
which we calculate numerically.

The fourth projected moment is used together with $\sigma_{\rm los}^4$ in order to
obtain the reduced projected kurtosis:
\begin{equation}
\label{eq8}
\kappa_{\rm los} (R) = \frac{\overline{v_{\rm los}^4} (R)}
{\sigma_{\rm los}^4 (R)}-3,
\end{equation}
which takes the value of 0 for a Gaussian distribution.
The advantage of using the kurtosis is that it can be easily derived from a
discrete radial velocity distribution such as
our PN data set (see Section \ref{sec:higher}).

One useful corollary of this formalism is that if $\sigma_{\rm los}$ is
observationally constant with the radius $R$, and one assumes that the
anisotropy is constant, then the internal kurtosis $\kappa(r)=0$, and
one can calculate a
one-to-one relation between $\kappa_{\rm los}$ and $\beta$.
Therefore one can estimate the anisotropy as a direct deprojection of
the observations, with no dynamical modelling necessary.
This approach is of course not applicable to NGC~4494 with its
declining VD profile.
For reference, the formulae are the following:
\begin{equation}
\kappa_{\rm los} (R)= 3 \left(\frac{I_0 I_4}{I_2^2} - 1\right) ,
\end{equation}
\begin{equation}
I_i \equiv \int_R^\infty \frac{f_i j_* r dr}{\sqrt{r^2-R^2}} ,
\end{equation}
\begin{equation}
f_i = \left \{
\begin{array}{ll}
\displaystyle
1 & \hbox{(i=0)} \ , \\ \\
\displaystyle
1-\beta\frac{R^2}{r^2} &  \hbox{(i=2)} \ , \\
\displaystyle
1-2\beta\frac{R^2}{r^2}+\frac{\beta(1+\beta)}{2}\frac{R^4}{r^4} & \hbox{(i=4)} \ .
\end{array}
\right.
\end{equation}

\section[]{Flattened model}\label{app:flatt}
Here we present an alternative model needed for \S\ref{sec:comp},
wherein the galaxy is assumed to be axisymmetric and intrinsically flattened but apparently round
because it is seen near face-on.
To simplify the analysis, we study the dynamics along the galaxy's {\it observed minor axis only}, where the
motions in the galaxy's equatorial angular direction $\phi$
do not contribute to the observed line-of-sight velocity dispersion $\sigma_{\rm los}$.

Our starting point for the Jeans analysis is an approximate relation between the vertical velocity
dispersion and potential in a flattened galaxy:
\begin{equation}
\label{eqz}
\sigma_Z^2(R,Z) =
    \frac{1}{j_*} \int_Z^\infty  j_* \frac{\partial \Phi}{\partial Z} \ dZ ,
\label{eq:sigz}
\end{equation}
where $R$ is the equatorial radius and $Z$ is the vertical height
in the galaxy's reference frame
(cf. \citealt{1987gady.book.....B}, eqn. 4-36).
Since we are observing along the minor axis, we will not be concerned with the azimuthal dispersion
$\sigma_\phi$, while for the radial dispersion $\sigma_R$ we adopt an anisotropy parametrisation:
$\beta_{RZ}\equiv 1-\sigma_Z^2/\sigma_R^2$.

\begin{figure}
\centering
\epsfig{file=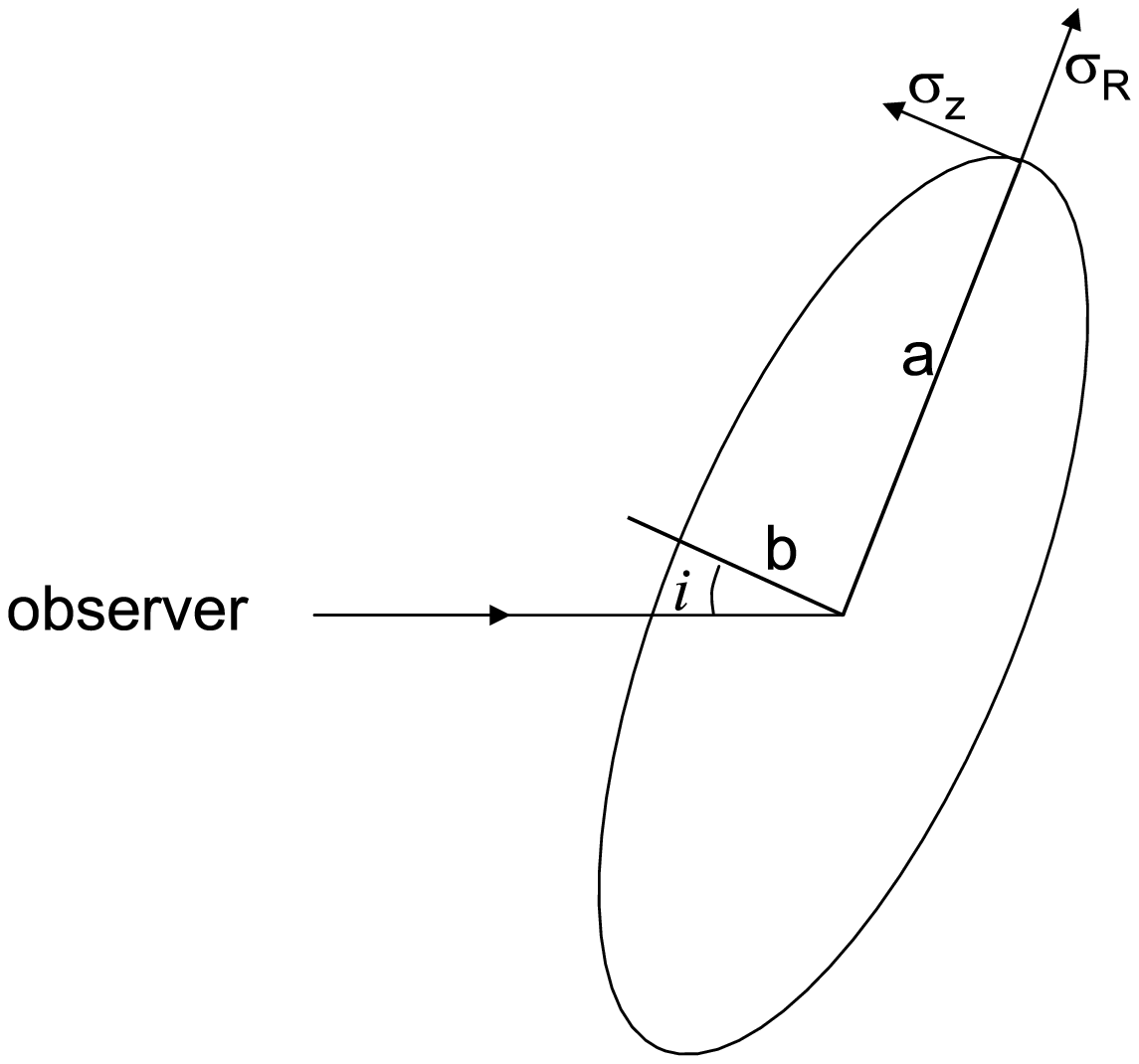,width=7.5cm,angle=-90}
\caption{Schematic geometry of the flattened galaxy model, viewed from a line-of-sight in the
plane of the sky.
Shown are the inclination $i$, the intrinsic major and minor axis lengths $a$ and $b$,
corresponding to radial and vertical velocity dispersions $\sigma_R$ and $\sigma_Z$.
The galaxy is axisymmetric in the equatorial plane (i.e. appears circular when $i=0^\circ$).}
\label{fig:incl}
\end{figure}

The local contribution to the dispersion along the line-of-sight $z$ is:
\begin{equation}
\sigma_z^2=\sigma_R^2~\sin^2 i ~\sin^2 \phi + \sigma_{\phi}^2~\sin^2 i ~\cos^2 \phi + \sigma_{\rm Z}^2 ~\cos^2 i ,
\label{eq:siglos}
\end{equation}
where $i$ is the galaxy's inclination relative to $z$
(see Fig.~\ref{fig:incl} and c.f. \citealt{1997MNRAS.287...35R} eqn. 17).
Along the minor axis, $\phi=90^\circ$ and we have:
\begin{equation}
\sigma_z^2 = \sigma^2_R \sin^2 i + \sigma_Z^2 \cos^2 i  = \sigma_Z^2 \left(\frac{\sin^2 i}{1-\beta_{RZ}} + \cos^2 i\right) .
\label{eq:siglos2}
\end{equation}
The observed dispersion in the sky plane $(x,y)$ is then calculated by:
\begin{equation}
\sigma^2_{\rm los}(x,y) = \frac{1}{I(x,y)}\int_{-\infty}^\infty j_* \sigma_z^2 dz .
\label{eq:siglosZ}
\end{equation}
With a luminosity model $j_*(R,Z$) and some assumptions for $i$ and $\beta_{RZ}$, we now have sufficent
equations to link $\Phi$ to $\sigma_{\rm los}$.

We derive the intrinsic ellipticity $\epsilon_{\rm 3D}$ from the observed ellipticity
$\epsilon_{\rm 2D}$ and an assumed inclination $i$ using the following equation:
\begin{equation}
(1-\epsilon_{\rm 2D})^2=(1-\epsilon_{\rm 3D})^2~\sin^2i+\cos^2i
\label{eq:ellapp}
\end{equation}
(\citealt{1998gaas.book.....B}, eqn 4.29).
For a razor-thin disc ($\epsilon_{\rm 3D}=1$), the minimum inclination allowed by the
observed flattening of NGC~4494 ($\epsilon_{\rm 2D} \simeq 0.14$ for $R \gsim 0.1 \Re$)
is $i=31^\circ$.
More realistically, the most flattened ETGs observed have $\epsilon_{\rm 2D}\sim0.75$
\citep{1998gaas.book.....B},
so as an extreme alternative to the spherical case, we adopt
$\epsilon_{\rm 3D}=0.75$~$(b/a=0.25)$ and $i=32^\circ$.
For equation~\ref{eq:siglos2}, we approximate this to $i=30^\circ$ and find:
\begin{equation}
\sigma_z^2 =  \sigma_Z^2 \left[\frac{1}{4(1-\beta_{RZ})} + \frac{3}{4}\right] .
\label{eq:siglos3}
\end{equation}

We next assume the vertical anisotropy $\beta_{RZ}$ to be constant throughout the galaxy,
and therefore make use of the SAURON dynamical results for galaxies' central regions
to make an {\it a priori} estimate for $\beta_{RZ}$.
From C+07 Fig.~7 we find that $\beta_{RZ}\sim0.5$ is typical for galaxies with
the flattening of NGC~4494 ($\epsilon_{\rm 3D}=0.75$).
From equation~\ref{eq:siglos3}, this implies $\sigma_z = 1.1 \times \sigma_Z$,
i.e. the line-of-sight dispersion is closely related to the intrinsic vertical dispersion.
The anisotropic flattening of the galaxy means $\sigma_Z$ is relatively low, which is
exactly the scenario we are attempting to explore for explaining the low observed values of $\sigma_{\rm los}$.
Note also that higher values for $\beta_{RZ}$ will for the same data
imply {\it lower} intrinsic $\sigma_Z$ and thus lower mass since the contribution
to $\sigma_z$ from $\sigma_R$ will increase.

As a consistency check on our model assumptions,
we consider the rotation dominance parameter within \Re, $(V/\sigma)_{\rm e}$.
For the SAURON galaxies, this parameter correlates well with the ellipticity when both
parameters are deprojected (C+07 Fig.~3).
For an NGC~4494 observed value of $(V/\sigma)_{\rm e} \approx 0.26$ (\S\ref{sec:rot}),
we search for values of
$i$ and $\beta_{RZ}$ that bring the galaxy into agreement with the
correlation (cf. C+07 eq. 12).
We find $i \sim 40^\circ$ and $\beta_{RZ}\sim 0.2$ with intrinsic
$\epsilon_{\rm 3D} \sim 0.4$ and $(V/\sigma)_{\rm e} \sim 0.4$,
suggesting that the true parameters of NGC~4494 could well be intermediate to the
extreme spherical and flattened cases that we are considering.
Note also that advanced axisymmetric modelling of the comparable galaxies
NGC~3379 and NGC~4697 found that $\beta_{RZ}$ increases from $\sim 0.3$
in the central regions to $\sim 0.5$ in the outer parts
\citep{2008MNRAS.385.1729D,2008arXiv0804.3350D}.

Next we follow the same steps as in the spherical case (\S\ref{sec:dynmeth}) to explore a
series of galaxy+DM mass models and compare them to the data.
For $j_*(z)$ in Eqn.~\ref{eq:siglosZ} we adopt a \citet{1990ApJ...356..359H} profile for simplicity
with the approximation that the Hernquist radius is the usual $r_{\rm H}=0.55 \Re$ and
$r=\sqrt{R^2+(z/q)^2}$, with $q=0.25$, in order to account for the flattening along the
line-of-sight.

Recall that we are now comparing to $v_{\rm rms}$ data along the minor axis only, with
no attempt to fit the kurtosis.
The best-fit dark matter halo has $\Ystar=3.9\Ysol$, $\log M_{\rm DM}=11.91\pm0.20$ (the dark mass), a concentration parameter $c_{\rm d}=2\pm1$, and a total virial ratio $f_{\rm vir}=8\pm5$. The latter is lower than the lowest value obtained for the spherical case. As a comparison, we found $\Upsilon(R\vir)=35$ and $\Upsilon_{B5}=5\pm1$, much lower than the values found in Table \ref{tab:jeanssumm}.
We therefore find that a face-on flattened configuration is not sufficient to produce
a very massive DM halo solution for NGC~4494.

\bibliography{napolitano_PNS_r2}

\end{document}